%% file: SpecNC.tex
\documentclass[rivista]{cimento} 

\title{New Hadronic Spectroscopy}
\author{N.~Drenska[ab], R.~Faccini[ab], F. Piccinini[c], A.~Polosa[b], F.~Renga[ab],
       \atque C. Sabelli[ab]}
\instlist{\inst[A]{Department of Physics, Universit\`a di  Roma `Sapienza', Piazzale A. Moro 2, I-00185 Roma, Italy},\inst[B]{INFN Roma, Piazzale A. Moro 2, I-00185 Roma, Italy},\inst[C]{INFN Pavia,  Via A. Bassi 6, I-27100 Pavia, Italy}}
\PACSes{\PACSit{14.40.Rt}{Exotic mesons}
\PACSit{14.40.Pq}{Heavy quarkonia}

}

\usepackage[english]{babel}
\usepackage{amssymb}
\usepackage{amsfonts}
\usepackage{amsthm}
\usepackage{multirow}
\usepackage{axodraw}
\usepackage{fancyhdr}
\usepackage{empheq}
\usepackage{mathbbol}
\usepackage{wasysym}
\usepackage{pstricks}
\usepackage{pstricks}
\usepackage{color}
\usepackage{bbm}
\usepackage{graphicx}
\usepackage{epsfig}
\RequirePackage{xspace}

\usepackage{pifont}
 \usepackage{lscape}
 \usepackage{rotating}
 \usepackage[format=hang]{subfig}
\def\qq{\mathbbmss{q}}

\def\qu{{q_{_1}}}
\def\qd{{q_{_2}}}

\newcommand{\ud}{\mathrm{d}}

\def\BXK{\ensuremath{B\to X K }}

\def\epem {\ensuremath{e^+e^-}\xspace}
\def\bbbar {\ensuremath{b\overline b}\xspace}

\def\invpb {\ensuremath{\mbox{\,pb}^{-1}}\xspace}

\def\invfb   {\ensuremath{\mbox{\,fb}^{-1}}\xspace}
\def\invab   {\ensuremath{\mbox{\,ab}^{-1}}\xspace}
\newcommand{\dedx}{\ensuremath{\mathrm{d}\hspace{-0.1em}E/\mathrm{d}x}\xspace}
\def\mum  {\ensuremath{{\,\mu\rm m}}\xspace}
\def\pipi  {\ensuremath{\pi^+\pi^-}\xspace}

\def\qqbar {\ensuremath{q\overline q}\xspace}
\def\mumu       {\ensuremath{\mu^+\mu^-}\xspace}
\def\Bbar    {\kern 0.18em\overline{\kern -0.18em B}{}\xspace}
\def\BB      {\ensuremath{B\Bbar}\xspace}

\begin{document}

\maketitle

\begin{abstract}
In the past few years the field of hadron spectroscopy has seen renewed interest due to the pubblication, initially mostly from $B$-Factories, of evidences of states that do not match regular spectroscopy, but are rather candidates for bound states with additional quarks or gluons. 
A huge effort in understanding the nature of this new states and in building a new spectroscopy is ongoing. This report reviews the experimental and theoretical state of the art on heavy quarkonium exotic spectroscopy, with particular attention on the steps towards a global picture.
\end{abstract}

\tableofcontents

\clearpage

\input{introduction}
\input{theory}

\input{experiments}

\input{charmonium}

\input{bottomonium}
\input{conclusion}

\emph{{\bf Aknowledgements}}.
We wish to thank C.~Bignamini and B.~Grinstein for fruitful collaboration, and T.~Burns for comments and suggestions on the manuscript.

\bibliography{SpecNC}                                                                                              
                         
\end{document}

%% file: introduction.tex
\section{Introduction}

The discovery of the $J/\psi$ in 1974 has been of pivotal importance for the development of the Standard Model of particles. On the one hand its existence confirmed the GIM~\cite{gim} prediction of a fourth quark, the charm. With the addition of the charm to the $u,d,s$ quarks, the problem of flavor changing neutral currents in the Cabibbo theory was removed and a  $2\times 2$ mixing matrix of quarks was found for the first time. On the other hand the impressive narrowness of the $J/\psi$ (about $90$~keV) was realized to be the footprint of the asymptotic freedom of quantum chromodynamics. Thus the $J/\psi$ discovery opened a new route to the theory of weak interactions and pointed to the most striking feature of the theory of strong interactions, namely  the quark freedom on very short timescales.
  
The bound state of two charm quarks, usually named {\it charmonium}, is the strong interaction analog of an atomic system. The reduced mass system of the two charm quarks is subject to  a central potential of the form 
$V(r)=a_1/r+a_2r$, where the first term is  the Coulomb-like potential between color charges and the linear term is the phenomenological implementation of the confining force between quarks. 

In the renormalization of  quantum chromodynamics, a scale, $\Lambda_{_{\rm QCD}}$, arises by `dimensional transmutation' of the ultraviolet cutoff. 
In the pure gauge theory, $\Lambda_{_{\rm QCD}}$ enters in the definition of the running coupling constant $\alpha_s(\mu)=2\pi/[11 \ln(\mu/\Lambda_{_{\rm QCD}})]$ showing that perturbation theory is valid only at distances $1/\mu$ smaller than $1/\Lambda_{_{\rm QCD}}$.
A priori $\Lambda_{_{\rm QCD}}$ is an arbitrary scale which can take any value.
Taken over the full set of available data (from deep inelastic scattering, $\Upsilon$ decay, $e^+e^-\to$~jets {\it etc}) one finds  $100\leq\Lambda_{_{\rm QCD}}\leq 400$~MeV. The fact that it turns out to be $\Lambda_{_{\rm QCD}}>>m_u,m_d$ is at the heart of the reason why light quarks are confined in hadrons of the $1$~fm size.

On dimensional grounds we can state that a restoring force  between quarks could be written as $F_{q\bar q}\sim \Lambda_{_{\rm QCD}}^2$, thus resulting from a potential $V(r)\sim \Lambda_{_{\rm QCD}}^2 r$. Fitting  $a_2$ on experimental data one gets $a_2=1/a^2$ with $a=2.3$~GeV$^{-1}$. Observing that $a_2\sim \Lambda_{_{\rm QCD}}^2 $ one finds indeed $\Lambda_{_{\rm QCD}}\sim 400$~MeV.  This in turn means that the effects of confinement are apparent at  a distance  $1/\Lambda_{_{\rm QCD}}=0.5$~fm. But the $J/\psi$ is three times heavier, and thus smaller than the proton. Thus one can imagine that the quarks in the $J/\psi$ are so close, {\it i.e.} far from the confinement distance $1/\Lambda_{_{\rm QCD}}$,  to be considered as almost free. The mass of the charm $m_c$ is therefore just half of the mass of the $J/\psi$, $m_c\simeq 1530$~MeV, and non-relativistic methods can be used to describe charmonia.  In this respect the 
discovery of the $J/\psi$ can be considered as the first evidence of the {\it existence} of quarks.

Since the discovery of the $J/\psi$, the charmonium system has been thoroughly studied~\cite{Brambilla:2004wf}. Very precise determinations of spectra have been provided including hyperfine effects. The decay mechanisms have been investigated in greater detail and no one  expected surprises from charmonia.
Particles like the $\eta_c$ and $\eta_b$, predicted and searched since many years, have eventually been detected, with no striking unexpected features.

It was therefore particularly impressive to discover in 2003 a charmonium-like system, the $X(3872)$, decaying into $J/\psi \pi\pi$, thus  behaving as a charmonium excitation, but surprisingly with almost  equal decay rate into $J/\psi \,\rho$ and $J/\psi \,\omega$.  Not to speak of the fact that the $X(3872)$ was soon realized not to have the very well known radiative decay pattern of the closest in mass standard charmonium state: the radial excitation $2^1P_2$.

After the discovery of the $X(3872)$ more and more similar  narrow resonances have been discovered and confirmed at electron-positron and proton-antiproton colliders. Namely, since the discovery of the $X(3872)$, about twenty new unexpected charmonium-like particles have been found. Most of them have definitely clear clashes with standard charmonium interpretations.

These facts opened a debate about the nature of the newly discovered particles. The mainstream thought has been that of identifying most of these resonances as {\it molecules}
of open charm mesons. In particular the $X(3872)$ happens to have a mass very close to the $D\bar D^*$ threshold. The binding energy left for the $X$ is consistent with  ${\mathcal E} \sim 0.25\pm 0.40$~MeV, thus making this state very large in size: order of ten times bigger than the typical range of strong interactions. Relying on intuition, this feature, per se, would discourage to pursue the molecular picture of the $X$: how is that possible that such a loosely bound state can rearrange its quarks  to produce a $J/\psi \rho$ final state? It must occur that far apart charm quarks must coalesce in a very compact object as we described the $J/\psi$ to be. How is that possible that the $X$ has such a large prompt production cross section at CDF as $30\div70$~nb? In a high energy collision environment one can think that it is rather difficult to form a molecule with practically zero binding energy. Why is the decay width of the $X$ in 
$DD\pi$ not exactly equal to the known decay width of the $D^*$ meson component of the state?  In particular what would be accelerating the decay rate of the $D^*$, as data seem to suggest, inside an almost unbound state? 

These questions apply to other near-to-threshold hypothetical molecules and have induced to think to alternative explanations for the $X$ and its relatives.
A rather natural, and more fundamental possibility, would be that of thinking to other forms of aggregation of quarks in hadrons, like diquarks.
A diquark is a colored quark-quark state which could neutralize its color, binding with an antidiquark.  The resulting meson is a particular realization 
of a  tetraquark. Indeed as for the color, the diquark is like an antiquark and the antidiquark is just like a quark. It follows that the resulting tetraquarks 
are kind of standard $q\bar q$ mesons but with the notable difference that at the end of the electric color string there are diquarks instead of quarks. 

What would happen then when stretching the color string at a distance $1/\mu$ larger than $1/\Lambda_{_{\rm QCD}}$?  The standard lore for conventional  mesons is that the string, which provides the restoring force $F_{q\bar q}$,  would break at some point. By then it stored enough energy  to 
excite the vacuum to produce a pair of light quarks at the breaking point. These in turn  saturate the color of the string endpoint quarks, turning the decaying meson into two lighter mesons. This pictorial way to describe meson decays has  a quantitative implementation in all the hadronization algorithms used in  Monte Carlo simulations of collider physics events. 

Following the same line of reasoning, a tetraquark is expected to decay into {\it two baryons} upon color string breaking. Indeed diquark-quark configurations  are possible in standard baryons as shown by a very simple interpretation of the ratio of deep inelastic scattering (DIS) neutron and proton structure functions in the $x\to 1 $ limit. Indeed it is known that experimentally the ratio $F_2^{(n)}(x)/F_2^{(p)}(x)\to1/4$ as the fraction $x$ of the momentum carried by the parton involved in DIS tends to $1$.  It results that $F_2^{({\rm nucleon})}=\sum_q e_q^2 f_q(x)$, where $f_q$ is the parton distribution function for the $q$ quark species. As the quark participating to DIS gets closer to carry the entire nucleon momentum, the remaining quarks are frozen in their lowest energy state. If diquark bound states exist, these will further lower the energy. But, as shown by lattice simulations, diquarks of the form $[uu]$ or $[dd]$ cannot be formed. Thus the lowest energy configurations for the spectator quarks in DIS are reached when we have a $[ud]$ diquark in the neutron with a $d$ quark involved in  DIS and again a $[ud]$ diquark in the proton with a $u$ quark in DIS. Since $e_d=-1/3$ and $e_u=2/3$, the experimental fact $F_2^{(n)}(x)/F_2^{(p)}(x)\to1/4~(x\to 1/4)$ is understood as due to the ratio of charges $e_d^2/e_u^2=1/4$. 

The $X(3872)$ on the other hand has not enough mass to decay into two charmed baryons because of phase space.  This forces the diquark-antidiquark system to rearrange itself into a $J/\psi \,\rho$ or $J/\psi \,\omega$ configuration. Contrary to the picture given in  molecular models, such a rearrangement happens inside the boundaries of $1/\Lambda_{_{\rm QCD}}$.  But some of the newly discovered hadrons have enough mass to decay into two baryons. The  $Y(4660)$, for example, appears to decay prominently into $\Lambda_c \bar \Lambda_c$ baryons, as expected for a tetraquark. 

The tetraquark model has more challenging predictions though. 
Charged states are expected, such as $[cu][\bar c\bar d]$ and even doubly charged states as
$[cu][\bar d \bar s]$. The model does not predict light doubly charged states though (since diquarks like $[uu]$ would be involved). 
In particular, in one of the first papers on the tetraquark model~\cite{Maiani:2004vq}, it was predicted  that narrow resonances decaying into $J/\psi \pi^+$ 
should be observed.  These kind of particles have indeed been observed by Belle about three years later. 

The $Z^+(4430)$, the first of a series of three newly discovered charged particles,  decays in a charmonium state and a charged pion. Even if Belle observes it at more than $6\sigma$ significance, BaBar has only $3.5\sigma$ and CDF has not yet confirmed it. Moreover the tetraquark model predicts charged partners of the $X(3872)$ with a very similar mass. The latter have not yet been observed. The doubly charged states, as for the time being, have not been searched.

In order to explain in the tetraquark model the peculiar decay pattern of the $X(3872)$ which dissociates with equal rate into $J/\psi\rho$ and $J/\psi \omega$  one needs to consider two neutral states with a difference in mass of few MeV. One can call them $X_u=[cu][\bar c \bar u]$ and $X_d=[cd][\bar c \bar d]$. Once these states are mixed by $u\bar u\leftrightarrow d\bar d$ annihilations we can end up with isospin pure mixtures of the kind $(X_u\pm X_d)/\sqrt{2}$. But if we suppose that annihilation contributions are suppressed by the smallness of $\alpha_s$ at the charm mass, we could align the eigenvectors of the mass matrix in the quark basis rather than in the isospin basis. In that extreme case we have two unmixed states $X_u=[cu][\bar c \bar u]$ and $X_d=[cd][\bar c \bar d]$: each of them can contain both $I=0$ and $I=1$. The difference in mass between the two must be $\Delta m\sim m_d-m_u$.  As of today it seems that the $3.5\sigma$ evidence of both a $X(3872)$ and a $X(3875)$ resonances have faded out by Belle and BaBar data analyses.
(At some stage we thought that  the former was decaying preferably in $D\bar D^*$ and the latter in $J/\psi\rho$. The decay pattern of these two neutral states was explained in~\cite{Maiani:2007vr}).

The main drawback of the tetraquark picture, as it will be discussed in this review, is the proliferation of expected states.  We do not have any clue of selection rules which could limit the production of tetraquark particles at the fragmentation level. Therefore, as it will be shown in the following, we will merely enumerate all the possible bound states trying an estimate of their mass values. Unfortunately we cannot predict the fragmentation probabilities of tetraquark states, so it is also difficult to estimate their production rates. 

The mass spectra are computed relying on the non-relativistic constituent quark model where the quark masses (constituent masses) incorporate the bulk of the non perturbative  interactions and basically are fitted from baryon and meson spectra. Corrections to the simple algebraic sums of those masses come from chromomagnetic  (spin-spin)  interactions among constituents. Since we will be dealing with diquark-antidiquark mesons, spin-spin interactions will be of two kinds: those occurring inside single diquarks and those between the two diquarks. As the relative orbital angular momentum of the tetraquark grows ($L>0$), point-like spin-spin interactions will be switched off. Actually, in this review, we propose to treat the orbital excitations and the spin-orbit corrections in the framework of a relativistic string model. The resulting model has been table tested on standard charmonia and has a rather good behavior at determining mass spectra.

One should be aware that the `realm' of the constituent quark model is actually the one of very small $\Lambda_{_{\rm QCD}}$, actually  $\Lambda_{_{\rm QCD}}<<m_u,m_d$. In this hypothetical world the pion would weight $600$~MeV and there would be a clear clash between the Nambu-Goldstone interpretation of light pseudoscalar mesons and their constituent quark nature. Yet, it is surprising to observe that the constituent quark model is a powerful tool in hadron spectroscopy.
An explanation of this from first principles likely requires new ideas in non-perturbative chromodynamics.

The scope of this short review is to give a summary of the tetraquark model expectations and of the methods used to  compute masses and decays in this framework.  We will briefly discuss the molecular interpretations of the  $X,Y,Z$ mesons, and give a comprehensive overview of the experimental situation. We will clearly stress the drawbacks of the various models. 
In particular, in our discussion we will show that the $X(3872)$ has serious difficulties both with the tetraquark and  the molecule interpretation.

We aim at  providing a thorough discussion of what has been made and what is still waiting to be done on the experimental side in order to help the
understanding of the new hadrons.

%% file: theory.tex
\section{Theoretical Models }
\label{sec:theory}
Quark, anti-quarks and gluons do not necessarily bind only in the $q\bar{q}$ pairs or $qqq$ triplets. The quark model from its beginning~\cite{GellMann:1964nj} contemplate the existence of other aggregations. The states recently discovered suggest in particular the possibility to be observing states with two or three  quarks and two or three anti-quarks or a quark, an anti-quark and valence gluons.
While the latter states (called hybrid) appear in only one configurations, there are two possibilities to form bound states out of two quarks and two anti-quarks:
\begin{itemize}
\item{binding the two quarks in a colored configuration called diquark $[qq]_{\alpha}$, ot anti-diquark $[\bar{q}\bar{q}]^{\alpha}$. Color charge is neutralized by the interaction diquark-antidiquark: this configuration is what will be called tetraquark;}
\item{binding each quark to an anti-quark $[q_{\alpha} \bar{q}^{\alpha}]$ and allowing interaction between the two color neutral pairs $[q_{\alpha} \bar{q}^{\alpha}]$$[q_{\beta}\bar{q}^{\beta}]$. This configuration is very close to the one with two interacting mesons and it is usually called molecular bound state.}
\end{itemize}

This section details the differences between the approaches in terms of predictions and interpretations of the currently observed states. Besides the exotic interpretations the possibility is still open of at least a fraction of them being ordinary charmonium with an incorrect $J^{PC}$ assignment
or for which the potential model predictions do not hold. It is to be considered indeed that, as shown in Fig.~\ref{fig:charmon}, while below the open charm threshold all the narrow states have been observed and match predictions, above it almost all conventional charmonium states are missing and the few candidate ones have masses significantly different from predictions.
Another lurking possibility is that the observed states are not strong resonances but the effect of opening 
thresholds. For a discussion of this aspect we remit to other reviews as for instance Refs.~\cite{Bugg:2008wu,Bugg:2009zz}.

Finally we briefly review the picture, recently proposed, in which some of the observed states are described as a standard charmonium stuck inside a light hadron, the so-called hadrocharmonium. Some hints about pentaquark and hexaquark structures are given in the end of this section.

\subsection{Tetraquarks}
\label{sec:theory:tetraquarks}

Tetraquarks are bound states of a diquark and an anti-diquark, combinations of two quarks and two anti-quarks respectively.  The combination of two $3_c$ color states can yield a $\bar{3}_c$ or a $6_c$.
Which of these state are bound can be estimated in the one gluon exchange model\cite{Jaffe:2004ph},\cite{Jaffe:1999ze}.
As in electromagnetism, the discriminant between attractive and repulsive forces is the sign of the product of the charges: in the case of the color interaction, the sign of the product of $SU(3)_c$ generators $T^a$ for the A(B) representation is computed as
\begin{equation}
I \propto \sum_a T^a_A T^a_B =\frac{1}{2}\sum_a \Bigg({{T^a}^2}-{T^a_A}^2 - {T^a_B}^2\Bigg )\qquad \qquad a=1,\cdots,8; \qquad A,B = \bf{3} \text{ or } \bar{\bf{3}}
\label{eq:oper}
\end{equation}
where 
\begin{equation}
T^a = T^a_A + T^a_B
\end{equation}
is the generator on the tensor product $A\otimes B$ space. Since ${T^a}^2 = \sum_a T^a T^a$ commutes with the $SU(3)$ generators it is a number on each irreducible representation (Casimir). Using the Shur lemma, the Casimir operator is proportional to the identity ${\bf{I}}$:
\begin{equation}
 \sum_a {T^a}^2 = C(D){\bf{I}},
\end{equation}
where $C$ is a constant depending on $D$, the dimension of the tensor product space generated by $T^a = T^a_A + T^a_B$.  The discriminator then becomes 
\begin{equation}
I \propto \frac{1}{2}\Bigg ( C(A\otimes B)-C(A)-C(B) \Bigg)
\end{equation}
Imposing the  normalization of the Casimir operator one obtains
\begin{equation}
C(D) = \frac{8 k_D}{dim(D)},
\end{equation}
where
\begin{eqnarray}
&&k_{A\oplus B} = k_A +k_B \\ \notag
&&k_{A\otimes B} = dim(A)k_B + dim(B)k_A
\end{eqnarray}
Tab.~\ref{tab:i}  shows the estimated  $k_D$ coefficients as evaluated in the  one gluon exchange model analysis. 
\begin{table}
\caption{One gluon excange model results.}
\label{tab:i}
\begin{tabular}{lccccc}
\hline
$D=A\otimes B$ & \qquad $\bf{1}$  \qquad & \qquad $\bf{\bar{3}}$ \qquad  & \qquad  $\bf{6}$ \qquad  &
\qquad  $\bf{8}$ \qquad \\
\hline
$k_{D}$&\qquad 0  \qquad & \qquad 1/3 \qquad & \qquad  5/2 \qquad  & \qquad 3 \qquad \\
\hline
 C(D) &\qquad 0 \qquad  &	 \qquad 4/3 \qquad & \qquad 10/3 \qquad & \qquad 3 \qquad   \\
\hline
$I=\frac{1}{2}(C(D)-C(A)-C(B))$ & \qquad -4/3 \qquad   & \qquad    -2/3 \qquad & \qquad 1/3 \qquad & \qquad 1/6 \qquad\\
\hline
\end{tabular}\\
\end{table}

These results show that only the $3_C$ state can be bound and therefore this is the only one considered hereafter. It is to be noted that the same reasoning, applied to a quark anti-quark pair, shows that only the color singlet is bound, as known.

The other relevant quantum number of the diquarks is the spin, which can be $S=0$ or $S=1$. 
Lattice simulations~\cite{Alexandrou:2006cq} show that scalar diquarks made of light quarks are more stable than vector ones: for this reason they are usually called ``good'' and  ``bad''  diquarks respectively.

Predictions of the tetraquark model are obtained with three different approaches:
\begin{itemize}

\item{QCD sum rules: calculations are based on the correlation functions of two hadronic currents with the quantum numbers of the hadron under investigation \cite{Shifman:1978bx,Nielsen:2009uh}. At the quark level the two-point correlation function is expanded as a series of local operators (generalized Wilson OPE):

\begin{equation}
\prod(q^2) = i \int{\ud^4x e^{iq \cdot x}\langle 0 | T[j(x)j^\dag (0)] |0\rangle} = \sum_n{C_{n}(Q^2)\hat{O}_n},
\label{eq:q2}
\end{equation}
where the coefficients $C_n(Q^2)(Q^2=-q^2)$ include by construction only the short-distance domain and can therefore be evaluated perturbatively, while the set $\{\hat{O}_n\}$ include all local gauge invariant operators expressible in terms of gluon and quark fields. 
The fundamental hypothesis of the sum rules approach is that the current-current correlation function in Eq.~(\ref{eq:q2})can be matched to the dispersion relation 
\begin{equation}
\prod(q^2) = - \int \ud s \frac{\rho(s)}{q^2-s+i\epsilon}+\cdots
\end{equation}
To extract physical quantities, the spectral function $\rho(s)$ is studied by parameterizing it as a single sharp function with a pole representing the lowest physical resonance and a  smooth continuum representing higher mass states. 
A modern introduction to the method of QCD sum rules can be found in \cite{Colangelo:2000dp}}; 
\item{Relativistic quasipotential model \cite{Ebert:2005nc,Ebert:2008se}: the interaction between two quarks in a diquark and between diquarks and antidiquarks in a tetraquark are described by the diquark wave function $\Psi_d$ of the bound quark-quark state and by the tetraquark wave function $\Psi_T$ of the bound diquark-antidiquark state respectively. They must satisfy the Schr\"odinger equation:
\begin{equation}
\Bigg ( \frac{b^2(M)}{2\mu_R} - \frac{\mathbf{p}^2}{2\mu_R} \Bigg)\Psi_{d, T}(\mathbf{p}) =\int {\frac{{\ud}^3 q }{(2\pi)^3} V_{d, T}(\mathbf{p}, \mathbf{q}; M)\Psi_{d, T}(\mathbf{q})  },
\label{eq:schro}
\end{equation}
where the relativistic reduced mass is:
\begin{equation}
\mu_R=\frac{E_1 E_2}{E_1 + E_2}
\end{equation}
and $E_1$, $E_2$ are given by:
\begin{equation}
E_1 = \frac{M^2 - m_2^2 + m_1^2}{2M}, \qquad \qquad E_2 = \frac{M^2 - m_1^2 + m_2^2}{2M}. 
\end{equation}
Here $M = E_1 + E_2$ is the bound state mass (diquark or tetraquark), $m_{1,2}$ are the masses of the quarks which form the diquark or of the diquark and antidiquark which form the tetraquark; $\mathbf{p}$ is their relative momentum. In the center of mass system the relative momentum squared on mass shell is indicated as $b^2(M)$. The kernel $V_{d, T}(\mathbf{p}, \mathbf{q}; M)$ is the potential operator of the quark-quark or diquark-antidiquark interaction and depends on the quark composition of the object under investigation.

The potential given by Eq.~(\ref{eq:schro}) is then solved numerically for the diquark wave function $\Psi_d$, the diquark masses and form factors.

Next, the masses of the tetraquarks are calculated introducing a kernel for the diquark-antidiquark bound state that involves the diquark parameters found in the first step}.

\item{A generalization of the constituent quark model: hadron masses are described by an effective Hamiltonian that takes as input the constituent quark masses and the spin-spin couplings between quarks. By extending this approach to diquark-antidiquark bound states it is possible to predict tetraquark mass spectra. This model will be discussed in detail in the next section}.
\end{itemize}

\subsubsection{Spectra}
\label{sec:th:spectra}

We estimate the tetraquark mass spectra in the framework of the non-relativistic constituent quark model combined with a hadron string model. In the following section the two models are defined and the mathematical details for both are given. The  numerical inputs required are discussed and derived and the results for the mass spectra determination are summarized for different quantum numbers.

In order to define the energy of a tetraquark state we have to consider all the possible interactions between the quarks. In the ground state the two diquarks interact only by spin couplings because the angular momentum is zero ($L=0$). An effective non-relativistic Hamiltonian can be written including spin-spin interactions within a diquark and between quarks in different diquarks.

If the angular momentum is different from zero ($L \neq 0$) the pointlike spin interactions are suppressed as the average distance between diquarks grows. 
Tetraquark in orbitally excited states are studied in a model of color flux tubes first proposed by Selem and Wilczek \cite{Selem:2006nd} in order to emphasize the importance of diquark correlations in hadronic physics. We will generalize the model to describe a four quark state as a spinning string with a given angular velocity $\omega$ connecting two massive diquarks $m_1$, $m_2$ at distances $r_1$ and $r_2$ away from the rotation axis.

{\bf  \emph{{The non-relativistic approach for $L=0$ }}}

The most general Hamiltonian describing the $L=0$ tetraquark state of flavor composition $[q_1q_2][\bar{q}_3\bar{q}_4]$ can be written as:  
\begin{equation}
H={m_\qq}_{1}+{m_\qq}_{2} + H_{_{SS}}^{(qq)} + H_{_{SS}}^{(q\bar q)}
\label{eq:h}
\end{equation}
where  ${m_\qq}_{i}$ are the diquark masses and:
\begin{eqnarray}
\:\:&&H_{_{SS}}^{(qq)}= 2{\kappa_\qq}_1 (\vec S_\qu\cdot \vec S_\qd)+2{\kappa_\qq}_2( \vec S_{\bar{q}_3}\cdot \vec S_{\bar{q}_4})  \\ 
\:\:&&H_{_{SS}}^{(q\bar q)}=  2\kappa_{q_1\bar{q}_3}(\vec{S}_{q_1} \cdot \vec{S}_{\bar{q}_3} ) +  2\kappa_{q_1\bar{q}_4}(\vec{S}_{q_1} \cdot \vec{S}_{\bar{q}_4} )+  2\kappa_{q_2\bar{q}_3}(\vec{S}_{q_2} \cdot \vec{S}_{\bar{q}_3} )+  2\kappa_{q_2\bar{q}_4}(\vec{S}_{q_2} \cdot \vec{S}_{\bar{q}_4}).
\label{hams}
\end{eqnarray}
the first term refers to spin-spin interactions between quarks bound to give a diquark and the second term refers to interactions between quarks in different diquarks.
 The coefficients $\kappa_{q_i q_j}$ depend on the flavor of the constituents $q_{i,j}$ and on the particular color state of the pair. We remind that the coefficients $\kappa_{q_i q_j}$ include the diquark masses dependence and the point-like behaviour of the spin-spin interactions:
\begin{equation}
\kappa_{q_i q_j} \vec{S}_{q_i}\cdot \vec{S}_{q_j} \sim \frac{{{\kappa}^{'}}_{q_i q_j}}{m_{q_i} m_{q_j}}\vec{S}_{q_i}\cdot \vec{S}_{q_j} \delta^{3}(\vec{r_i}-\vec{r_j})
\label{eq:delta}
\end{equation}  
It follows that the coefficients $\kappa_{q_i q_j}$ have the dimention of energy. The constituent quark masses and the ${\kappa_{q\bar{q}}}_{_1}$ couplings for color singlet combinations are determined from the  scalar and vector light mesons. 
They can then be translated into the tetraquark couplings using
\begin{equation}
\kappa_{q_1{\bar{q}}_3}([q_1 q_2 ][\bar{q}_3 \bar{q}_4]) = \frac{1}{4} (\kappa_{q_1{\bar{q}}_3})_{{\bf{1}}} 
\end{equation}
as detailed in Ref.~\cite{Maiani:2004vq}.
The $\kappa_{qq}$ couplings are instead determined from the masses of the 
$qqq$ baryons ground ($J=1/2$) and excited ($J=3/2$) states.

The numerical values for the free parameters useful for the determination of tetraquark mass spectra are summarized in Tab.~\ref{tab:qaq}-\ref{tab:massq}. 


\begin{table}
  \caption{Numerical values of quark-antiquark spin couplings.}
  \label{tab:qaq}
  \begin{tabular}{lcccccccccc}
    \hline
    & $q\bar{q}$ & $s\bar{q}$ & $s\bar{s}$ & $c\bar{q}$ & $c\bar{s}$ & $c\bar{c}$  & $b\bar{q}$  & $b\bar{s}$   & $b\bar{c}$  & $b\bar{b}$    \\
    \hline
    ${\kappa_{q\bar{q}}}_{\bf{_1}}$ (MeV)         & 315 & 195 & 121 & 70 & 72  &59& 23  & 23   & 20  & 36    \\
    $\kappa_{qq}$ (MeV)         & 103 & 64 & 22 & 25 & 72  & 6 & 8     \\
    \hline
  \end{tabular}
\end{table}

\begin{table}
  \caption{Constituent quark masses as determined from meson.}
  \label{tab:massq}
  \begin{tabular}{lcccc}
    \hline
    & q & s & c & b     \\
    \hline
    $m_i$ (MeV) & 305 & 490 & 1534 & 4720 \\
    \hline
  \end{tabular}
\end{table}

To obtain the diquark masses one state must be assumed as tetraquark and the rest of the spectrum can be derived accordingly. In this review will start assuming that the 
 X(3872) is a  $[cq]_{S=1}[\bar{c}\bar{q}]_{S=0}$ tetraquark (see Sec.~\ref{sec:x3872}). By diagonalizing the Hamiltonian in Eq.~(\ref{eq:h}) and, using the spin couplings values derived above, we obtain the $m_{[cq]}$ diquark mass value. In order to reduce the experimental information needed we estimate the remaining diquark masses by substituting the costituent quark forming the diquark. We have:
\begin{eqnarray}
m_{[cs]} &=& m_{[cq]} - m_q + m_s\\ \notag
m_{[bq]} &=& m_{[cq]} - m_c + m_b\\ \notag
m_{[bs]} &=& m_{[bq]} - m_q + m_s
\end{eqnarray}
The numerical values for the diquark masses are given in Tab.~\ref{tab:massqq}.

\begin{table}
  \caption{Diquark masses.}
  \label{tab:massqq}
  \begin{tabular}{lcccc}
    \hline
     & $[cq]$ & $[cs]$ & $[bq]$ &  $[bs]$     \\
    \hline
    $m_{q_iq_j}$ (MeV) & 1933 & 2118&5119&5304 \\
    \hline
  \end{tabular}
\end{table}
 
The last step for the mass spectrum determination is the diagonalization of the Hamiltonian in  Eq.~(\ref{eq:h}).

As done in \cite{Drenska:2009cd} we label the particle states with the notation $|S_\qq, S_{\bar \qq};S_{ \qq\bar\qq}, J \rangle $ 
where $\qq$ and $\bar \qq$ represent diquark and antidiquark states respectively and $S_{ \qq\bar\qq}$ is the total spin of the diquark-antidiquark system. In the most general case $[q_1q_2][\bar{q}_3 \bar{q}_4]$ we have:

\begin{itemize}
\item{two positive parity states with $J^{P}$ = $0^{+}$
\begin{eqnarray}
{|0^{+}\rangle}_1 & = & |0_{q_1 q_2}, 0_{\bar{q}_3\bar{q}_4};0_{\rm \qq\bar\qq}, J=0 \rangle \\ 
{|0^{+}\rangle}_2 & = & |1_{q_1 q_2}, 1_{\bar{q}_3\bar{q}_4};0_{\rm \qq\bar\qq}, J=0 \rangle 
\label{eq:zp}
\end{eqnarray}}
\item{three states with $J=1$ and positive parity
\begin{eqnarray}
{|1^+\rangle}_1  & = & |1_{q_1q_2}, 0_{\bar{q}_3\bar{q}_4};1_{\rm \qq\bar\qq}, J=1 \rangle  \\
{|1^+\rangle}_2  & = & |0_{q_1q_2}, 1_{\bar{q}_3\bar{q}_4};1_{\rm \qq\bar\qq}, J=1 \rangle  \\
{|1^+\rangle}_3  & = & |1_{q_1q_2}, 1_{\bar{q}_3\bar{q}_4};1_{\rm \qq\bar\qq}, J=1 \rangle 
\label{eq:up}
\end{eqnarray}
}
\end{itemize}

We observe that if the diquark and the antidiquark have the same flavor composition, $\it{i.e.}$ $[q_1q_2][\bar{q}_1\bar{q}_2]$, the charge conjugation is a symmetry of the state. Thus we define:
\begin{itemize}
\item{two positive parity states with $J^{PC}$ = $0^{++}$
\begin{eqnarray}
{|0^{++}\rangle}_1 & = & |0_{q_1q_2}, 0_{\bar{q}_1\bar{q}_2};0_{\rm \qq\bar\qq}, J=0 \rangle \\
{|0^{++}\rangle}_2  & = &|1_{q_1q_2}, 1_{\bar{q}_1\bar{q}_2};0_{\rm \qq\bar\qq}, J=0 \rangle 
\label{eq:zpp}
\end{eqnarray}}
\item{one state with positive parity and positive charge conjugation and $J=1$
\begin{equation}
|1^{++}\rangle = \frac{1}{\sqrt{2}} \left (|1_{q_1 q_2},0_{\bar{q}_1\bar{q}_2};1_{\rm \qq\bar\qq}, J=1 \rangle   +|0_{q_1 q_2}, 1_{\bar{q}_1\bar{q}_2};1_{\rm \qq\bar\qq}, J=1 \rangle  \right )
\label{eq:upp}
\end{equation}}
\item{two states with positive parity and negative charge conjugation
\begin{eqnarray}
{|1^{+-}\rangle}_1 & = &\frac{1}{\sqrt{2}} \left (|1_{q_1 q_2}, 0_{\bar{q}_1\bar{q}_2};1_{\rm \qq\bar\qq}, J=1 \rangle  - |0_{q_1 q_2 }, 1_{\bar{q}_1\bar{q}_2};1_{\rm \qq\bar\qq}, J=1 \rangle  \right ) \\
{|1^{+-}\rangle}_2  & = &|1_{q_1 q_2}, 1_{\bar{q}_1\bar{q}_2};1_{\rm \qq\bar\qq}, J=1 \rangle 
\label{eq:upm}
\end{eqnarray}}
\end{itemize}

As introduced in \cite{Maiani:2004vq} the action of the  spin operators in Eq.~(\ref{hams}) can be evaluated using matrix representaion for the states given above. Individual diquark spins are expressed by the $2\times 2$ Pauli matrices
\begin{equation}
\Gamma^0 = \frac{1}{\sqrt{2}}\sigma_2 \qquad \qquad \Gamma^i = \frac{1}{\sqrt{2}}\sigma_2 \sigma^i
\end{equation} 
for spin 0 and spin 1, respectively. The matrices $\Gamma$ are normalized in such a way that:
\begin{equation}
\rm{Tr} [(\Gamma^{\alpha})^{\dag}\Gamma^{\beta}] = \delta^{\alpha \beta}
\end{equation}
It is straightforward to write:

\begin{eqnarray}
\left\vert 0_{\qq},0_{\bar{\qq}};0_{\qq \bar{\qq} }, J=0\right\rangle &=&\frac{1}{2}\sigma
_{2} \otimes \sigma _{2} \nonumber \\
\left\vert 1_{\qq},1_{\bar{\qq}};0_{\qq \bar{\qq} },J=0\right\rangle &=&\frac{1}{2\sqrt{3}}%
\left( \sigma _{2}\sigma ^{i}\right) \otimes \left( \sigma _{2}\sigma
^{i}\right)   \nonumber \\
\left\vert 0_{\qq},1_{\bar{\qq}}; 1_{\qq \bar{\qq} }, J=1\right\rangle &=&\frac{1}{2} \sigma
_{2} \otimes \left( \sigma _{2}\sigma ^{i}\right)   \nonumber \\
\left\vert 1_{\qq},0_{\bar{\qq}};1_{\qq \bar{\qq} }, J=1\right\rangle &=&\frac{1}{2}\left( \sigma
_{2}\sigma ^{i}\right) \otimes  \sigma _{2}   \nonumber \\
\left\vert 1_{\qq},1_{\bar{\qq}};1_{\qq \bar{\qq} }, J=1\right\rangle &=&\frac{1}{2\sqrt{2}}%
\varepsilon ^{ijk}\left( \sigma _{2}\sigma ^{j}\right) \otimes \left( \sigma
_{2}\sigma ^{k}\right)   \label{notations}
\end{eqnarray}%

\begin{figure}[bht]
\begin{center}
\epsfig{file=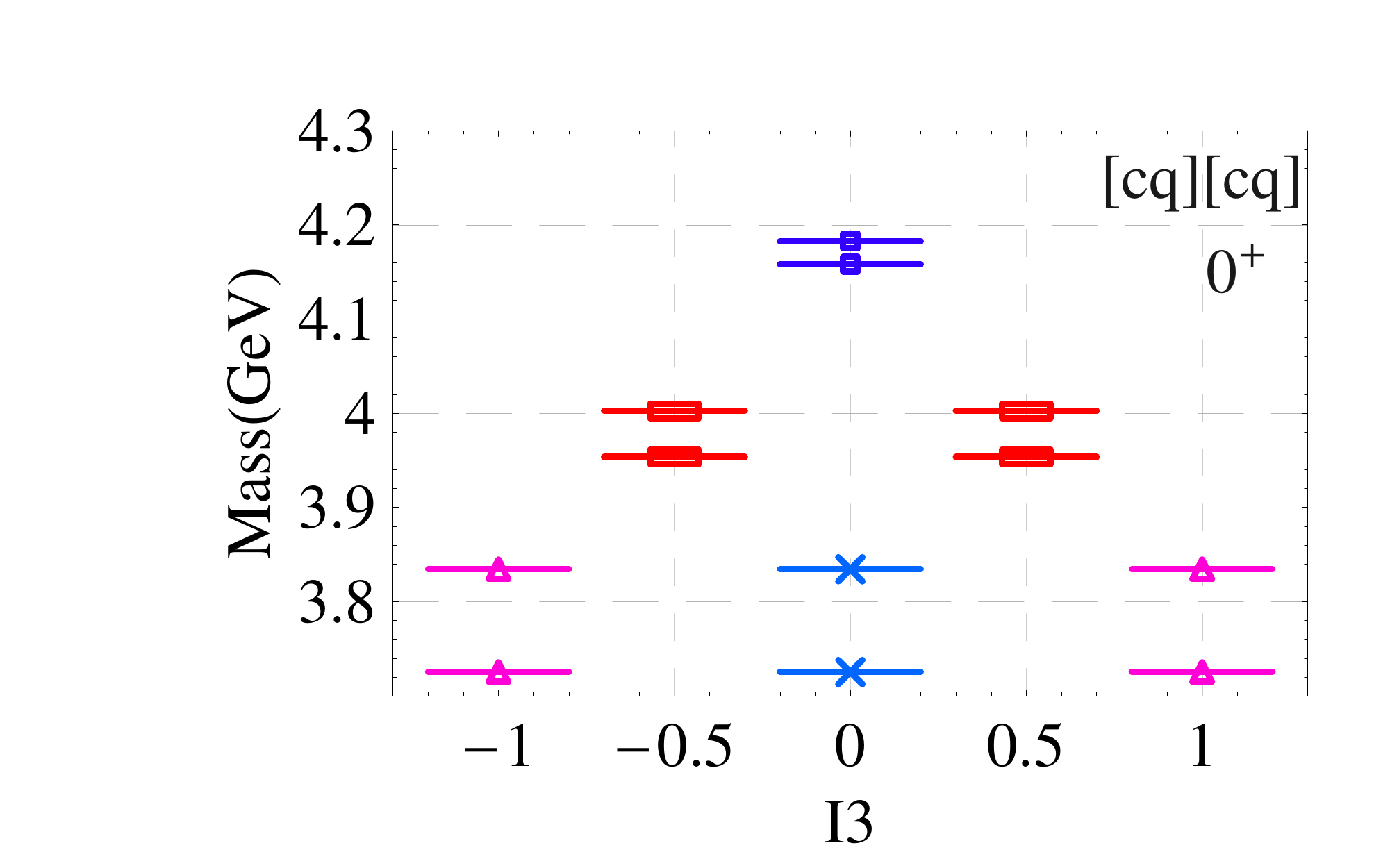,width=6.6cm} 
\epsfig{file=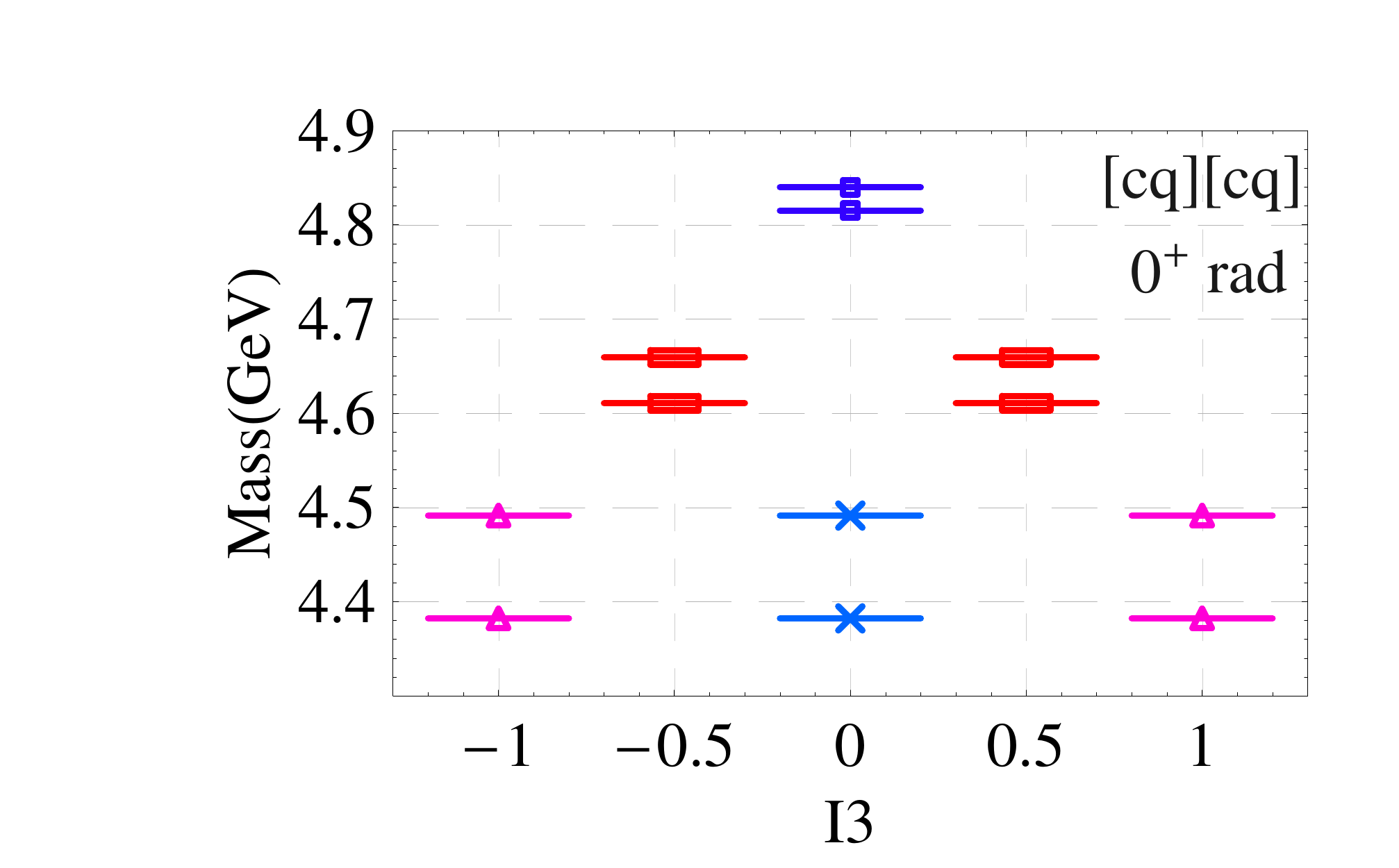,width=6.6cm} 
\epsfig{file=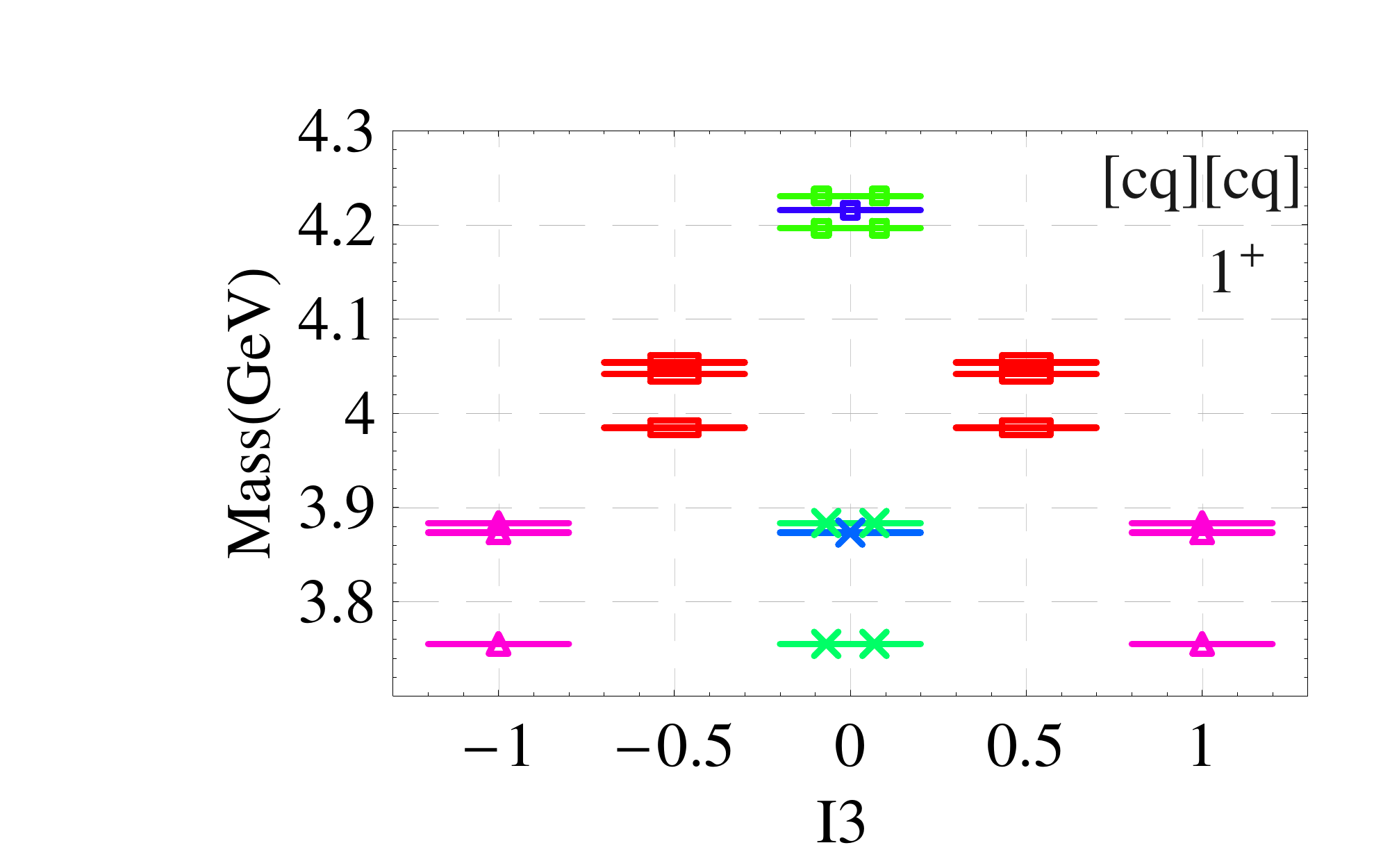,width=6.6cm} 
\epsfig{file=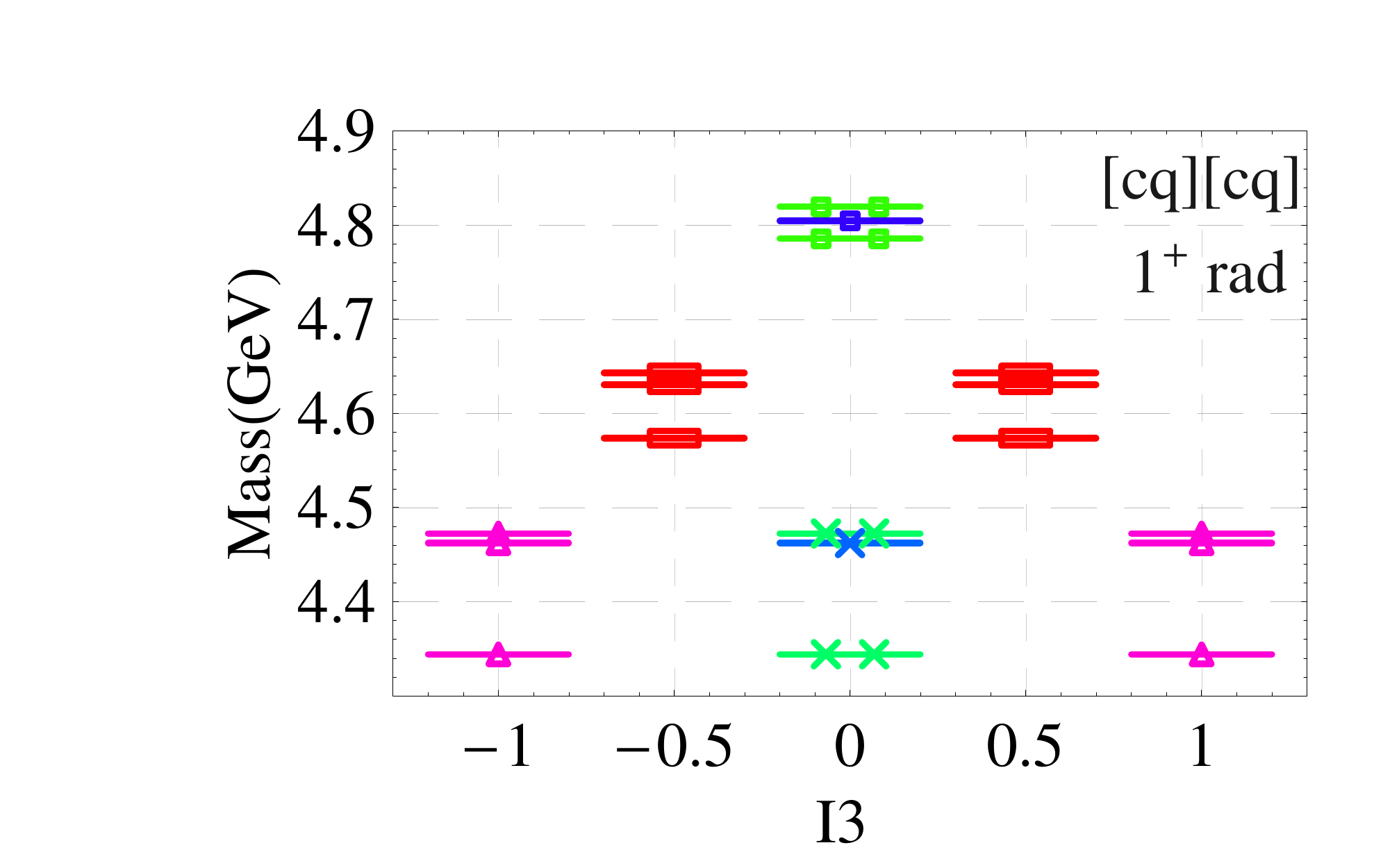,width=6.6cm}
\end{center}
\epsfig{file=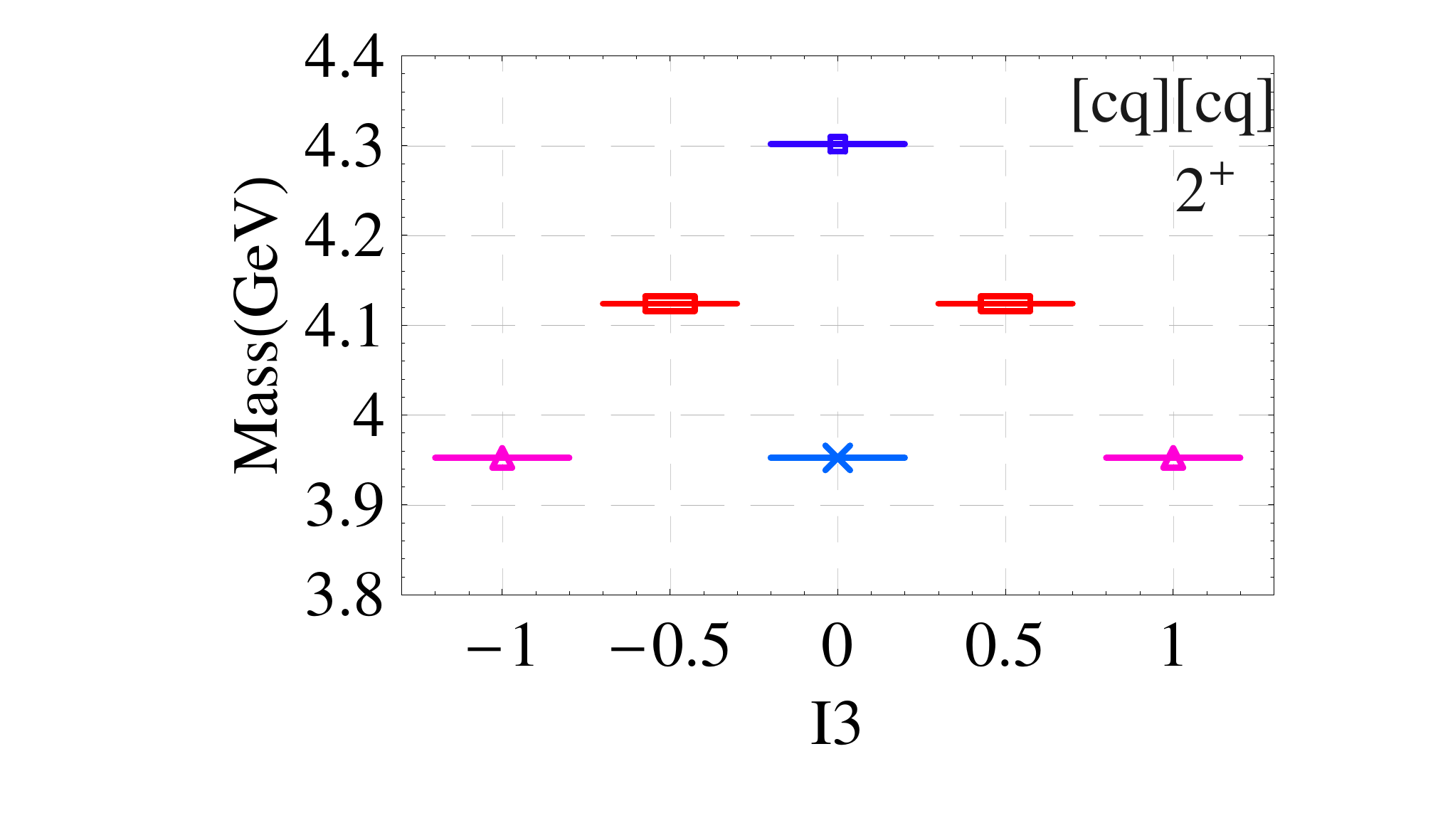,width=6.9cm} 
\caption{\it Hidden charm multiplets with zero orbital excitation ($L=0$): on left the first radial excitation, on right the second radial excitation.}
\label{fig:cqcq}

\end{figure}

The states we want to focus on are hidden charm $[cq][\bar{c}\bar{q}]$ and hidden bottom $[bq][\bar{b}\bar{q}]$ states, where $q$ is a light quark $u$, $d$ or $s$. A useful way to visualize the possible tetraquark states is to organize them in multiplets with defined $J^{P(C)}$ quantum numbers showing the mass value versus the third isospin component. All the plots shown in Fig.~\ref{fig:cqcq} have the same range in both axis in order to better visualize the mass splittings for the different configurations. Since the diquarks $[cu]$ and $[cd]$ are assumed to have the same mass, we will also have degenerate states. In Fig.~\ref{fig:cqcq} the following multiplets are shown:
\begin{description}
\item[$J^P = 0^+:$] contains 18 states. These states are taken as $L=0$ and $S=0$ states. This configuration requires the diquarks and the antidiquarks to have the same spin.   

Starting from the lightest masses, we find two iso-triplets, one with both diquarks in spin 1 and one with both diquarks with the spin 0. The corresponding two iso-singlets are degenerate in mass. We observe that the pink(triangle) states in the figure are charged, while the light blue(star) are neutral and have positive charge conjugation quantum number $C=+1$ (see Eq.~(\ref{eq:zpp})). 

The four red(rectangle) states shown are open strange iso-doublets: here the same mass value for $[cu]$ and $[cd]$ leads to the same mass value for the charged states (like $[cu][\bar{c}\bar{s}]$) and the neutral one (like $[cd][\bar{c}\bar{s}]$). 

The two states shown in dark blue(box) are the two iso-singlets $[cs][\bar{c}\bar{s}]$ with definite $J^{PC} = 0^{++}$ quantum numbers.        

\item[$J^P = 1^+:$] there are 27 states. Assuming $L=0$, there are three different spin configurations that give $J=1$ as defined in Eq.~(\ref{eq:up}). 

Starting from the lightest states, we find three iso-triplets and the corresponding iso-singlets degenerate in mass. Here again we have charged states shown in pink(triangle), while the neutral states are drawn in light blue(single star) if $C=+1$ (see Eq.~(\ref{eq:upp}))and in green(double star) if $C=-1$ (see Eq.~(\ref{eq:upm})).

In the middle of the plot, in red(rectangle), the three open-strange iso-doublets are shown, each of which has a corresponding charge state degenerate in mass.

The three heaviest states (hidden-charm hidden-strange) are drawn in dark blue(single box) for $J^{PC} = 1^{++}$, in green(double box) for  $J^{PC} = 1^{+-}$.

\item[$J^P = 2^+:$] contains 9 states. These states are taken as $L=0$ and $S=2$ states.  

Starting from the lightest masses, we find one iso-triplet (both diquarks are in spin 1) and the corresponding iso-singlet is degenerate in mass. The pink(triangle) state in the figure is charged, while the light blue(star) is neutral and has positive charge conjugation quantum number $C=+1$. 

The two red(rectangle) states shown are open strange iso-doublets: each one is double degenerate in mass since the assumption $m_{[cu]} = m_{[cd]}$.

The state shown in dark blue(box) is the iso-singlet $[cs][\bar{c}\bar{s}]$ with definite $J^{PC} = 2^{++}$ quantum numbers.

\item[$J^P = 0^+$ radial excitation:] the same as the $J^P = 0^+$ mass spectrum is here reproduced adding to each mass value a correction to obtain the second radial excitation. The only information we make use of is the mass spilitting $\Delta$M = 0.657 GeV between the $\eta_{c}(2S)$ and $\eta_c(1S)$. Having this states the same quantum number as the tetraquark multiplet under investigation we simply evaluate the second radial excitation adding the mass splitting $\Delta$M = 0.657 GeV to each $J^P = 0^+$ ground state. 

\item[$J^P = 1^+$ radial excitation:] the radial excitation splitting between the states $\psi(2S)$ and $J/\psi(1S)$ is used to evaluate the mass spectrum for the tetraquark multiplet with these quantum numbers. We evaluate numerically the mass spectrum for the $J^P = 1^+$ second radial excitation adding a mass splitting $\Delta$M = 0.589 GeV to each  $J^P = 1^+$ ground state.

\end{description}

\begin{figure}[bht]
\begin{center}
\epsfig{file=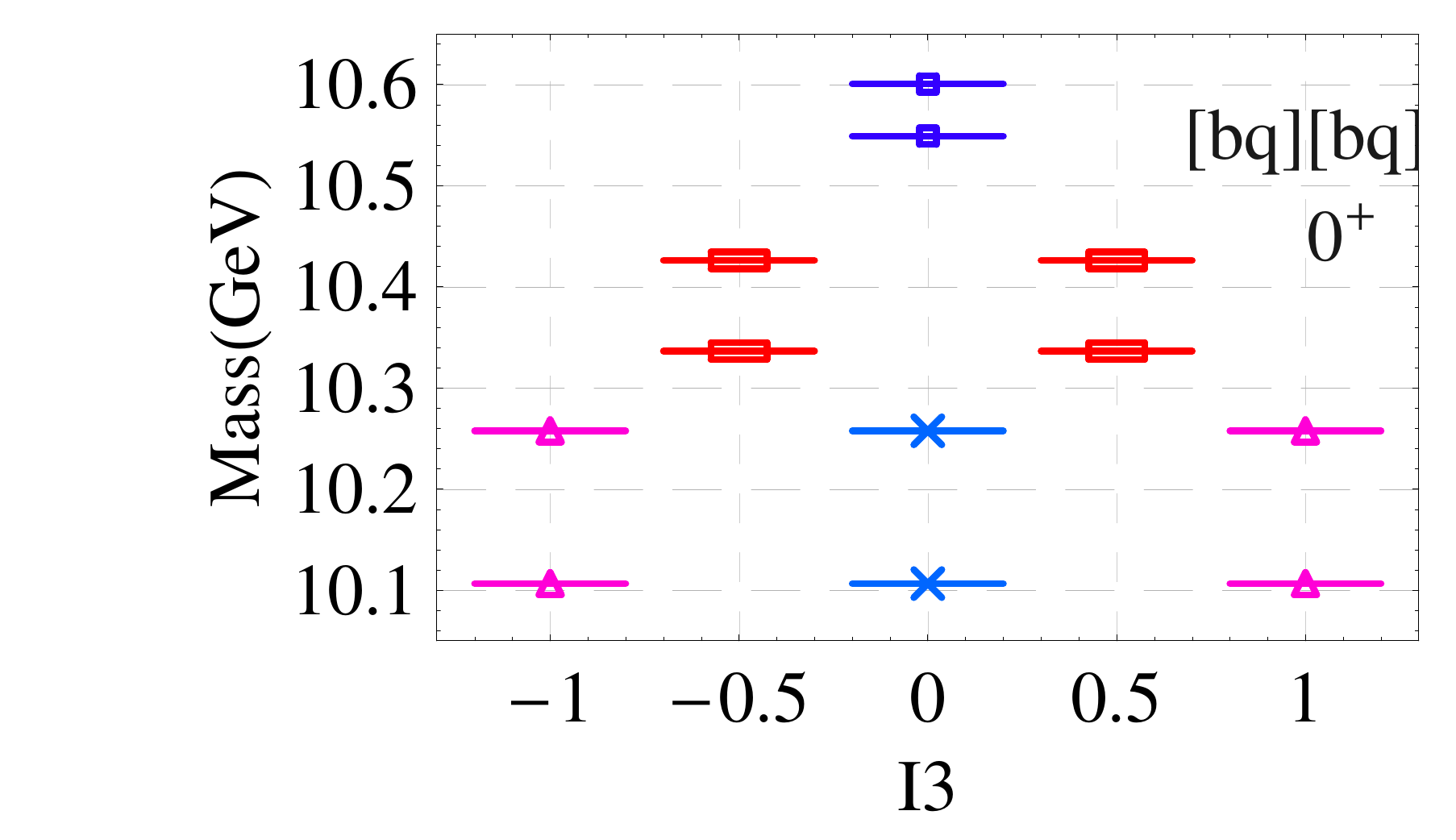,width=6.3cm} 
\epsfig{file=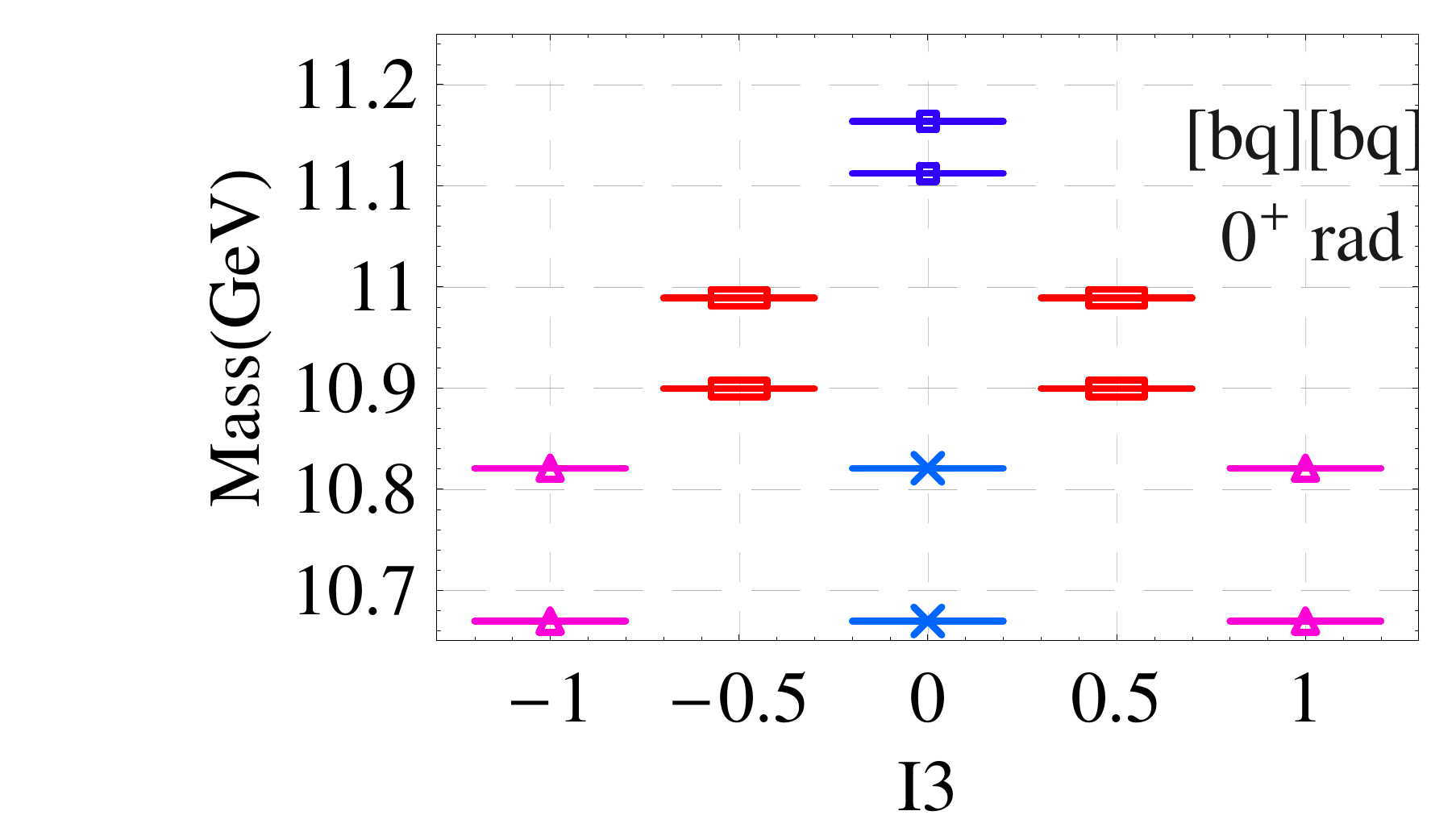,width=6.3cm} 
\epsfig{file=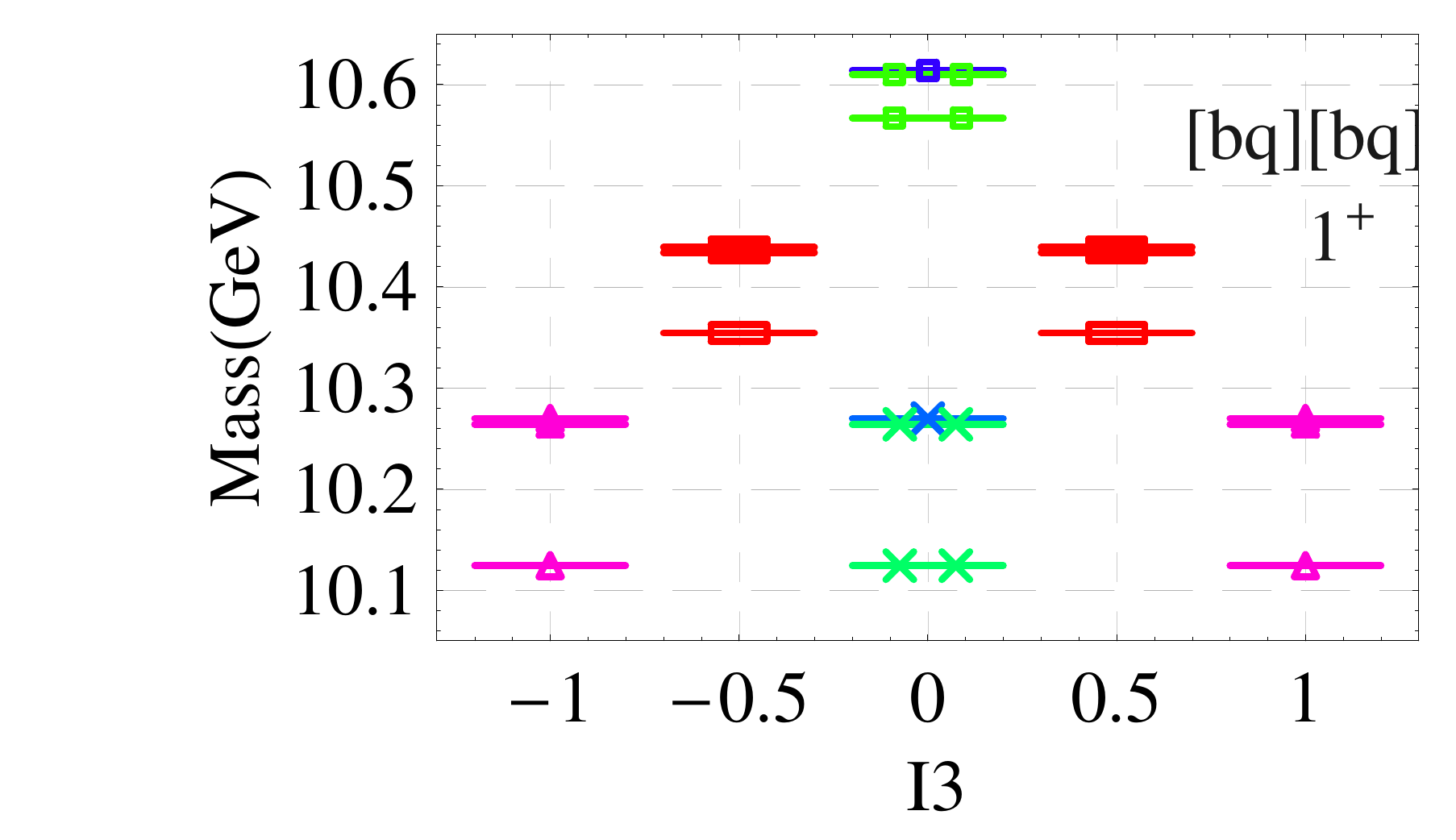,width=6.3cm} 
\epsfig{file=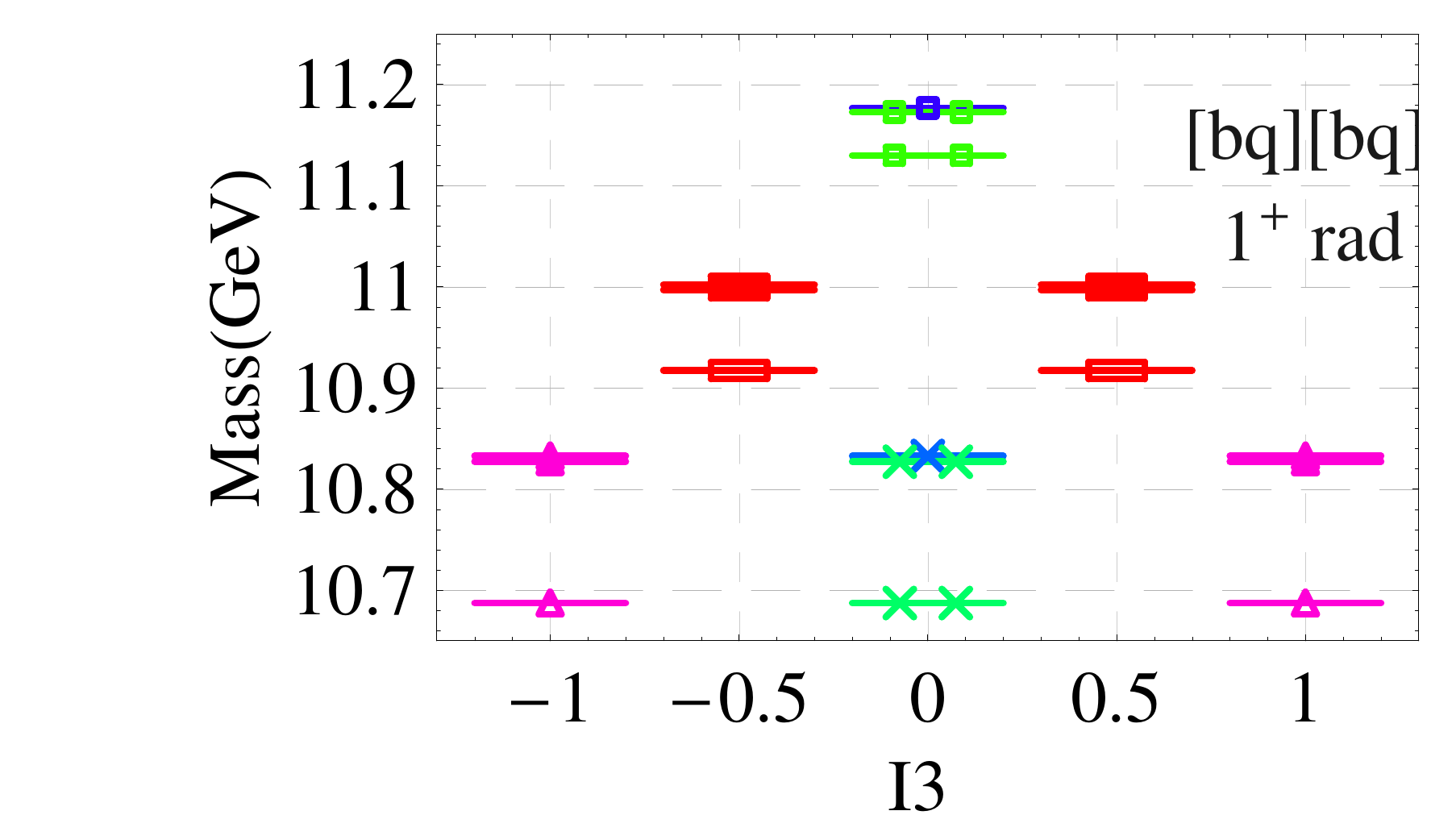,width=6.3cm} 
\end{center}
\epsfig{file=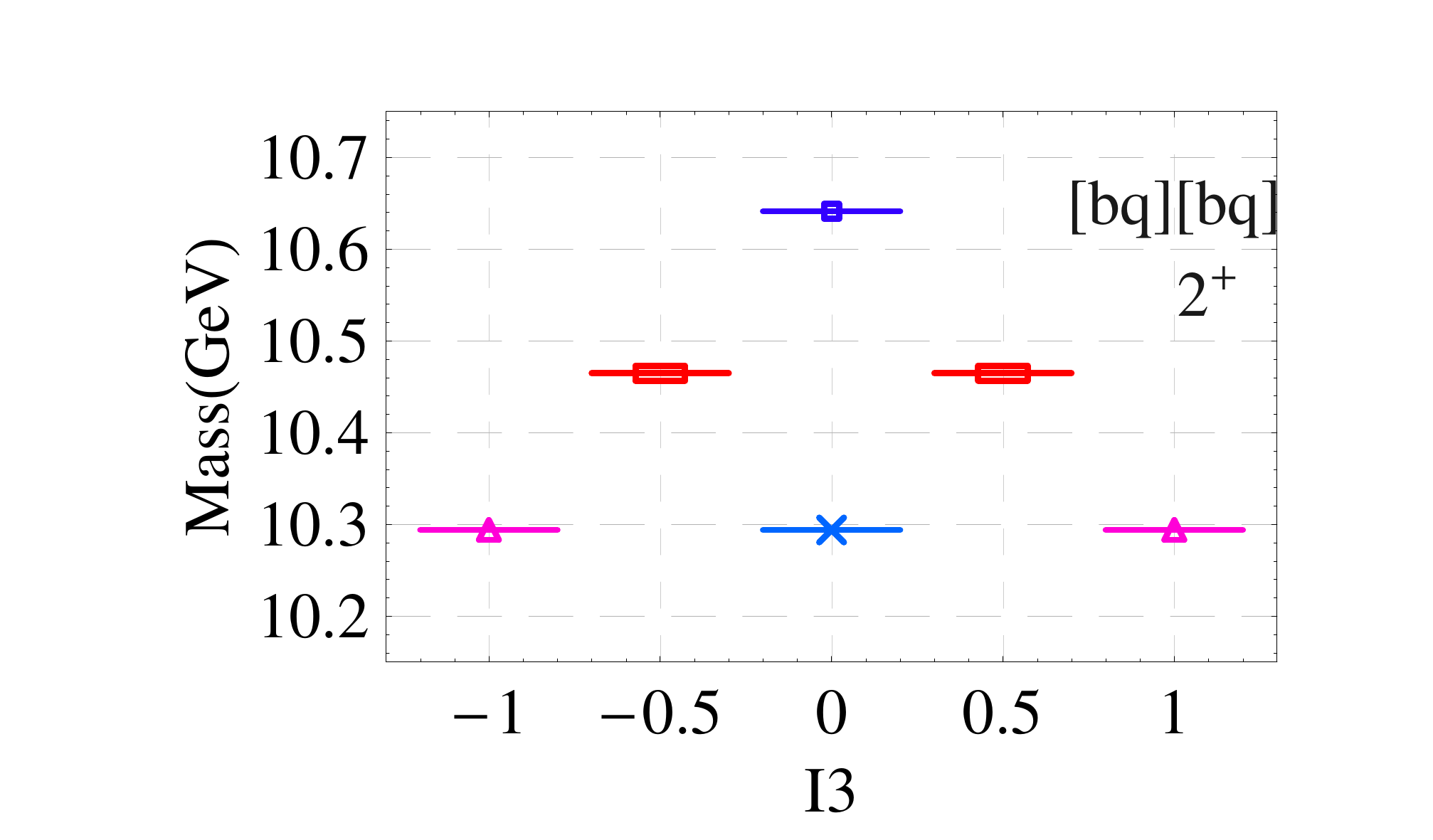,width=7.1cm} 
\caption{\it Hidden bottom multiplets with zero orbital excitation ($L=0$): on left the first radial excitation, on right the second radial excitation. }
\label{fig:bqbq}

\end{figure}

The same analysis can be done for the hidden-bottom tetraquark states and the resulting multiplets are shown in Fig.~\ref{fig:bqbq}. The conclusions discussed above hold for the $[bq][\bar{b}\bar{q}]$ system replacing charm with bottom. The only difference is that for the radial excitations where the only information we have to compute the radial excitation gap comes from the splitting $\Delta$M = 0.563 GeV between the states $\Upsilon(2S)$ and $\Upsilon(1S)$. This value is used independently on $J$.

    {\bf \emph{{The relativistic approach for $L\neq0$}}}

When the two diquarks are orbitally excited spin interactions are suppressed by the $\delta$-Dirac behaviour in Eq.~(\ref{eq:delta}). In this case we can study the mass spectra using a relativistic string model.

If orbital excitation is switched on the system can be visualized as a relativistic string spinning with angular velocity $\omega$ connecting two masses $m_1$, $m_2$ at distances $r_1$ and $r_2$ away from the center of rotation respectively. This is the generalization of the Regge formula relating the masses of the hadrons $M$ to their angular momenta $L$:
\begin{equation}
M^2 = \sigma L 
\label{eq:regge}
\end{equation}
Eq.~(\ref{eq:regge}) arises from solving the equations for a relativistic spinning string with a tension $T = \sigma /2 \pi$, terminated by the boundary condition that both ends move transversely at the speed of light. 


Let us consider a system composed of two masses $m_i$ connected by a string spinning with angular velocity $\omega = v_i/r_i$. Each single mass experiences a relativistic central acceleration, direct along the string axis and given by:
\begin{equation}
T = m_i \omega^2 r_i {\gamma_i}^2
\end{equation}
where $\gamma_i$ it the relativistic factor    
\begin{equation}
\gamma_i = \frac{1}{\sqrt{1-(\omega r_i)^2}}
\end{equation}
The tension force $T$ contributes to the energy of the system with
\begin{equation}
\ud E_i = T \gamma_i \ud r_i = \frac{T}{\omega}\frac{\ud v_i}{\sqrt{1-{v_i}^2}}
\end{equation}
where the relativistic $\gamma_i$ factor arises in the laboratory reference frame. Thus the total energy of a spinning string is the sum of the rest energy and the rotational energy of each mass $m_i$:
\begin{equation}
E_{string} = m_1 \gamma_1 + m_2 \gamma_2 + \frac{T}{\omega}\int_0^{\omega r_1} \frac{1}{\sqrt{1-v^2}} \ud v + \frac{T}{\omega} \int_0^{\omega r_2} \frac{1}{\sqrt{1-v^2}}\ud v. 
\label{hamstring}
\end{equation}   
In the same way the orbital angular momentum is evaluated. The string contribution is
\begin{equation}
\ud L_i = \omega \ud I_i = \omega {r_i}^2 \ud E_i = \frac{T}{\omega^2}\frac{{v_i}^2 }{\sqrt{1-{v_i}^2}}\ud v_i  
\end{equation} 
and adding the single mass terms we have the total angular momentum:
\begin{equation}
L = m_1\omega r_1^2 \gamma_1 + m_2 \omega r_2^2 \gamma_2 + \frac{T}{\omega^2}\int_0^{ \omega r_1} \frac{v^2}{\sqrt{1-v^2}} \ud v + \frac{T}{\omega^2} \int_0^{\omega r_2} \frac{v^2}{\sqrt{1-v^2}}\ud v.
\end{equation}
%

After solving the integrals 
\begin{eqnarray}
&& E_{string}  = m_1\gamma_1 + m_2 \gamma_2 + \frac{T}{\omega}(\arcsin[\omega r_1] + \rm{arcsin}[\omega r_2]),\\
&& L  =   m_1 \omega r_1^2 \gamma_1 +m_2 \omega r_2^2 \gamma_2 + {} \\ 
&& {} \frac{T}{\omega^2} \frac{1}{2} \Big (  -\omega r_1 \sqrt{1-(\omega r_1)^2} + \rm{arcsin}[\omega r_1] - \omega r_2 \sqrt{1- (\omega r_2)^2} + \rm{arcsin}[\omega r_2]  \Big) \nonumber
\end{eqnarray}

An important observation is that hadrons have not only orbital angular momentum but also spin angular momentum. In order to describe in more detail the hadrons mass spectra we can introduce spin-orbit interactions as:
\begin{equation}
E_{SL} = k \frac{1}{R} \bigg (  \frac{\partial E_{string}}{\partial R} \bigg ) \vec{L} \cdot \vec{S}
\end{equation}
where $k$ is a free parameter of the model and $R = r_1 + r_2$. The total energy of the state will then be given by the sum of the two contributions:
\begin{equation}
E = E_{string} + E_{SL}
\end{equation}

The free parameters of this model are: the masses $m_1$ and $m_2$, the string tension $T=\sigma/2 \pi$ and the spin-orbit coupling $k$. Given this input, for each fixed $J,  L,  S$ quantum numbers, the distances $r_1$ and $r_2$ and the angular velocity $\omega$ are defined.

The free parameters are evaluated applying this relativistic string model to  $q\bar{q}$ mesons with $L\neq0$. We use as input the $c\bar{c}$ and $b\bar{b}$ mesons for the hidden charm and hidden bottom tetraquarks states respectively. It is surprising how the fitted values for charm and bottom quarks in this model are the same as in the constituent quark model, within a few MeV. This allows us to use in the relativistic string model the diquark masses already estimated in the framework of a non relativistic constituent quark model. 

\begin{figure}[bht]
\begin{center}
\epsfig{file=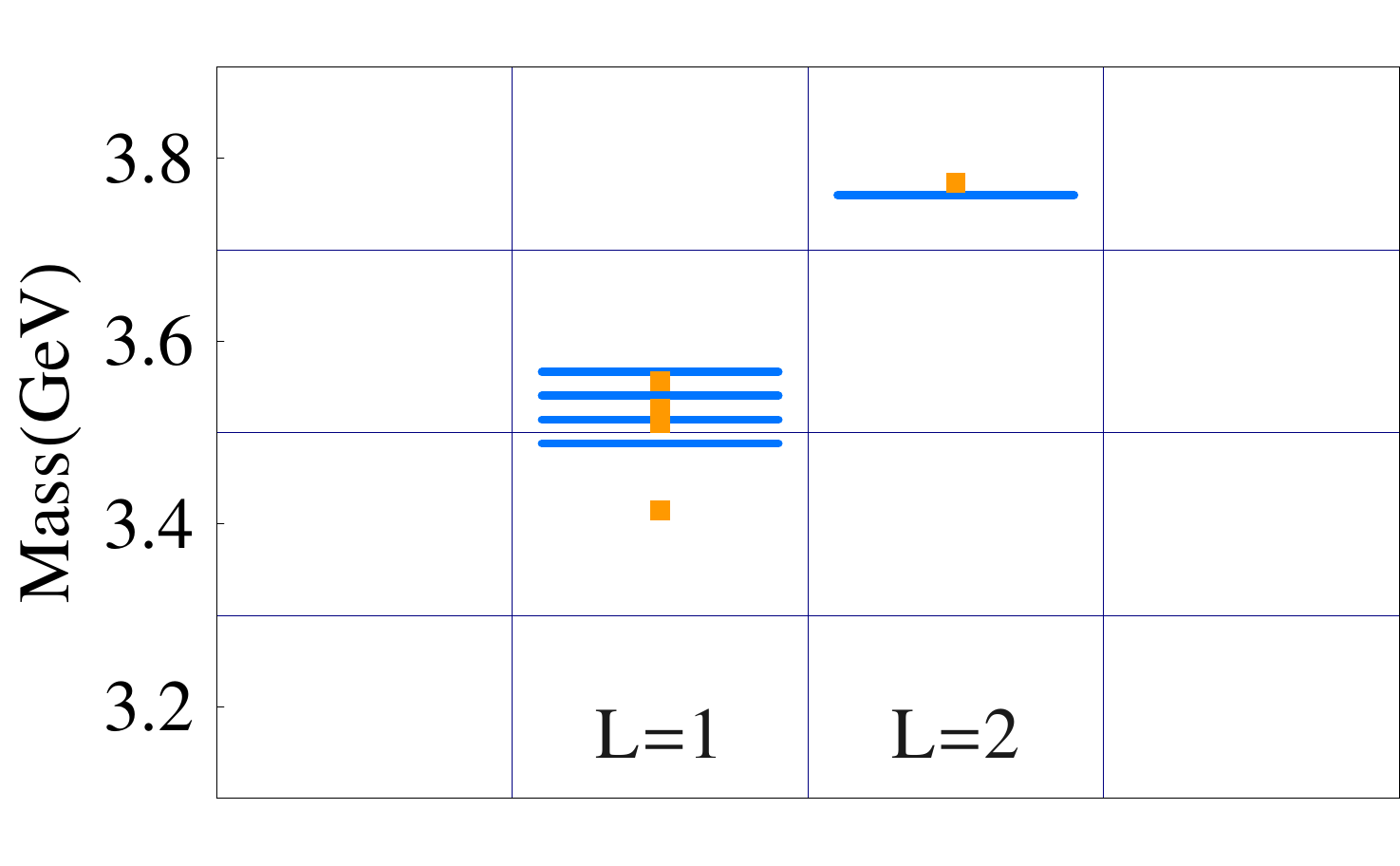,width=6.3cm} 
\epsfig{file=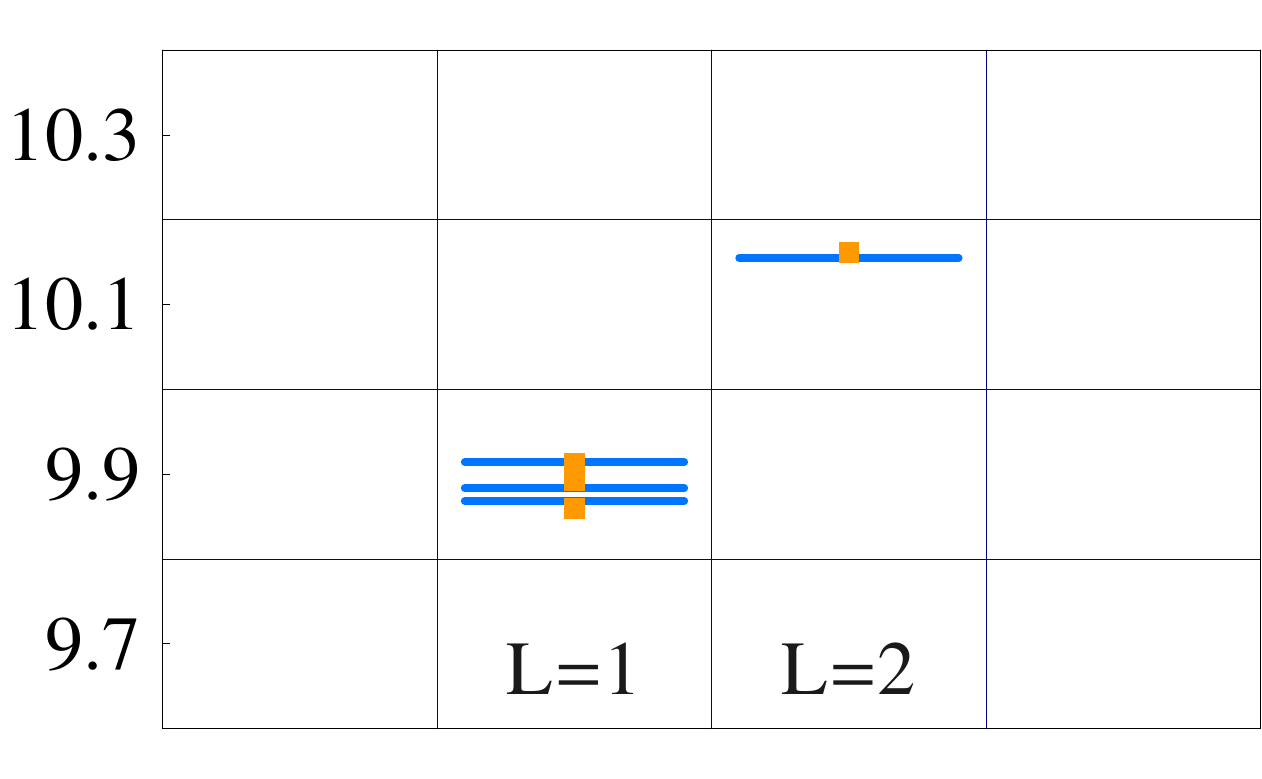,width=6.2cm} 
 \caption{\it Standard $c\bar{c}$ and $b\bar{b}$ mesons: agreement between experimental data and theoretical evaluation. }
    \label{fig:ccbb}
    \end{center}
\end{figure}

In Fig.~\ref{fig:ccbb} the agreement between the experimental data (orange box) and the corresponding fitted mass values is shown.  

Using as input the $\chi_c$ iso-triplet, the $h_c$ iso-singlet and the $\psi(3770)$ mesons we obtain the following parameters used for the determination of the hidden- charm $[cq][\bar{c}\bar{q}]$ tetraquark mass spectra ($L=1$):

\begin{equation}  \left.
\begin{array}{rl} T = 0.159 \, {\rm{GeV}}^2 \\ k = 0.214\, {\rm{GeV}}^{-2} 
\end{array} \right\} \text{hidden charm} \end{equation}

For the hidden bottom sector we have the $\chi_b$ iso-triplet and the $\Upsilon(1D)$ mesons. Fit results give:

\begin{equation}  \left.
\begin{array}{rl} T = 0.261  \, {\rm{GeV}}^2 \\ k = 0.046\, {\rm{GeV}}^{-2}  
\end{array} \right\} \text{hidden bottom} \end{equation}

It is remarkable that the fitted values for the string tension $T = \sigma/2\pi$ in both heavy systems are very close to what is found by Wilczek and Selem in their phenomenological analysis of Regge trajectory in light mesons states \cite{Selem:2006nd}. 

\begin{figure}[bht]
\begin{center}
\epsfig{file=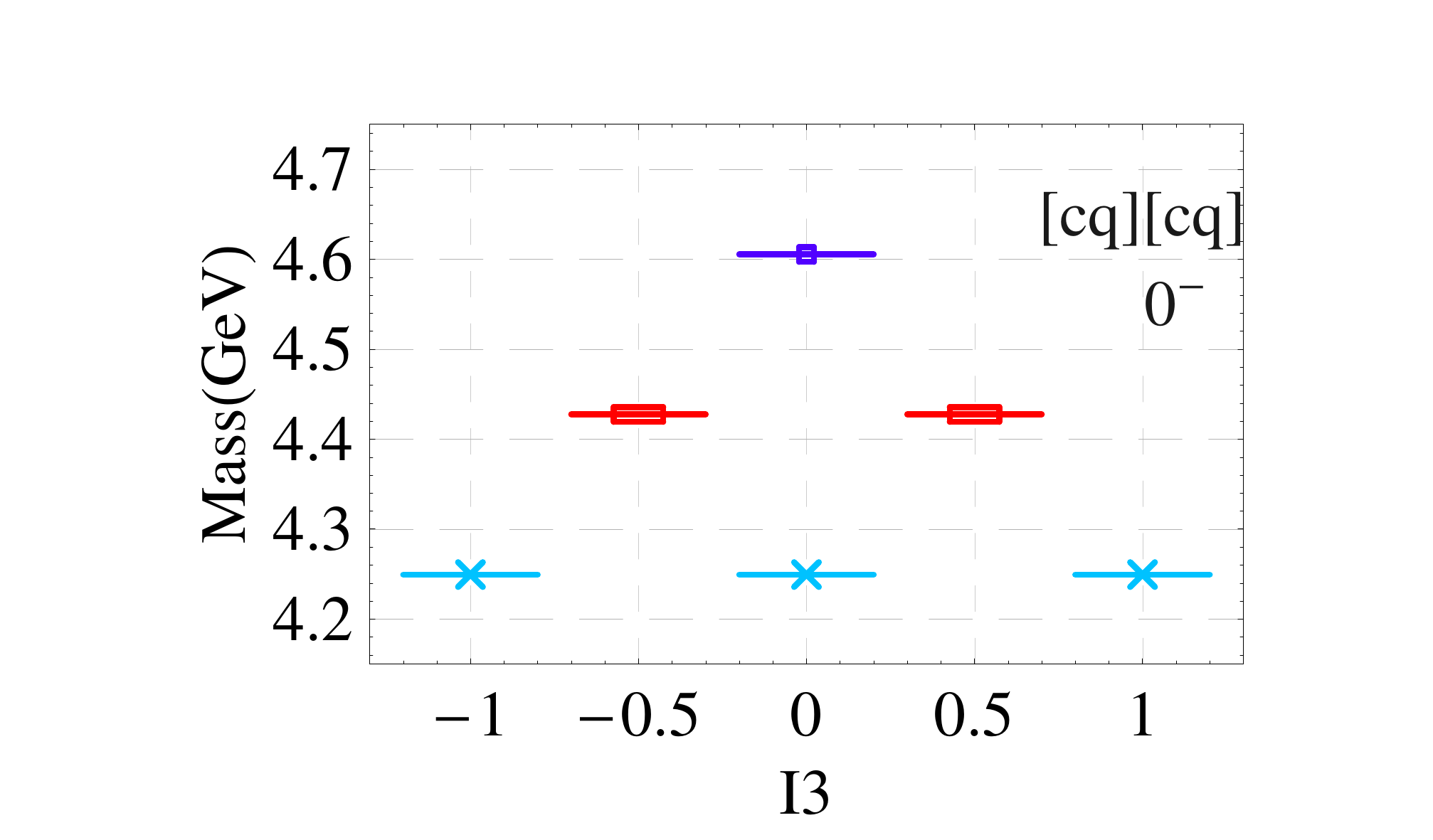,width=6.5cm} 
\epsfig{file=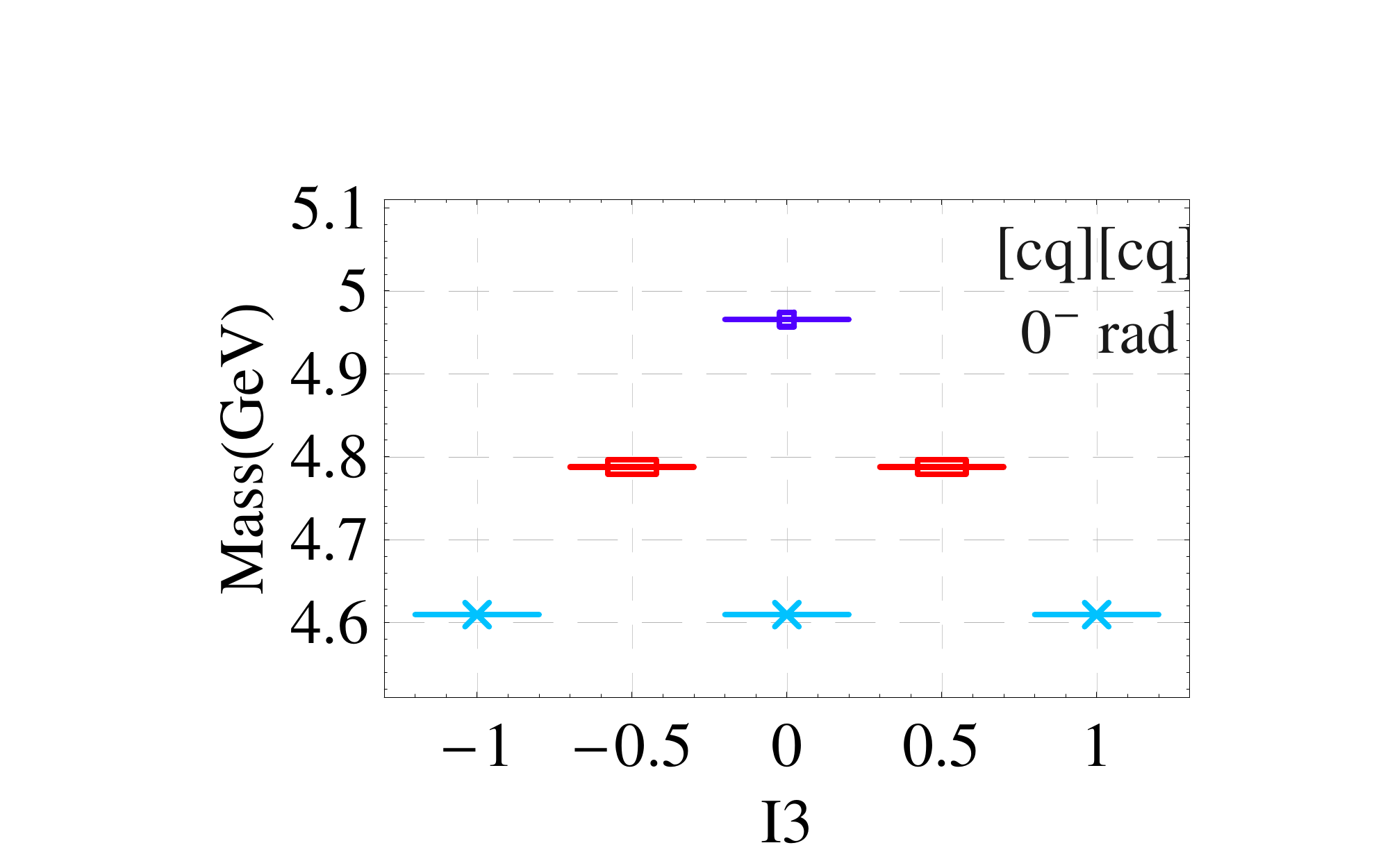,width=6.6cm} 
\epsfig{file=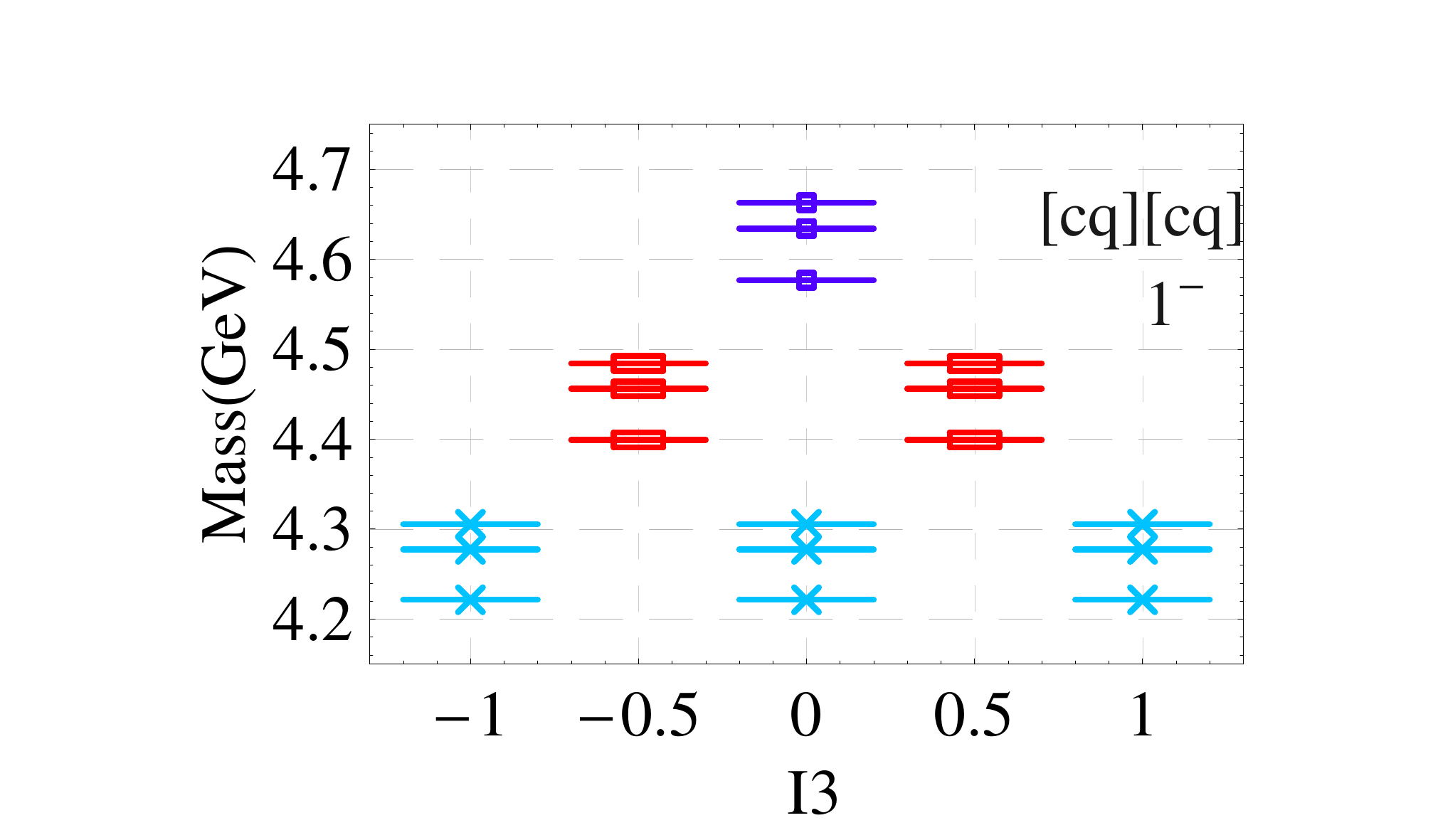,width=6.5cm} 
\epsfig{file=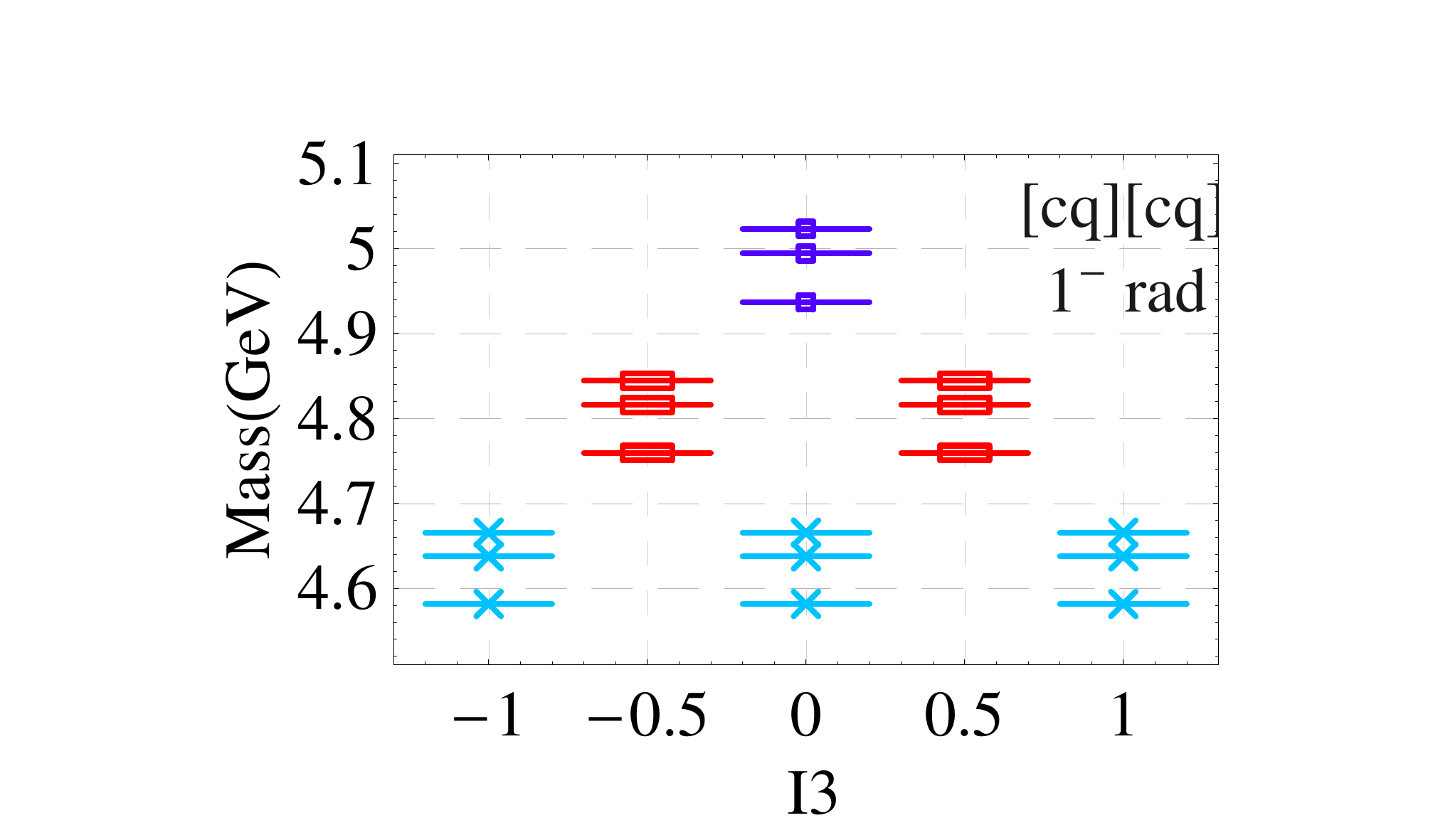,width=6.6cm} 
\end{center}
\epsfig{file=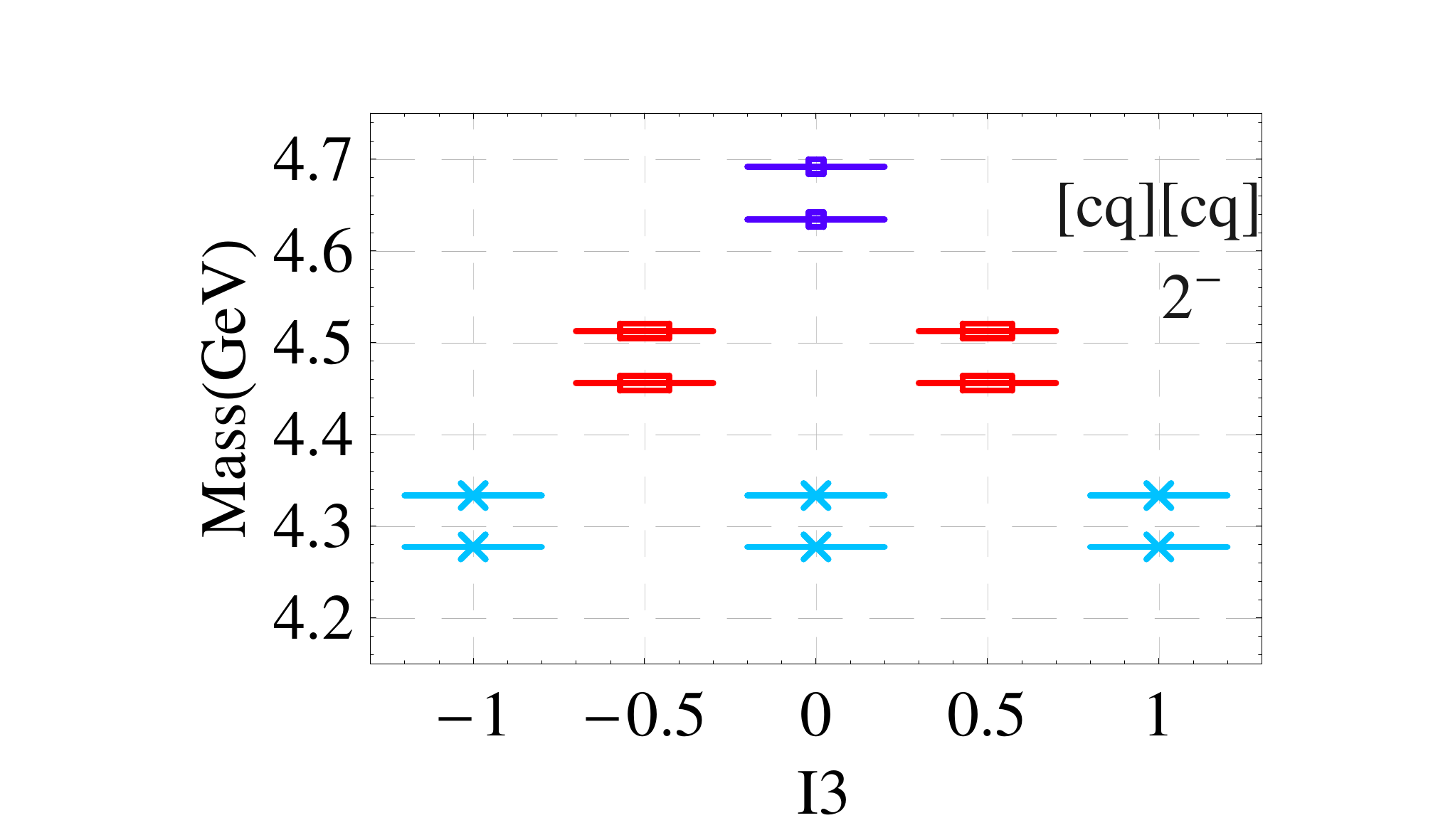,width=6.5cm}
\caption{\it Hidden charm multiplets with L=1: on left the first radial excitation, on right the second radial excitation. }
\label{fig:cqcql1}
\end{figure}

In Fig.~\ref{fig:cqcql1} are organized the $[cq][\bar{c}\bar{q}]$ multiplets. The orbital excitation $L=1$ leads to negative parity states. We have:
\begin{description}
\item[$J^P = 0^-$] multiplet is realized only if the total spin of the tetraquark is $S=1$. Since the model depends only on the total spin (and not on the single diquark spins) we have the same mass value for $[cq]_{S=0}[\bar{c}\bar{q}]_{S=1}$and $[cq]_{S=1}[\bar{c}\bar{q}]_{S=1}$states. For this reason each shown state is a double-degenerate state. Different colors(symbols) refers to different flavor compositions: starting from the lightest states we have in light blue(star) a iso-triplet with no strange quarks, at higher mass values we find an iso-doublet with open-strangeness drawn in red(rectangle)and the heaviest state is a hidden-strange iso-singlet drawn in dark-blue(box). 

\item[$J^P = 1^-$] multiplet is realized with three different total spin angular momenta $S = 0$(double-degenerate states), $S = 1$(double-degenerate states), $S = 2$ and the corresponding mass splitting is shown for each different flavor composition: in light-blue(star) we have no-strange states, in red(rectange) we have open-strange states, in dark-blue(box) we have hidden-strange states. We observe that, as in the $h_c$-$\chi_{c1}$ system, the mass values increase with decreasing tetraquark total spin.       

\item[$J^P = 2^-$] multiplet is realized with two different total spin angular momenta: $S = 1$(double-degenerate states) and $S=2$. The same color legend as for $J^P = 1^-$ is used and again we observe that the mass values increase with decreasing tetraquark total spin.   

\item[radial excitation] multiplets with $L=1$ are estimated assuming the splitting $\Delta$M = 0.360 GeV observed in the $\chi_b$ multiplets for both hidden charm and hidden bottom tetraquark mesons.  
\end{description}

\begin{figure}[bht]
\begin{center}
\epsfig{file=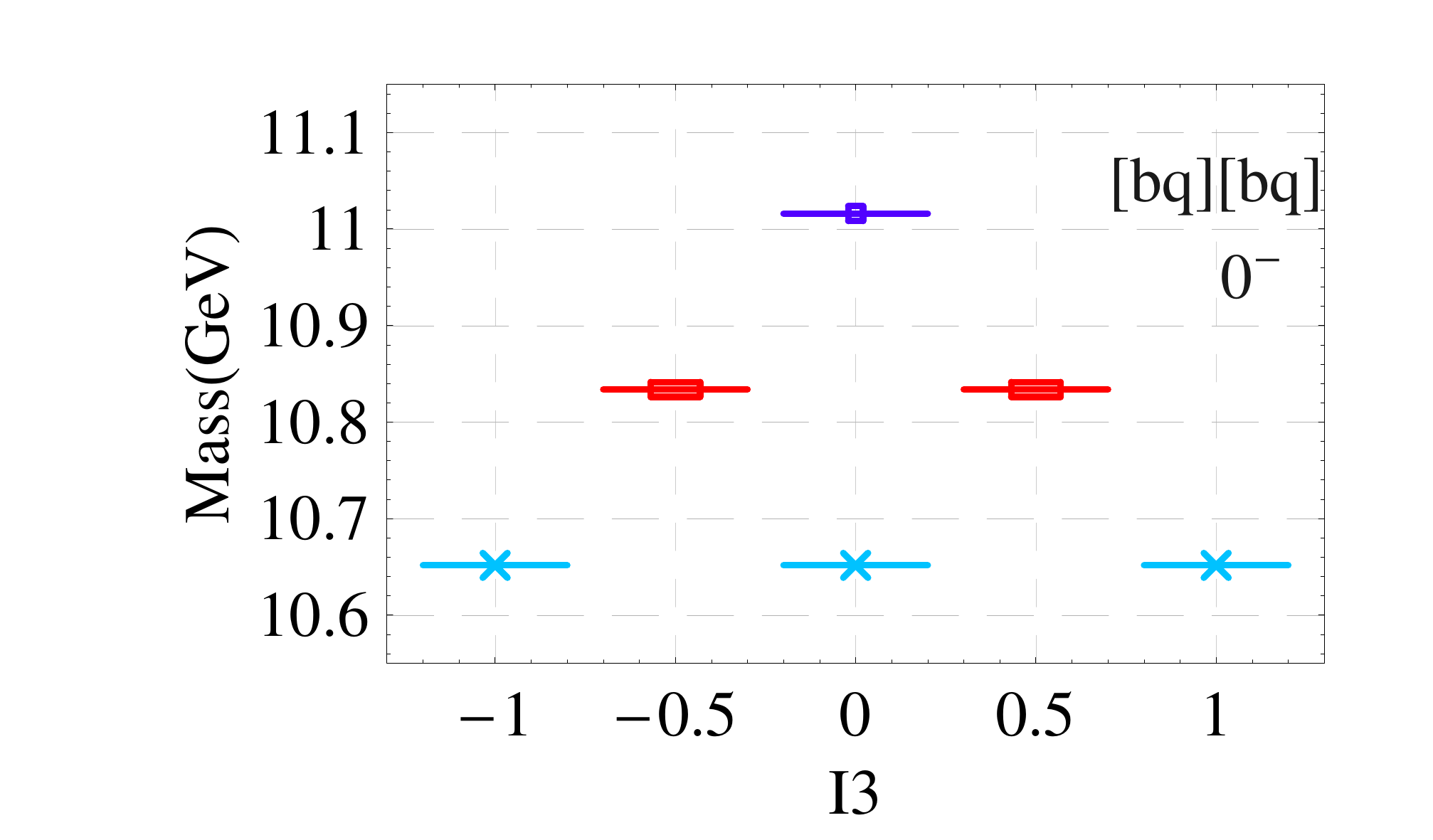,width=6.5cm} 
\epsfig{file=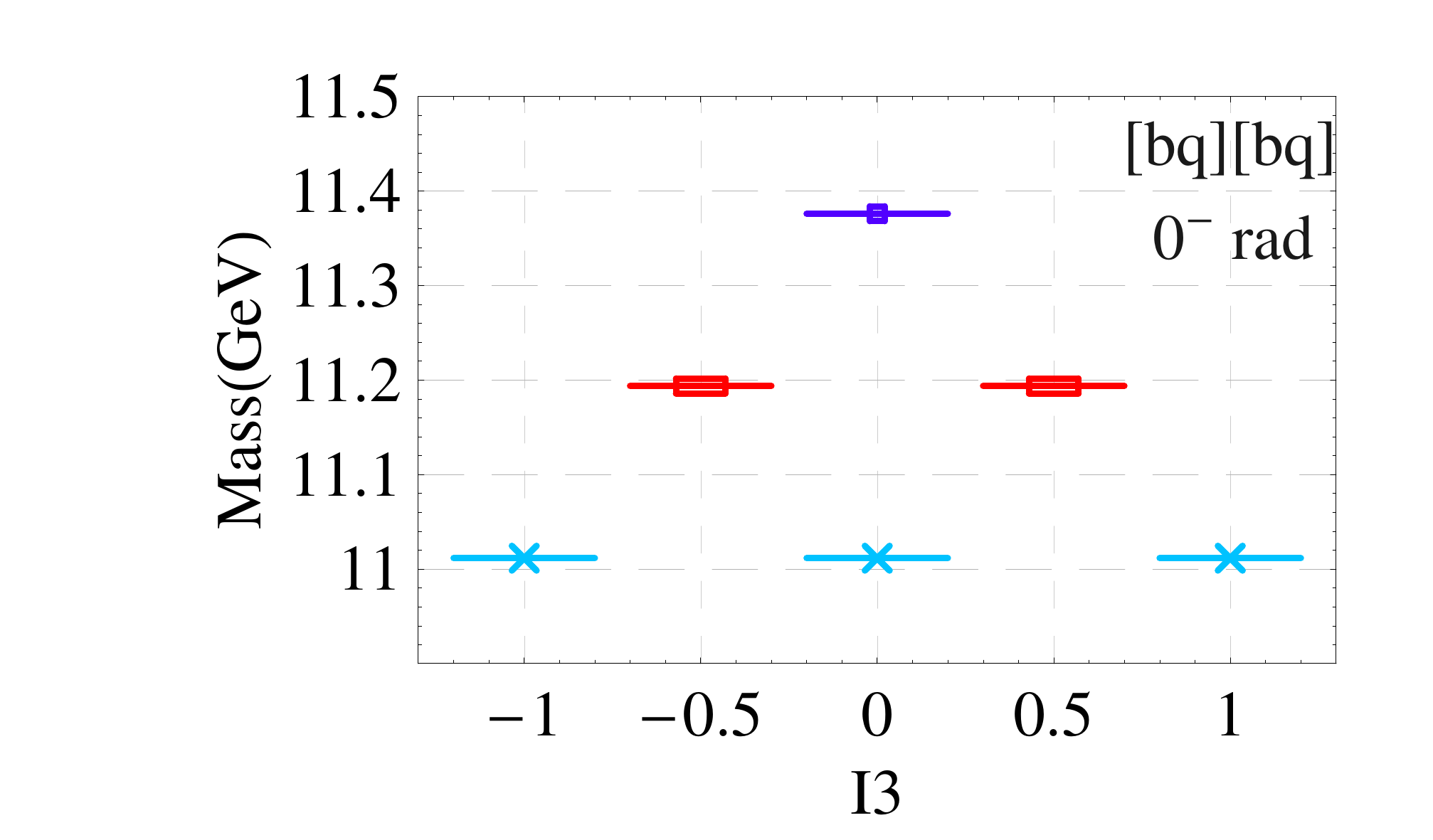,width=6.6cm} 
\epsfig{file=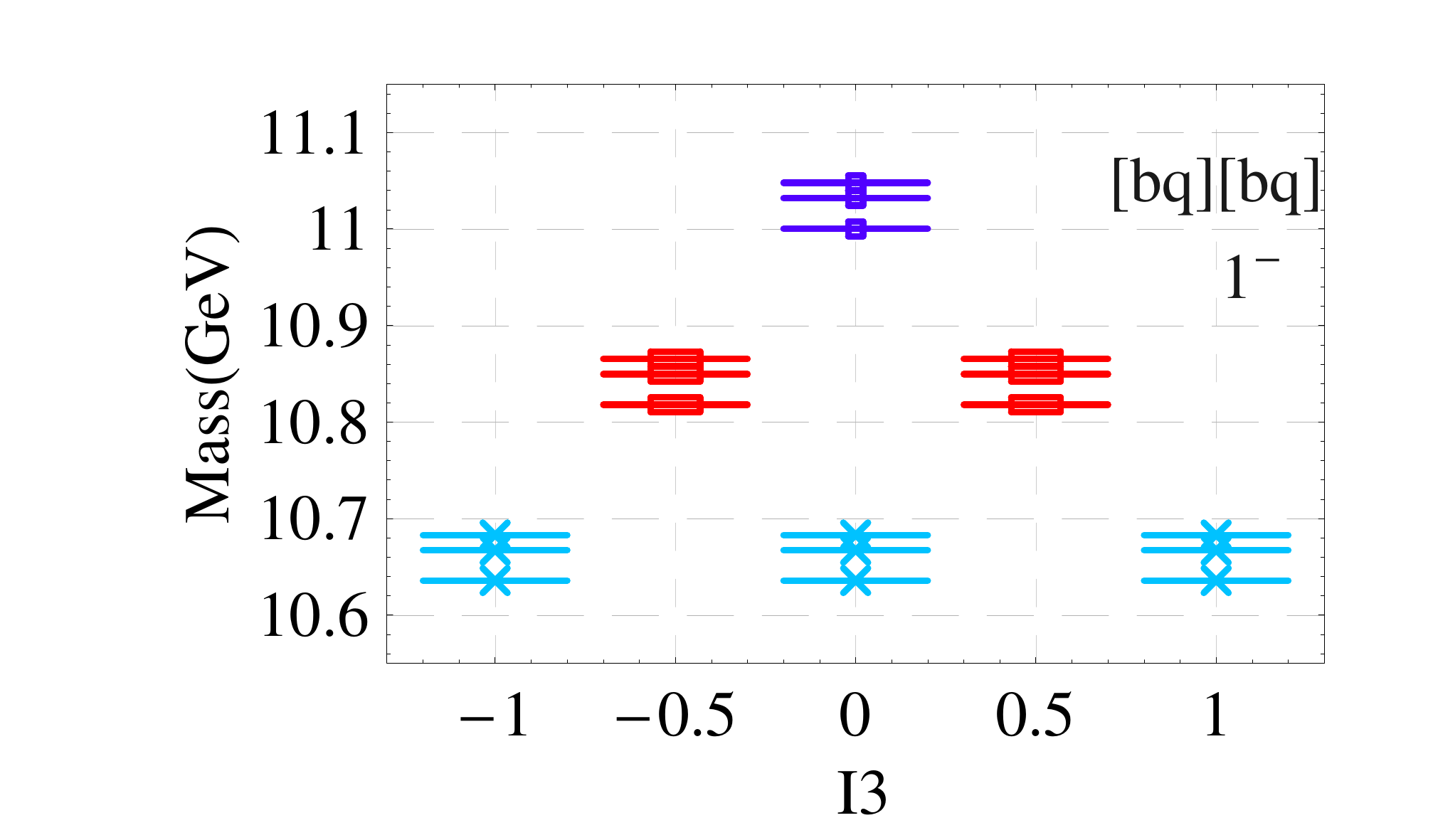,width=6.5cm} 
\epsfig{file=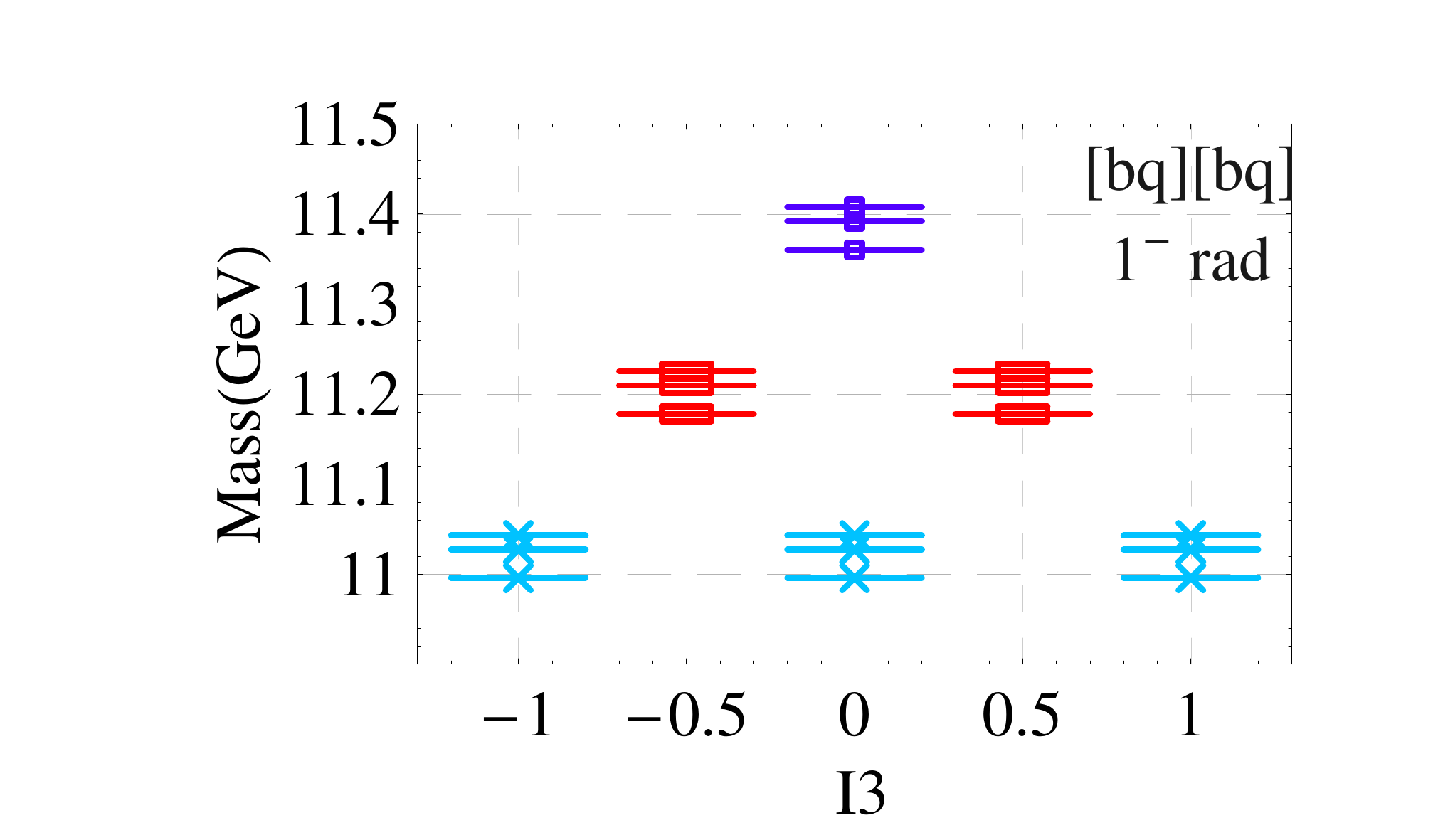,width=6.7cm} 
\end{center}
\epsfig{file=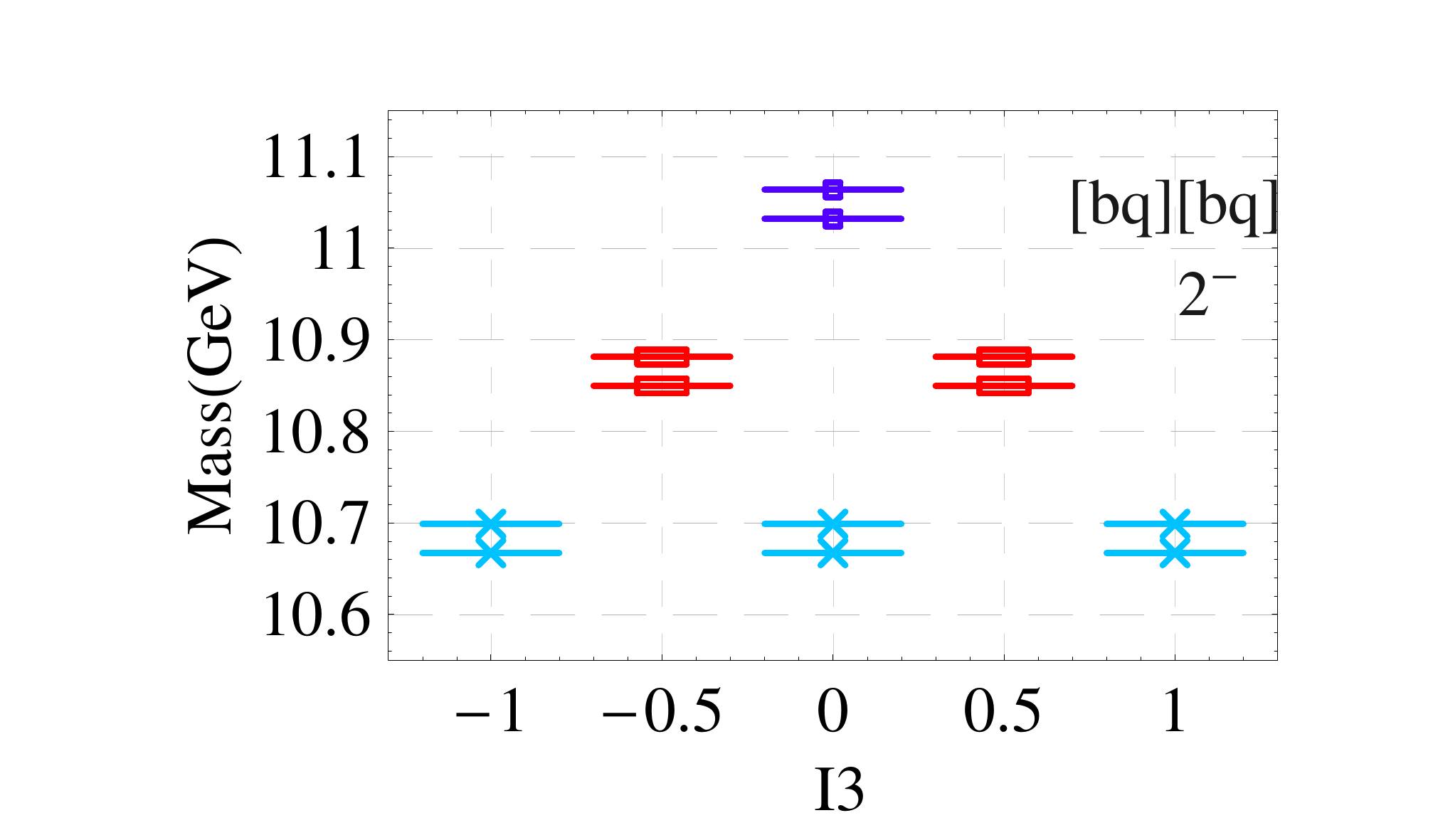,width=6.5cm}
\caption{\it Hidden bottom multiplets with L=1: on left the first radial excitation, on right the second radial excitation.}
\label{fig:bqbql1}
\end{figure}



\subsubsection{Production and Decays}

As shortly discussed in \cite{Drenska:2008gr} there are different decay modes for a given tetraquark state. For heavy states, like hidden charm and hiddem bottom, the most important are:
\begin{itemize}
\item{quark-antiquark pair production: the final state will be a two particle state, a tetraquark and a $q\bar{q}$ meson;}
\item{quark-antiquark exchange diagram: depending on which quarks exchange we have different two-mesons final states;}
\item{stretching the color string that bind the diquark to the antidiquark, a quark-antiquark pair is formed, letting two baryons in the decay products.}
\end{itemize}

\subsection{Molecules}
\label{subsec:molecules}
Among the possible multiquark states, the main alternative to tetraquarks are hadron molecules, bound states of two (or more) hadrons. 
They were proposed a long time ago to describe the deuteron as a bound state of nucleons. 
Afterwards, many of the resonances which do not fit in the standard $q\bar{q}$ and $qqq$ scenario have been given a molecular assignment. 
Nevertheless, it is challenging to identify possible multiquark states in an environment of many broad and often overlapping conventional states. 
More reliable experimental indications of the existence of exotic hadronic candidates came recently when searching for more $c\bar{c}$ mesons. 
The first hadron molecule interpretation in the charm meson sector was proposed for the $\psi(4040)$, observed in $e^{+}e^{-}$ annihilations~\cite{okun,derujula}.

The main difference between molecules and tetraquarks is that the hadrons bound in the molecule preserve their ``atomic integrity". 
This obviously leads to different predictions for spectra, production mechanisms and decay rates.
Before analyzing  the details of the dynamics we give some distinctive signatures of hadronic-pair candidates~\cite{Barnes:1994pd}:

\begin{enumerate}

\item in the light sector $J^{PC}$ and flavor quantum numbers should be compatible with those of an $L=0$ hadron pair 
(nuclear forces which bind the two mesons together are short ranged so that a $P$-wave bound state seems unlikely).

In the heavy mesons sector there is the possibility of $P$-wave bound states but this would imply the existence of more deeply bound $S$-wave molecules~\cite{Close:2009ag}.

\item binding energies should be of the order of $50~{\rm MeV}$ for the light mesons and of the order of $10~{\rm MeV}$ for the heavy ones. 
This is because the minimum distance required for the hadrons to maintain separate identities is $R\thicksim1~{\rm fm}$ and the binding energy is  $E_{B}\thicksim 1/(2\mu R^{2})$ 
(where $\mu$ is the reduced mass).
As an example $K\bar{K}$ molecules have $\mu\thicksim 500~{\rm MeV}$ and thus $E_{B}<50~{\rm MeV}$, 
whereas $D\bar{D}$ molecules have $\mu\thicksim 1~\rm{GeV}$ and thus $E_{B}<10~{\rm MeV}$, 
and finally for $B\bar{B}$ molecules $\mu\thicksim 5~{\rm GeV}$ leading to $E_{B}<4~{\rm MeV}$.

\item Large branching ratios into final states containing the constituent mesons despite the reduced phase space available.

\item Anomalous electromagnetic coupling 
with respect to the ordinary charmonium states.

\end{enumerate}

The spectrum of hadron bound states can be predicted once a model for the interaction between the hadrons has been proposed. 
The interaction between two hadrons varies with the distance: at short distance quarks inside the hadron interact with each other through the exchange of gluons, 
at long distance the exchange of mesons ($\pi$, $\omega$, $\rho$, ...) between the hadrons themselves is dominant. We now discuss separately these two dynamical regimes. 

\textit{Gluon exchange}.
Hadrons are bound states of quarks, which interact with each other through electric and color charge. 
Thus, one can study the problem of hadron-hadron interactions and consequently hadron-hadron binding by looking at their constituents interactions. 
This interactions is mediated by the gluons, which are the vector boson particles associated with the gauge fields of the SU(3) non abelian gauge field theory
 which quantitatively describes the strong interactions.

At first order in perturbation theory using $g_{s}$ (the strong coupling constant) as perturbative parameter, 
we are able to compute the non relativistic potential associated with one gluon exchange diagrams. 
This results in a Coulombic term, to which a linear confining term must be added to account for the confinement in the strong coupling limit 

\begin{equation}
\label{conf}
V^{c}_{ij}= \left(-\frac{\alpha_{s}}{r_{ij}}+\frac{3}{4}b r_{ij}\right)\frac{\lambda^{a}_{i}}{2}\left(-\frac{\lambda^{a*}_{j}}{2}\right)
\end{equation}

\noindent where $r_{ij}$ is the distance between the $i$-th and $j$-th quarks or antiquarks which constitute the hadrons and $b$ is a string tension parameter.
The confinement is indeed phenomenologically implemented through a string model. The string is responsible for the attractive force
between two quarks at large distances. The energy associated to a string with two masses at its ends is $\propto br$, $b$ is the string tension.

Relativistic corrections to the potential can be included  as in the case of the hydrogen atom, just introducing in the calculation the SU(3) generators $\lambda^{a}$.
Through a perturbative expansion in $(p/m)^2$ one is able to derive the fine and hyperfine structure equivalent, in the atomic model,  to the spin-orbit and Darwin terms and
contact spin-spin interactions, respectively. The two terms, in the case of strong interactions read:
\begin{equation}
\label{fine}
V^{r}_{ij}=-\frac{8\pi}{3}\frac{\alpha_{s}}{m_{i}m_{j}}\left(\frac{\sigma^{3}}{\pi^{3/2}}e^{-\sigma^{2}r^{2}_{ij}}\right)\left(\vec{s}_{i}\cdot\vec{s}_{j}\right)\frac{\lambda^{a}_{i}}{2}\left(-\frac{\lambda^{a*}_{j}}{2}\right)+V^{SO}
\end{equation}

The hyperfine term can be understood as a dipole-dipole interaction: to each quark is associated a chromo-magnetic dipole moment which is proportional, 
just as in the electron case, to the coupling $g_{s}$ and to the inverse of its mass. This spin-spin term is weighted by a smeared delta function, 
usually a gaussian, which accounts for the short distance nature of the interaction. 
The interaction potential is, thus, the same as that used successfully to describe the conventional meson/baryon spectrum.
The parameters of the interaction hamiltonian can be thus derived from ordinary meson spectroscopy, and there is generally a quite broad agreement in the 
literature, as summarized in Tab.~\ref{tab:ogepar}.

\begin{table}[h!]
\caption{One Gluon Exchange parameters.}
\label{tab:ogepar}
\begin{tabular}{lcccccccc}
\hline
& $m_{u}$(GeV) & $m_{s}$(GeV) &$m_{c}$(GeV) &$\alpha_{c}$ & $C$ (GeV)& $b$(GeV) & $\alpha_{h}$ & $\sigma$(GeV) \\
\hline
\cite{weinstein3} & 0.375 &0.600 & - & 0.748 & -0.777 & 0.178 & - & - \\
\cite{swanson} & 0.375 &0.650 &- & 0.857 & -0.4358 & 0.154 & 0.840 & 0.70 \\
\cite{Swanson:2003tb} & 0.335 &0.550 & 1.600 & 0.590 & - & 0.162 & 0.590 & 0.9 \\
\hline
\end{tabular}
\end{table}



\textit{Meson exchange}.
The existence of bound states of two mesons due to one pion exchange was proposed for the first time by Tornqvist in \cite{tornqvist}. 
The inspiration came from the deuteron, a bound state of two nucleons which interact through the long range One Pion Exchange potential:

\begin{equation}
\label{ope}
V_{\pi}(\vec{r})=\frac{g^{2}_{\pi N}}{3}\left(\vec{\tau}_{1}\cdot \vec{\tau}_{2}\right)\left[\left(3(\vec{\sigma}_{1}\cdot \hat{r})\; (\vec{\sigma}_{2}\cdot \hat{r}) -\vec{\sigma}_{1}\cdot \vec{\sigma}_{2}\right)W(r)+\left(\vec{\sigma}_{1}\cdot \vec{\sigma}_{2}\right) \right]\frac{e^{-m_{\pi}r}}{r}
\end{equation}

\noindent with

\begin{equation}
W(r)=1+\frac{3}{(m_{\pi}r)^{2}}+\frac{3}{m_{\pi}r}\\
\end{equation}

\noindent where $\sigma_{1,2}$ are the spins of the nucleons, and $\tau_{1,2}$ their isospins. 
The potential contains a scalar term proportional to the spin-ispospin factor with a pure Yukawa potential, and a tensor term, which is a higher order correction.
Let's discuss the deuteron case, since it will be useful for comparison later on in our discussion. 
The spin-isospin factor gives an indication of the potentially binding channels:

\begin{equation}
\label{rbn}
\left(\vec{\tau}_{1}\cdot \vec{\tau}_{2}\right)\left(\vec{\sigma}_{1}\cdot \vec{\sigma}_{2}\right)=\frac{1}{4}\left[I(I+1)-\frac{3}{2}\right]\left[S(S+1)-\frac{3}{2}\right]=\begin{cases}&+9/16 \; {\tiny (S=0,I=0)}\\ &-3/16\; {\tiny (S=0,I=1)}\\ &-3/16 \;{\tiny (S=1,I=0)} \\ &+1/16 \;{\tiny (S=1,I=1)}\end{cases}
\end{equation}

The binding is expected in the $(S=0,I=1)$ channel and in the $(S=1,I=0)$ channel,  the deuteron one.
The first possibility is ruled out when including also the tensor term, which instead strengthens the attraction in the second configuration adding a {\it D}-wave component.

Tornqvist speculated on the possibility that such a potential could bind pairs of mesons, calling  \textit{deusons} these deuteron-like bound states.
Since the pion is very light, deusons can be very large, much larger than ordinary $q\bar{q}$ mesons. 
For ground state mesons there are two possibilities: pseudoscalar-vector mesons and vector-vector mesons,
since bound states of pseudoscalar-pseudoscalar mesons are forbidden by parity.  
When considering {\it S}-wave bound states, the relative strength and sign of the potential for different spin and isospin channels is given by the so-called
{\it relative binding number}, firstly introduced by Tornqvist in \cite{tornqvist}, which is the analog of (\ref{rbn}) for meson pairs. 
Thus the relative binding number gives an indication of the binding channels.
If one allows for higher angular momentum the spectroscopy becomes more complicated and more detailed calculation are needed. 
Moreover in \cite{Close:2009ag,Close:2010wq} the possibility for {\it S}-wave pions is considered.

Manohar and Wise studied in \cite{Manohar:1992nd} the interaction between two heavy-light mesons $Q\bar{q}$ and considered the possibility that a $BB$
bound state could exist. 
Their derivation of the potential proceeds from the Heavy Quark Effective Theory formalism combined with the chiral Lagrangian approach to describe the light pseudoscalar mesons. 
Their potential agrees with (\ref{ope}) .
The same calculation scheme is applied by Tornqvist in \cite{tornqvist2} to meson-antimeson bound states (remind that the meson-antimeson interaction potential is the opposite of the meson-meson one).
The only parameter of the model, besides the mesons masses, is the pion-meson-meson coupling  $g$. Its value can be deduced from the $\pi N$ coupling exploiting the relation

\begin{equation}
\label{pionnu}
\frac{g^{2}_{\pi N}}{4\pi}=\frac{25}{9}\frac{g^{2}}{f^{2}}m^{2}_{\pi}
\end{equation}

\noindent where $f$ is the decay constant of the pion. 
This relation arises when one uses chiral perturbation theory to describe the pseudoscalar octet mesons.
In this way one can derive the effective quark-pion interaction
\begin{equation}
\mathcal{L}_{int}=\frac{g}{f}\bar q \gamma^\mu\vec{\tau}q\,\partial_\mu\pi
\end{equation}
\noindent Treating the quark fields as constituent quarks and using the non relativist quark model to derive 
hadronic matrix elements gives the relation~(\ref{pionnu}) between the $\pi N$ coupling and the $\pi D\bar{D}^{*}$ one.
The measured value $g^{2}_{\pi N}\thicksim 1$ leads to $g^{2}\thicksim 0.6$.
A striking confirmation of this estimate is the prediction of the $D^{*}$ width of $\sim$70~KeV which was later confirmed.
The potential is computed for each of these states paying attention to the fact that there will be some coupled channels 
(states with same $J$ and $S$ but different $L$ will mix with each other).

However, because of the singular nature of the tensor part of the potential, {\it i. e. } the first term in (\ref{ope}), a regularization procedure is needed. 
The most natural method is to introduce a form factor at the $\pi N$vertex, which gives to the pion source a spherical extension
with radius $R\sim 1/\Lambda$, with $\Lambda$ an ultraviolet cutoff.
Even if the phenomenological knowledge of the cutoff $\Lambda$ is rather poor, it may be fixed by comparison with nuclear physics.
From $NN$ interactions $\Lambda$ must be in the range 0.8-1.5 GeV, while to reproduce the deuteron binding energy one needs $\Lambda \thicksim$0.8 GeV. 
The value employed in \cite{tornqvist2} and in the literature in general is $\Lambda\thicksim$1.2 GeV, which seems appropriate for $D$ mesons. 
It is crucial here to emphasize that the existence or otherwise of meson-meson bound states depends strongly on the value chosen for $\Lambda$ \cite{tornqvist,swanson}. 

\subsubsection{Spectra}We briefly review the result for the spectra of possible molecular states as obtained in the two approaches.

\textit{Gluon exchange}. 
This interaction scheme has been used to study a number of systems with different approaches from case to case.
 
Weinstein and Isgur, by using the potential terms of Eq.~(\ref{conf},\ref{fine}), showed that 
$f_{0}(975)$ and $a_{0}(980)$ could be interpreted as a $K\bar{K}$ molecule, \cite{weinstein1,weinstein2,weinstein3}. 
They find the full four-quark wavefunction of the bound states using a variational Gaussian basis. 
Then they invert the Schroedinger equation to obtain an effective potential and integrate this potential to obtain the observed phase shifts in $\pi\pi$ scattering for $a_{0}$ and $f_{0}$.
This model predicts that in general the $qq\bar{q}\bar{q}$ (with $q=u,d,s$) ground states are two unbound mesons except for a $K\bar{K}$ molecule.

Another case in which the one gluon exchange has been successfully exploited is that of $f_{1}(1420)$ as a $K^{*}\bar{K}$ bound state  \cite{caldwell,longacre}.

Furthermore in a series of papers by Swanson and Barnes \cite{swanson,Barnes:1991em,Swanson:2001sb} meson-meson interactions
have been obtained extracting an effective potential between the two mesons from the Born order scattering amplitude computed from Eq.~(\ref{conf},\ref{fine}). 
Such interactions necessarily involve the exchange of quarks between mesons. This is because the application of the one gluon exchange hamiltonian induces transitions 
among different color configurations. For example  $\langle 1_{i}\otimes 1_{j}|\lambda^{a}_{i}\lambda^{a*}_{j}|8_{i}\otimes 8_{j}\rangle\neq 0$. 
When searching for molecules one wants to deal always with two color-singlet objects and thus it is necessary to exchange quarks or antiquarks 
between the two mesons to obtain again a $1\otimes 1$ configuration. Nevertheless this consideration has a conceptual drawback, as explained 
in \cite{swanson},  since if quark exchange must occur the range of the effective potentials is limited by the mesonic radii to roughly 1~fm.
Thus a two-meson bound state obtained from one gluon exchange has a spatial extension which cannot exceed 1~fm: at this point the difference between 
hadronic molecule and tetraquark state seems to be just a matter of language, the only difference between the two being the way in which color is saturated.

\textit{Meson exchange}. 
Bound states of mesons due to one pion exchange have been studied in \cite{Manohar:1992nd,tornqvist2} and many others.

In \cite{Manohar:1992nd} the authors compute the binding energy of an eigenstate of the potential using a variational calculation. 
They assume that the short range potential, which is dominated by vector meson exchange, cannot be repulsive and find that the 
$BB^{*}$ in the ($I=0$, $S=0$) channel leads to a binding energy of 8.3 MeV (for $g^{2}=0.5$). The authors state that the value of 
the binding energy is sensitive to the precise value of $g^{2}$ and to the value chosen for the potential at short distances. 
They neglect the contribution of heavier mesons exchange, arguing that they are less important than in the nucleon-nucleon case.
On the other hand no $DD$ or $DB$ bound state is found because of the positive contribution from the kinetic energy which overwhelms the attraction due to the potential.

In \cite{tornqvist2} the method used to find bound states is to solve numerically the Schroedinger equation with the one pion exchange potential of Eq.~(\ref{ope}). 
The author considers flavor non exotic states in the heavy-heavy (Tab.~\ref{tab:tornneh}) and light-light sectors (Tab.~\ref{tab:tornnel}), 
and the flavor exotic heavy-light and light-light states (Tab.~\ref{tab:torne}). 
The general pattern is that for pseudoscalar-vector mesons the binding is more likely in $J^{PC}=0^{-+},1^{++}$, 
while for vector-vector ones in $J^{PC}=0^{-+},0^{++},1^{+-},2^{++}$. In both cases the isospin configuration is $I=0$.

\begin{table}[ht!]
\caption{Flavor non exotic bound states in the heavy sector found in \cite{tornqvist2}. 
The $D^{(*)}\bar{D}^{(*)}$ are predicted to be almost at threshold, compatible with zero binding energy, 
while for the $B^{(*)}\bar{B}^{(*)}$ molecule the typical binding energy is of the order of $50~{\rm MeV}$.}
\label{tab:tornneh}
\begin{tabular}{lccc}
\hline
Constituents & $M$~(MeV) & $J^{PC}$ & $\Lambda$(GeV) \\ 
\hline
\multirow{2}{*}{}
$D\bar{D}^{*}$ & $\thicksim3870$ &$0^{-+}$ & $>1.5$ \\ 
& $\thicksim 3870$ & $1^{++}$ &$1.2$\\ 
\hline
\multirow{2}{*}{}
$D^{*}\bar{D}^{*}$ & $\thicksim4015$ &$0^{-+}$ & $1.5$ \\ 
& $\thicksim4015$ & $0^{++}$ &$1.2$\\ 
& $\thicksim4015$ & $1^{+-}$ &$1.3$\\ 
& $\thicksim4015$ & $2^{++}$ & $1.2$\\ 
\hline
\multirow{2}{*}{}
$B\bar{B}^{*}$ & $\thicksim10545$ &$0^{-+}$ &$1.2$ \\ 
& $\thicksim 10562$ & $1^{++}$ &$1.2$\\ 
\hline
\multirow{3}{*}{}
$B^{*}\bar{B}^{*}$ & $\thicksim10590$ &$0^{-+}$ &$1.2$ \\ 
& $\thicksim10582$ & $0^{++}$ & $1.2$\\ 
& $\thicksim10608$ & $1^{+-}$ & $1.2$ \\ 
& $\thicksim10602$ & $2^{++}$ &$1.2$ \\
\hline
\end{tabular}
\end{table}








\begin{table}[htb]
\caption{Flavor non exotic bound states in the light sector found in \cite{tornqvist2}. ``waa'' stands for ``weak attraction added'', which means that the binding is obtained if a weak short-range interaction potential ($\thicksim v_{0}e^{-(r/r_{0})^{2}}$) is added to (\ref{ope}).}
\label{tab:tornnel}
\begin{tabular}{lccccc}
\hline
Constituents & Threshold (MeV)&  Candidates &$J^{PC}$ & $\Lambda$(GeV) & $BE_{expected}$(MeV) \\
\hline
\multirow{2}{*}{}
$K\bar{K}^{*}$ &$1394$& $\eta(1440)$ & $0^{-+}$ & waa & $\thicksim 50$ \\
&& $f_{1}(1420)$ & $1^{++}$ & waa & $\thicksim 30$ \\
\hline
\multirow{2}{*}{}
$K^{*}\bar{K}^{*}$ &$1792$  &$\eta(1760)$ &$0^{-+}$ & waa & $\thicksim 30$ \\
& &$f_{0}(1710)$ &$0^{++}$ &waa& $\thicksim 80$ \\
&  && $1^{+-}$ &waa & \\
&  &$f_{2}(1720)$& $2^{++}$ & \? & $\thicksim 70$\\
\hline
$(\rho\rho+\omega\omega)/\sqrt{2}$ & $1550-1566$ & $\eta(1490)$ &$0^{-+}$ &  &$\thicksim 70$ \\
$(\rho\rho-\omega\omega)/\sqrt{2}$ & $1550-1566$ & $\eta(1515)$ &$0^{++}$ &  & $\thicksim 40$\\
$(\rho\rho+\omega\omega)/\sqrt{2}$ & $1550-1566$ & $f_{2}(1520)$ &$2^{++}$ & &$\thicksim 40$ \\
$(K^{*}\rho-K^{*}\omega)/\sqrt{2}$ & $1671-1679$ & &$0^{++}$ &  & \\
\hline
\end{tabular}
\end{table}

\begin{table}[htb]
\caption{Flavor exotic bound states found in \cite{tornqvist2}. They are expected in the isoscalar $0^{-}$ and $1^{+}$ channels and in the isovector $1^{-}$ channel.}
\label{tab:torne}
\begin{tabular}{lc}
\hline
Constituents & Threshold (MeV) \\
\hline
$KK^{*}$ & $1394$ \\
$DD^{*}$ & $3871$ \\
$BB^{*}$ & $10605$ \\
$KD^{*}+K^{*}D$ & $2505-2761$\\
$K\bar{B}^{*}+K^{*}\bar{B}$ &$5823-6176$\\
$D\bar{B}^{*}+D^{*}\bar{B}$ & $7190-7286$\\
\hline
\end{tabular}
\end{table}

In \cite{swansonrept} Swanson obtained the same results for the vector-vector heavy-heavy mesons bound states, 
except for the fact that he finds only one $D^{*}\bar{D}^{*}$ bound state with $J^{PC}=0^{++}$.

In Tab.~\ref{tab:global} the possible molecular assignments for exotic meson candidates are summarized.
\begin{sidewaystable}[bth]
\caption{Possible molecular assignment for exotic meson candidates.}
\label{tab:global}
\begin{tabular}{ccccc|cc}

\hline
 \textbf{State} & $I^{G}(J^{PC})$& \textbf{Observed\?} & \textbf{Production} &\textbf{Decays} & \textbf{Molecular content}& \textbf{References}  \\
\hline

$f_{0}(975)$  & $0^{+}(0^{++})$ & \ding{51} &&&$K\bar{K}$ & \cite{weinstein1,weinstein2,weinstein3,swanson,Tornqvist:1995kr,Guo:2007mm,Branz:2008ha, Lemmer:2009ze,Branz:2009jt}\\

$a_{0}(980)$& $1^{-}(0^{++})$  & \ding{51}& & &$K\bar{K}$& \cite{weinstein1,weinstein2,weinstein3,swanson,Tornqvist:1995kr,Guo:2007mm,Branz:2008ha, Lemmer:2009ze,Branz:2009jt}\\

$f_{1}(1420)$ &   $0^{+}(1^{++})$ &\ding{51} && &$K\bar{K}^{*}(t.e.)$ & \cite{swanson,caldwell,longacre,tornqvist}\\

$f_{0}(1710)$ &  $0^{+}(0^{++})$ &\ding{51} &&&$K^{*}\bar{K}^{*}$& \cite{swanson,dooley} \\

$X(1812)$ & $(0^{+})$ & \ding{51} \cite{Ablikim:2006dw}  &$J/\psi\to \gamma X$&$X\to \omega\phi$&$K^{*}\bar{K}^{*}$  &\cite{Chen:2007zz} and Ref. therein\\

\hline

\multirow{4}{*}{}

$Y(2175)$ &$1^{--}$ &\ding{51}   \cite{Aubert:2006bu}  &$e^{+}e^{-}\to\phi(1020)f_{0}(980)$&&$\phi(1020)f_{0}(980)$ & \cite{AlvarezRuso:2009xn} \\
&&\ding{51} \cite{ Aubert:2007ym}&$e^{+}e^{-}\to\phi(1020)\eta$& &&\\
&&\ding{51}  \cite{ 2007yt}&$J/\psi\to\eta\phi(1020)f_{0}(980)$&&&\\
&&\ding{51}   \cite{Shen:2009zze}&&&&\\

\hline

$D^{+}_{s}(2317)$ & $0(0^{+})$   &\ding{51}  \cite{Aubert:2003fg} &$\Upsilon(4S)\to D_{s}(2317)+all$ & $D^{+}_{s}(2317)\to D_{s}^{+}\pi^{0}$ & $DK^{\pm}$ & \cite{Barnes:2003dj}\\

$D^{+}_{s}(2460)$ & $0(1^{+})$ & \ding{51}  \cite{Besson:2003cp} &$\Upsilon(4S)\to D_{s}(2460)+all$&$D^{+}_{s}(2460)\to D_{s}^{*+}\pi^{0}$& $DK^{*\pm}$ &\cite{Barnes:2003dj} \\

\hline

$\psi(4040)$ & $0^{-}(1^{--})$&\ding{51}&&& $D\bar{D}$ & \cite{okun,derujula} \\

$\eta_{c}(3870)$ &$(0^{-+})$ & \ding{55} &&&$D\bar{D}^{*}$&\cite{tornqvist2} \\

$X(3872)$ & $(1^{++})$ & \ding{51} &&&$D\bar{D}^{*}$ & \\

$Y(3940)$& $(1^{++})$ & \ding{51} \cite{Abe:2004zs,Aubert:2007vj}& $B^{0,+}\to Y(3940) K^{0,+}$ &$Y(3940)\to J/\psi \omega$&$D^{*}\bar{D}^{*}/D^{*-}D^{*+}$ & \cite{Branz:2010qw}\\

$\chi_{c0}(\thicksim4015/4019)$ &  $(0^{++})$ & \ding{55} &&& $D^{*}\bar{D}^{*}$ &\cite{tornqvist2}/ \cite{swansonrept}, \cite{Ding:2009zq} \\

$\eta_{c}(\thicksim4015)$ & $(0^{-+})$ & \ding{55} &&&$D^{*}\bar{D}^{*}$ &\cite{tornqvist2,Ding:2009zq} \\

$h_{c0}(\thicksim4015)$ & $(1^{+-})$ & \ding{55} & &&$D^{*}\bar{D}^{*}$ &\cite{tornqvist2} \\

$\chi_{c2}(\thicksim4015)$ &$(2^{++})$&\ding{55} &&& $D^{*}\bar{D}^{*}$ & \cite{tornqvist2,Ding:2009zq} \\

$(\thicksim 4015)$  & $(1^{--})$ &\ding{55} &&&  $D^{*}\bar{D}^{*}$ &  \cite{Ding:2009zq} \\

$Z_{1}^{+}(4050)$ &&\ding{51}   \cite{Mizuk:2008me}  &$\bar{B}^{0}\to K^{-}Z^{+}_{1}(4050)$&$Z^{+}_{1}(4050)\to\pi^{+}\chi_{c1}$&$D^{*}\bar{D}^{*}$ &  \cite{Liu:2008tn}\\

$Y(4140)$ & $$ & \ding{51}\cite{Aaltonen:2009tz} &$B^{+}\to Y(4140) K^{+}$ &$Y(4140)\to J/\psi \phi$&& \cite{Branz:2010qw,Ding:2009vd,Mahajan:2009pj} \\

$Z^{+}_{2}(4250)$& $$ &\ding{51}  \cite{Mizuk:2008me}&$\bar{B}^{0}\to K^{-}Z^{+}_{2}(4250)$&$Z^{+}_{2}(4250)\to\pi^{+}\chi_{c1}$&$D_{1}\bar{D}$&  \cite{Liu:2008tn,Ding:2008gr}\\

$Z^{+}(4430)$ &  &\ding{51} \cite{:2007wga} &$B^{0}\to K^{0}Z^{+}(4430)$&$Z^{+}(4430)\to\psi^{'}\pi^{+}$& $D^{*}\bar{D}_{1}$ &\cite{Meng:2007fu,Ding:2007ar,Rosner:2007mu,Liu:2007bf,Liu:2008xz,Ding:2008mp} \\

\hline

\multirow{3}{*}{}

$Y(4260)$ & $(1^{--})$ &\ding{51}  \cite{Aubert:2005rm} &$e^{+}e^{-}\to\gamma_{ISR}Y(4260)$&$Y(4260)\to J/\psi \pi^{+}\pi^{-}$& $D^{*}(2010)\bar{D}_{1}(2420)$& \cite{Ding:2008gr,Meng:2007fu,Rosner:2006vc,Swanson:2005tq,Rosner:2006sv,Close:2007ny,Close:2009ag,Close:2010wq}\\
&&&&&$\omega/\rho-\chi_{cJ}$&\cite{Yuan:2005dr,Liu:2005ay}\\
&&&&&$J/\psi K\bar{K}$&\cite{MartinezTorres:2009xb}\\

$(4616)$ & $$ &\ding{51}  \cite{:2007sj} &&& $\psi^{'}f_{0}(980)$& \\

\multirow{2}{*}{}
$Y(4360)$ & $(1^{--})$ &\ding{51} \cite{Aubert:2006ge} & $e^{+}e^{-}\to\gamma_{ISR}Y(4360)$&$Y(4360)\to \psi^{'} \pi^{+}\pi^{-}$& $D^{*}(2010)\bar{D}_{1}(2420)$& \cite{Close:2009ag,Close:2010wq}\\
&&&&$Y(4360)\to\Lambda_{c}\bar{\Lambda}_{c} $  \cite{Pakhlova:2008vn} & &\\

$Y(4660)$ &  $(1^{--})$ &\ding{51} \cite{:2007ea} &$e^{+}e^{-}\to Y(4660)$&$Y(4660)\to\psi^{'}\pi^{+}\pi^{-}$& $\psi^{'}f_{0}(980)$ &\cite{Guo:2008zg} \\

\hline

$\eta_{b}(10545/10543)$ & $0^{-+}$ & \ding{55} &&& $B\bar{B}^{*}$ &\cite{tornqvist2}/\cite{swansonrept} \\

 $\chi_{b1}(10562/10561)$ & $1^{++}$ &\ding{55}& &&$B\bar{B}^{*}$&  \cite{tornqvist2}/\cite{swansonrept} \\

$\chi_{b0}(10582/10579)$ & $0^{++}$ & \ding{55}&&& $B^{*}\bar{B}^{*}$ & \cite{tornqvist2}/\cite{swansonrept}\\

$\eta_{b}(10590/10588)$ & $0^{-+}$ &  \ding{55}  & && $B^{*}\bar{B}^{*}$ & \cite{tornqvist2}/\cite{swansonrept}\\

$h_{b}(10608/10606)$ &$1^{+-}$ & \ding{55}&&& $B^{*}\bar{B}^{*}$ &  \cite{tornqvist2}/\cite{swansonrept},  \cite{Barnes:1999hs} ($I=0$)\\

$\chi_{b2}(10602/10600)$ & $2^{++}$ & \ding{55}  &&&  $B^{*}\bar{B}^{*}$ & \cite{tornqvist2}/\cite{swansonrept} \\

$$ & $$ & &&& $B^{(*)}\bar{B}^{(*)}$ & \cite{Meng:2007fu,Ding:2007ar,Cheung:2007wf}\\

 \\
\end{tabular}
\end{sidewaystable}


\subsubsection{Production and decays}
The production of molecular bound states proceeds necessarily through the simultaneous production 
of its constituent mesons in a suitable relative momentum configuration. 
On general grounds it is reasonable to assume that the relative momentum between the two mesons 
which are candidates for the binding is related to the binding energy of the molecule itself. 
This relation can be deduced using the uncertainty (or minimal uncertainty) principle: $\langle\Delta x\rangle \langle\Delta p\rangle \sim \hbar$. 
The binding energy gives an estimate of the size of the bound state $E_{B}=\hbar^2/2\mu \langle \Delta x\rangle^2$ 
and thus the average momentum spread is $\langle \Delta p\rangle\sim \sqrt{2\mu E_B}$. 
Each of the molecules predicted nearly at threshold, thus with a tiny binding energy, will allow for very small momentum spreads.
A rather special case is the $X(3872)$ which will be treated in detail later.

The decay modes of a meson-meson bound state can be divided into two classes: {\it long-distance} decay modes and {\it short-distance} decay modes.

The former class consists of the decaymodes of the constituent mesons. The partial decay widths of the molecular state in these modes are related to
those of its constituent mesons. If the binding energy is very small, nearly zero, the partial decay widths of the molecule in these modes will be almost 
equal to the ones of its constituents, whereas for deeply bound states one expects large deviation from the free meson widths. 
In particular a large binding energy tends to stabilize the meson when it is bound.

The latter class of decay modes, that of short-distance decay, is a manifestation of the existence of some inelastic channels in the meson-meson scattering mechanism. 
The associated partial decay widths are proportional to the probability that the two mesons come together at a point, 
which is given by the square modulus of the bound state wave function at the origin $|\psi(0)|^2$.

\subsection{Hybrids}

Both the molecular and tetraquark hadrons are built with quarks and antiquarks. However the QCD Lagrangian contains also the gluons,
which can act as dynamical degrees of freedom besides being the particles which mediate strong interactions. 
One can indeed suppose the existence of {\it gluonic hadrons}, bound states of gluons and quarks. 
There are two kinds of gluonic hadrons: the {\it glueballs}, which are bound states of only gluons, 
and the {\it hybrids} which are $q\bar q g$ bound state, that is a $q\bar q$ state with a gluonic excitation. 
This is is not surprising from the point of view of color, since $q\bar q\in 8\oplus 1$ and $g\in 8$ and one can pick a singlet component 
from the $8\otimes 8$ configuration. In this section we will focus on hybrid mesons.

The existence of hybrid mesons was suggested in 1976 by Jaffe and Johnson \cite{Jaffe:1975fd} and Vainsthein and Okun \cite{Vainshtein:1976ke}.
Hybrids have been studied using different approaches: 
({\it i}) the MIT bag model, 
({\it ii}) an adiabatic heavy-quark bag model, 
({\it iii}) constituent gluon models and 
({\it iv}) heavy quark lattice gauge theory.
({\it v}) the flux tube model, 
The average mass obtained for the lightest hybrid with light quarks is about 1.5-2~GeV. Hybrids can exhibit exotic $J^{PC}$ quantum numbers, 
and thus could be easily identified experimentally. 

The MIT \cite{Barnes:1982zs,Barnes:1982tx} bag model predicts  the existence of a lightest hybrid mesons mulitplet at $\sim$1.5~GeV 
and the presence of an exotic $1^{-+}$ state in this multiplet. 
The exotic $J^{PC}$ quantum numbers are due to the boundary conditions in the bag.

For the heavy quarks a spherical bag would be quite unrealistic, and thus an adiabatic bag model was introduced
by Hasenfratz, Horgan, Kuti and Richard in \cite{Hasenfratz:1980jv}. In this model the bag was allowed to deform in the presence of a fixed $Q\bar Q$ source. 
The resulting potential was used in a Schroedinger equation to compute the mass of the hybrids. 
The lightest hybrids for $c\bar c$ was found at $\simeq$3.9~GeV and at $\sim$10.5~GeV for $b\bar b$. 
For a recent result on adiabatic potentials in QCD string models  see~\cite{Kalashnikova:2002tg}.

Constituent gluon models treat the gluon as the constituent quark model treat the quarks. 
These models have been introduce by Horn and Mandula in \cite{Horn:1977rq} and later develped by 
Tanimoto, Iddir et al. \cite{Tanimoto:1982eh,Tanimoto:1982wy,Iddir:1988jd} 
and Ishida et al. \cite{Ishida:1991mx,Ishida:1989xh}.
The gluon has a fixed orbital angular momentum relatively to the $q\bar q$ pair, usually called $l_g$, 
and the $c\bar c$ is in a defined orbital configuration $l_{q\bar q}$ and spin configuration $s_{q\bar q}$. 
The quantum number of such a bound state are somewhat different from the ones predicted in other model: 
$P=(-1)^{l_g +l_{q\bar q}}$ and $C=(-1)^{l_{q\bar q}+s_{q\bar q}+1}$. 
The lightest hybrid states in this model has $l_{g}=0$ and thus non exotic quantum numbers such as 
$1^{--}$ are obtained using {\it p}-wave $q\bar q$ states with $s_{q\bar q}=1$, while exotic $1^{-+}$ state has $s_{q\bar q}=0$. 

Lattice QCD is supposed to give the most reliable predictions for absolute hybrid masses. 
In the heavy quark lattice QCD in which the $Q\bar Q$ pair is kept fixed while the gluonic degrees of freedom are allowed 
to be excited the lightest charmonium hybrids was predicted in \cite{Perantonis:1990dy} to have a mass of 4.2~GeV. for $c\bar c$ and 10.81~GeV for $b\bar b$.

Finally, the flux-tube model is  the most widely used model for hybrids. In lattice QCD two separated color 
sources are confined by approximately cylindrical regions of color fields if they are sufficiently far apart. The flux tube model  describes 
this confinement in a simple dynamical way, approximating the confining region between quarks by a string of massive points. 
This approach is motivated by the strong coupling expansion of lattice QCD. Since in a lattice gauge theory the flux line can be expanded only 
in transverse directions, in the flux tube models one allows only for transverse spatial fluctuations of the massive point positions. 
In the first studies with this model an adiabatic separation of the quark and gluon degrees of freedom was carried on, exploiting the fast dynamical 
response of the flux tube degrees of freedom with respect to the heavy quarks time scales. 
This separation allows to fix the $q\bar q$ separation at some value $R$ and compute the eigenenergy  
of the system in some fixed configuration of the flux tube: $E_{\Lambda}(R)$. The ground state $\Lambda=0$ 
gives the ordinary meson spectrum. Hybrids are obtained for $\Lambda>0$ and can be studied using the excited potential $E_{\Lambda}(R)$. 
The lightest hybrid state is the one in which the string has a single orbital excitation about the $q\bar q$ axis. 
In initial models the adiabatic potentials were determined in the approximation of small fluctuations relatively to the $q\bar q$ axis.
 This approximation was removed by Barnes, Close and Swanson in \cite{Barnes:1995hc}.

In the charmonium family hybrids are predicted in the mass region 4.3~GeV, with an estimated uncertainty of 100/200~MeV. 
As for the bottom sector hybrids are predicted in the region 10.7-11.0~GeV.

While the masses of hybrid mesons are computable in all the models listed above, and in particular in Lattice QCD, the decay dynamics is more difficult to study.

The only model which offers a description of the decay dynamics is the flux-tube model. 
Indeed in this context the decay occur when the flux-tube breaks at any point along its length, 
producing in the process a $q\bar q$ pair in a relative $J^{PC}=0^{++}$ state. 
A similar model has been applied to the ordinary $Q\bar Q$ mesons, since, as we stated before, 
the flux-tube model in its ground state describe ordinary mesons. 
The distance from the $Q\bar Q$ axis at which the light pair is created is controlled by the transverse distribution of the flux-tube. 
This distribution varies when going from the non-excited flux-tube to the first excited flux-tube configuration. 
Exploiting the empirical success of this model in describing the ordinary mesons decay dynamics  Close and Page in  \cite{Close:1994hc}
derived the decay pattern for hybrids. They found that in a two meson decays the unit of orbital angular momentum of the incoming hybrid around the $Q\bar Q$ axis
is exactly absorbed by the component of the angular momentum of one of the two outgoing mesons along this axis. 
In  \cite{Close:1994hc} they treated explicitly the light flavor case, but a generalization to hybrid charmonia is straightforward. 
The final state should be in this case $D^{(*,**)}\bar{D}^{*,**}$, where $D^{**}$ indicates $D$-meson which are formed from {\it p}-wave $c\bar q$ ($q=u,d$) pairs. 
However, since the masses predicted in the flux-tube model are about $\sim$4.3 GeV, {\it i.e.} below the $DD^{**}$ threshold, it is possible that this decay is kinematically 
forbidden giving a rather narrow resonance decaying in charmonium and light hadrons. 
These modes offer a clear experimental signature and furthermore should have large branching fractions if the total width is sufficiently small. 


\subsection{Hadrocharmonium}
Recently \cite{Dubynskiy:2008mq} a new interpretation has been proposed for the
states with $J^{PC}=1^{--}$ in the region $4.2-4.6~{\rm GeV}$, namely $Y(4260)$-$Y(4350)$-$Y(4660)$,
and for the only charged state observed at $4.43~{GeV}$, namely the $Z^{+}(4430)$.

These states show some common characteristics: they decay prominently in only one of the two charmonium
states $J/\psi$ and $\psi(2S)$ and furthermore the decay into open-charm mesons is highly suppressed.
These common features have been interpreted as the indication of an hadronic  structure in which a standard charmonium
state is stuck into a light hadrons. This picture is inspired by the much discussed case of charmonium states bound inside a nucleus.
This charmonium state embedded inside light hadronic matter is referred to as hadro-charmonium or hadro-quarkonium in general.
The light hadronic matter act as a spatial extended environment in which the more compact $J/\psi$ or $\psi(2S)$ moves. This picture is at least
able to explain why the decay into $J/\psi$ or $\psi(2S)$ is favored or suppressed, depending on which charmonium state is stuck
inside the hadron.

The reason why the $c\bar c$ state interacts with the light-hadronic stuff although being neutral with respect to color charge, is 
that it possesses a chromo-electric polarizability. 
Thus its chromo-electric dipole moment interacts with the chromo-electric field generated by the light hadronic matter. This interaction can be treated with the multipole expansion in QCD
used for the charmonium binding inside nuclei. 
The chromo-electric dipole moment is proportional
to the chromo-electric field $\mathbf{E}^a$ through the chromo-polarizability $\alpha$, resulting in an effective 
interaction hamiltonian of this form:

\begin{equation}
H_{eff} = -\frac{1}{2}\alpha \mathbf{E}^{a}\cdot \mathbf{E}^{a}
\end{equation}

The chromo-polarizability $\alpha$ can be deduced from the decay $\psi^{'} \to J/\psi\pi^+\pi^-$. The hadronic transition in quarkonium systems arise from the interaction of a quark or an antiquark with the gluons, which can the materialized in light hadrons, such as $\pi$ or $\eta$. This kind of
interactions can be treated within the multipole expansion, as it happens for the interaction of
heavy quarks and antiquarks with photons \cite{Voloshin:2007dx}. 
The radiative transition terms are 
proportional to the electric dipole moment and to the magnetic dipole moment associated to the heavy quarks, the so-called $E1$ and $M_1$ transitions:

\begin{equation}
\begin{split}
&H_{E1}=-e_c\,e \left(\vec{r}\cdot E\right)\\
&H_{M1}=-\mu_c \left(\vec{\Delta}\cdot E\right)\\
\end{split}
\end{equation}

\noindent where $\vec{\Delta}=\vec{\sigma}_1-\vec{\sigma}_2$, $\vec{\sigma}_1$ and $\vec{\sigma}_2$ being the spin of the quark and the antiquark respsectively.
In the same way one can treat the interaction of a chromo-electric dipole moment and a chromo-magnetic moment with the chromo-electric and chromo-magnetic fields:

\begin{equation}
\label{chromo}
\begin{split}
&H_{E1}=-\frac{1}{2} \xi^a \left(\vec{r}\cdot E^a\right)\\
&H_{M1}=-\frac{1}{2M} \xi^a \left(\vec{r}\cdot B^a\right)\\
\end{split}
\end{equation}

\noindent where $\xi^a=t^a_1-t^a_2$ is the difference between the color generators acting 
on the quark and antiquark, and $E^a$ and $B^a$ are the chromo-electric and chormo-magnetic
component of the gluon strength tensor.
The two pion transition $\psi'\to J/\psi\pi^+\pi^-$ is generated by two insertions of the operator
$H_{E1}$ in Eq. (\ref{chromo}):

\begin{equation}
H_{eff}=-\frac{1}{2}\alpha^{(12)}_{ij}E^a_i E^a_j
\end{equation}

\noindent where 

\begin{equation}
\alpha^{(12)}_{ij}=\frac{1}{16}\langle1S|\xi^a r_i \mathcal{G}r_j \xi^a|2S\rangle
\end{equation}

\noindent $\mathcal{G}$ being the two point Green function of the heavy quark pair in
a color octet configuration. In the leading non relativistic order for transitions in {\it S}-wave 
$\alpha^{(12)}_{ij}$ actually reduces to a scalar $\alpha^{(12)}$, wich is measured to be $\alpha^{(12)}\sim2~{\rm GeV}^{-3}$.
On the other hand the average value of the product of chromo-electric fields over the light hadron
$X$ can be estimated using the conformal anomaly relation in QCD:

\begin{equation}
\langle X|\frac{1}{2}\mathbf{E}^a\cdot \mathbf{E}^a|X\rangle\geq \frac{8\pi^2}{9}M_X
\end{equation}

In this way one is able to estimate the strength of the interaction between the light hadronic matter
and the quarkonium system bound inside it.
The possibility that such a bound state exists depends on the relation between the mass $M_X$
and the spatial extension of the light hadron. In particular in \cite{Dubynskiy:2008di} it has been
shown that a quarkonium state does form a bound state inside a sufficient highly excited light hadron. Furthermore the authors in \cite{Dubynskiy:2008di} find that for this kind of bound 
state the decay into open heavy falvour mesons is suppressed exponentially as ${\rm exp}\,(-\sqrt{\Lambda_{QCD}}/M_Q)$, which would explain the non observation of the decay of the $Y$ resonances into final states with pairs of charmed mesons.

\subsection{Pentaquarks and hexaquarks}
Beyond the tetraquark structure some other multiquarks hadrons have been considered
in the literature, such as baryon states built of 4$q$ and a $\bar{q}$, namely pentaquarks, or mesons constituted by 3$q$ and 3$\bar{q}$, the so-called hexaquarks.

The search for pentaquarks was originally triggered by the observation of a narrow exotic 
resonance $\Theta^+$ \cite{Nakano:2003qx}  and has motivated the study of $uudd\bar{s}$ state \cite{Jaffe:2003sg}.
The subsequent disproof of the observation of the $\Theta^+$ in photoproduction has not
been taken as the last word. Studies on pentaquarks are ongoing.

In particular the spectrum of the pentaquark \cite{Abud:2008cb} and hexaquark \cite{Abud:2009rk} states has been computed assuming that the main role is played by the chromo-magnetic interactions.
In \cite{Abud:2008cb} both positive and negative parity states have been considered
including also the splitting due to the $SU(3)_F$ breaking. 
In \cite{Abud:2009rk} the same authors have extracted the chromo-magnetic interaction parameters from the supposed tetraquark structures, such as $a_0$, $f_0$, $\sigma$ and $X(3872)$ and 
use this parameters to compute the masses of $qqq\bar{q}\bar{q}\bar{q}$ states. 
Hexaquarks should show a strong affinity to the baryon-antibaryon decay mode.

%% file: experiments.tex
\section{Experimental Primer }

Hadronic spectroscopy has experienced a renaissance in the last few years,
thanks to the opportunities provided by several experiments that allowed 
to improve our knowledge of standard hadrons and discover a rich 
zoology of exotica. In this section we briefly introduce the main 
apparatuses and methods that are behind the experimental results described in the paper.

\subsection{Experiments}

In the last two decades, three \epem colliders, CESR~\cite{cesr} at LEPP (Cornell, USA), PEP-II~\cite{pep} at SLAC (Stanford, USA) and
KEK-B~\cite{kekb}  at KEK (Tsukuba, Japan), have been mainly operated at a center of mass energy of 10.58 GeV,
corresponding to the mass of the $\Upsilon(4S)$ \bbbar resonance,
with the main purpose of discovering CP violation in the $B$ meson
sector and make precision studies of CKM physics.

Interesting flavour physics studies have also been performed at similar \epem facilities running around the threshold
for the production of $tau$ and charm, between 3.7 and 5.0 GeV: BEPC at IHEP (Beijing, China)
and CESR-c~\cite{cesrc} at LEPP.

Meanwhile the Tevatron $p \overline p$ collider at Fermilab (USA)
investigated the high energy frontier, running at a center of mass
energy of 1.96 TeV. 

All these facilities hosted experiments that demonstrated to be a perfect 
place for studying standard and exotic hadronic spectroscopy, through 
different production and decay mechanisms. 

\subsubsection{$B$-Factories}

CESR, PEP-II and KEK-B, commonly named $B$-Factories for the large
production cross section of $B$ meson pairs at the $\Upsilon(4S)$
resonance, operated in the last decades with a peak luminosity of
$1.2 \times 10^{33}$, $12 \times 10^{33}$ and $21 \times 10^{33}$
$cm^{-2} s^{-1}$, respectively. CESR, after collecting 15.5\invfb  of integrated 
luminosity, ceased operations as a $B$-Factory in 1999, 
when PEP-II and KEK-B started providing collisions. PEP-II was
operated until April 2008 and delivered 553\invfb, while KEK-B 
is still running and exceeded 1\invab. In the case of PEP-II and
KEK-B, the electron and positron beams collide with asymmetric
energies and the $\Upsilon(4S)$ resonance is produced with a Lorentz 
boost of $\beta\gamma = 0.53$ and 0.46 respectively.
PEP-II and KEK-B have been also run at the energies of
the $\Upsilon(2S)$, $\Upsilon(3S)$ and $\Upsilon(5S)$ resonances, in order to study the
decays of these states. The relevant parameters of the three colliders are quoted in Tab.~\ref{tab:exp:bfactories}.

\begin{table}
\caption{Design parameters of the three $B$-Factories.}
\label{tab:exp:bfactories}
\begin{tabular}{lccccc}
 \hline
  & CESR & \multicolumn{2}{c}{KEK-B}& \multicolumn{2}{c}{PEP-II}\\ 
  &  & LER & HER & LER & HER \\
\hline
Energy (GeV) & 5.29 & 3.5 & 8.0 & 3.1 & 9.0 \\ 
Collision mode & 2~mrad & \multicolumn{2}{c}{11~mrad}& \multicolumn{2}{c}{Head-on}\\
Circumference (m) & 768 & \multicolumn{2}{c}{3018}& \multicolumn{2}{c}{2199}\\ 
$\beta^*_x/\beta^*_y$ (cm) &100/1.8 & 100/1 & 100/1 & 37.5/1.5& 75/3\\ 
$\xi^*_x/\xi^*_y$ &0.03/0.06  & \multicolumn{2}{c}{0.05/0.05}& \multicolumn{2}{c}{0.03/0.03}\\ 
$\epsilon^*_x/\epsilon^*_y~(\pi \mathrm{rad}-\rm{nm})$ & 210/1 & 19/0.19 & 19/0.19 & 64/2.6& 48.2/1.9\\ 
relative energy spread ($10^{-4}$)& 6.0 & 7.7 & 7.2 & 9.5 & 6.1 \\ 
Total Current (A) & 0.34 & 2.6 & 1.1 & 2.14 & 0.98 \\ 
number of bunches & 45 & \multicolumn{2}{c}{5120}& \multicolumn{2}{c}{1658}\\ 
RF Frequency ($MHz$)/ Voltage ($MV$)&500/5 & 508/22 & 508/48 & 476/9.5 & 476/17.5\\
number of cavities & 4 & 28 & 60 & 10 & 20 \\ 
\hline
\end{tabular}
\end{table}
 
Three general purpose detectors were installed at these facilities:
the CLEO detector~\cite{cleo,cleo2,cleo3} at CESR (in three different configurations,
CLEO-II/II.V/III), the BaBar detector~\cite{babar} at PEP-II and the Belle detector~\cite{belle}
at KEK-B. 

The design of the the three detectors is quite similar. All of them are provided with
a multi-layer vertex tracker of double-sided silicon strip detectors
and a drift chamber, operating in a 1.5 T magnetic field . The silicon detectors 
allow a vertex resolution at the level of 100\mum and are also used for standalone 
tracking of low momentum particles ($p_t \leq 100$ MeV). The drift chambers
are used for the tracking of higher momentum tracks and also provide \dedx
measurements for particle identification (PID) below 700 MeV and above 1.5 GeV.
Momentum resolutions below the percent level are reached by these tracking systems.

Different technologies have been used for PID of charged hadrons above 700 MeV. 
A ring imaging \v{C}erenkov counter (RICH) is used by CLEO. LiF is used as the radiator and
photons are collected by a CH$_4$/TEA detector. A aerogel-based RICH was adopted
by Belle. Since the \v{C}erenkov threshold for pions in the aerogel is 1.5 GeV, a time-of-flight
(TOF) detector with 95 ps resolution is also used for $K/\pi$ separation below this threshold. 
Finally, BaBar developed a detector of internally reflected \v{C}erenkov light (DIRC), where
the light produced in quartz bars propagates by internal reflection along the bar itself 
and is finally detected by 11000 PMTs surrounding a water box at the end-cap of the detector.

The detection and energy measurement for photons and electrons is provided in all three
experiments by a CsI(Tl) electromagnetic calorimeter, with excellent energy resolutions, ranging
from 2 to 6\% in a wide energy range (from 20 MeV to 9 GeV).

The flux return of the magnetic field was finally instrumented for the detection and identification of
muons, by using plastic streamer counters in CLEO and resistive plate chambers (RPC) in BaBar and Belle.

\subsubsection{Tevatron}

The Tevatron $p \overline p$ collider at FNAL started its Run II in 2002 providing collisions
at $\sqrt{s} = 1.96$ TeV with a typical luminosity of $10^{32} cm^{-2} s^{-1}$ and a peak luminosity
exceeding $3 \times 10^{32} cm^{-2} s^{-1}$. A total integrated luminosity of 6.5\invfb have been delivered 
up to now.

Two general purpose detectors, CDF-II~\cite{cdf} and D0~\cite{d0}, were operated at the Tevatron during the Run II. Both
detectors are provided with a silicon vertex tracker. Seven and four layers of double sided silicon strips
are used by CDF and D0, respectively. An additional innermost layer is also included in both experiments
to improve the impact parameter resolution, down to 40\mum in CDF. The tracking system is completed in CDF
by a drift chamber for the reconstruction of tracks with $p_t > 400$ MeV. In D0, a tracker composed by scintillating 
fibers read by visible light photon counters. The tracking systems are operated inside a solenoidal field of 1.4 T in CDF and 
2 T in D0. The CDF tracker reached a transverse momentum resolution of $\sigma_{p_T}/p_T \sim 0.001 \cdot p_T\,$ [GeV] and
allows a $K/\pi$ separation on 1.4$\sigma$ for $p_T > 2.0$ GeV thanks to the \dedx measurement in the drift chamber.
PID for kaons and pions below 1.5 GeV is guaranteed in CDF by a TOF device with 110 ps resolution.

Energy measurement for electrons, photons and hadrons is provided by calorimeters. In CDF, a segmented electromagnetic
calorimeter is used, with alternate sheets of plastic scintillators and lead. In the hadronic calorimeter,
iron is used instead of lead. Resolutions of $13.5\%/\sqrt{E} \oplus 2\%$ and $50\%/\sqrt{E} \oplus 3\%$ 
are obtained in the central part of the electromagnetic and hadronic calorimeter, respectively. In D0, a
LAr sampling calorimeter is used, divided in three sections: a finely segmented electromagnetic section, a fine hadronic
section and a coarse hadronic section. Uranium is used as absorber in the first two sections, copper and stainless steel in the
third one. Energy measurement for electrons is complemented by a central preshower detector instrumented with scintillating fibers.
An energy resolution of $\sigma_E/E = 15\%/\sqrt{E} + 0.3\%$ and $\sigma_E/E= 45\%/sqrt{E} +  4\%$ is obtained for
electrons and pions respectively.

Both CDF and D0 are finally provided with muon detectors composed by tracking systems and scintillator trigger counters. 
In CDF, wire chambers are used for muon tracking. In D0, drift tubes are adopted and two out of three detector layers are
located inside a 1.8 T toroidal magnetic field.

\subsubsection{$\tau-$charm Factories}

Lepton colliders running slightly above 3.5 GeV allow to produce large quantities of $tau$ and charm hadrons, providing an 
ideal environment for studying $\tau$ physics and charm spectroscopy. The BEPC collider at IHEP covers the energy
range between 3.7 and 5.0 GeV, with a peak luminosity of $12.6 \times 10^{30} cm^{-2} s^{-1}$, with recent upgrades aimed to
reach $10^{33} cm^{-2} s^{-1}$ and host the BES-II experiment~\cite{bes,bes2}. CESR-c is the $\tau$-charm version of CESR and covers the energy range
between 3.97 and 4.26 GeV with a peak luminosity of $76 \times 10^{30} cm^{-2} s^{-1}$. A modified version of CLEO,
called CLEO-c~\cite{cesrc}, is operated at this facility. Other parameters of the two colliders are provided in Tab.~\ref{tab:exp:taucharm}.

\begin{table}
\caption{Design parameters of BEPC, CESR-c and the upgraded version of BEPC.}
\label{tab:exp:taucharm}
\begin{tabular}{lccc}
\hline
  & BEPC & CESR-c & BEPC-II \\ \hline 
Max. energy (GeV) & 2.2 & 2.08 & 2.3 \\
Collision mode & Head-on & $\pm 3.3$~mrad & 22~mrad \\
Circumference (m) & 240 & 768 & 240 \\
$\beta^*_x/\beta^*_y$ (cm) &120/5 & 94/1.2 & 100/1.5 \\ 
$\xi^*_x/\xi^*_y~(10^{-4})$ &350/350  & 420/280 & 400/400 \\
$\epsilon^*_x/\epsilon^*_y (\pi~\mathrm{rad}-\rm{nm})$ & 660/28 & 120/3.5 & 144/2.2 \\
relative energy spread ($10^{-4}$)& 5.8 & 8.2 & 5.2  \\
Total Current (A) & 0.04 & 0.072 & 0.91 \\ 
number of bunches & 1  & 24 & 93 \\
RF Frequency (MHz)/ Voltage (MV)&200/0.6-1.6 & 500/5 & 500/1.5 \\
number of cavities & 4 & 4 & 2 \\ \hline
\end{tabular}
\end{table}

At BEPC, the BES-II detector was composed by a straw tube vertex detector and a drift chamber for tracking, a scintillating
TOF device with 180 ps resolution for PID and a sampling electromagnetic calorimeter with streamer tubes and lead absorber. 
These detector were installed inside a 0.4 T solenoidal magnetic field, whose yoke was instrumented with proportional tubes
for muon detection. The upgraded version of the detector, BES-III, has a drift chamber for tracking, a
better TOF with 100 ps resolution, a CsI(Tl) calorimeter and RPCs for muon identification. The magnetic field is increased to 1 T.
BES-II collected 58 million $J/\psi$ events and 14 million $\psi(2S)$ events, while BES-III started operations in 2008 and 
so far has collected 10\invpb at the $\psi(2S)$.

The CESR-c detector, CLEO-c, derives from CLEO-II, adapted to the lower energy. The magnetic field was reduced to 1 T and and the
silicon detector was replaced by an inner drift chamber. CLEO-c collected 27 million of $\psi(2S)$ events, 818\invpb
at the $\psi(3770)$ and 602\invpb at a center of mass energy of 4.17 GeV, in order to study $D_s \overline D_s^*$ events.

\subsection{Analysis methods}

In this section, we briefly discuss the fundamental analysis approaches that are adopted in the different experiments to search
for new resonances and investigate their properties, like masses, widths and quantum numbers.

\subsubsection{Search and measurement of mass and width}

The experimental method adopted to look for hadron resonances depend on the 
production mechanism considered for the exotic particle ($Y$). The choice is among direct production, $e^+e^-\to Y$, the initial state radiation (ISR) production $e^+e^-\to Y\gamma_{ISR}$, $\gamma\gamma$ production $e^+e^-\to \gamma \gamma e^+e^-\to Y e^+e^-$, and $B$ decays, in particular $B\to YK^{(*)}$.

Direct production can only be studied by scanning the center-of-mass energy of the accelerator ($\sqrt(s)$) and requires a dedicated plan of the machine. After measuring the cross-section ($\sigma$) either of the inclusive hadronic production or of exclusive final states as a function of the center-of-mass energy, a new state would appear as a resonance in $\sigma(\sqrt(s))$. Scans dedicated to the exotic charmonia were performed by CLEO-c, while both Belle and BaBar investigated the regions above the $\Upsilon(4S)$ for exotic bottomonium exploiting energy scans.

In case of ISR production at \epem colliders, it is possible, in principle, to look for an almost monochromatic peak in the energy distribution 
of the ISR photon. This fully inclusive method has been used in the successful search for the bottomonium ground state $\eta_b$ 
at BaBar~\cite{etab}, but it can be applied only in very special cases. In fact, there is a typical photon detection
threshold of 30 MeV that prevents the use of this method for a mass near to $\sqrt{s}$, there is a very large background
for low photon energies, and photons are preferentially emitted along the beam axis, so that most of them escape the detection. 
Hence, most of the searches are performed by looking for fully reconstructed final states, requiring no additional charged 
particles in the event and without requiring the detection of the additional photon. 
In both inclusive and exclusive searches, the mass and width of the resonance can be measured. 
In the first case, they can be inferred from the position and width of the photon energy peak, and the resolution on the width
is limited by the photon energy resolution. In the second case, the invariant mass measurement is available from the 
reconstruction of the 4-momenta of the final state particles, and the resonance line shape is directly observed.

In the study of the line shape, the decay amplitude is usually described by a Breit-Wigner function:
\begin{equation}
A(m) \propto \frac{m_0 \Gamma_{tot}(m_0)}{m^2 - m^2_0 + i\Gamma_{tot}(m)m_0} \, ,
\end{equation}
that is possibly summed to a non resonant contribution to get the total amplitude and finally extract the cross section:
\begin{equation}
\sigma(m) \propto \Phi^f_{PS}(m)|A(m)|^2 \, ,
\end{equation} 
where $\Phi^f_{PS}(m)$ is a phase space factor depending on the final state (for 2-body decays, 
$\Phi^f_{PS}(m) = \left[ p(m)/p(m_0) \right]^{\alpha_f}$, where $p(m)$ is the momentum
of the decay products in their center of mass frame and $\alpha_f$ depends on the final state). 
Background models depend on the specific decay channel and will be described,
where needed, in the following sections.

Different approaches are used in the search for resonances produced in the decay of other particles. At the $B$-Factories,
standard and exotic hadrons can be produced in the $B \to K X$ decays. Also in this case an inclusive or exclusive
approach can be adopted. In the inclusive approach, one of the two $B$'s produced in the $\Upsilon(4S)$ decay
is fully reconstructed in a hadronic final state, and it allows to infer the 4-momentum of the other $B$. This information,
along with the 4-momentum of the kaon, allows to reconstruct the $X$ mass without considering its decay products.
In the exclusive approach, resonances are searched for in specific decay channels and the $K X$ pair is required to 
be consistent with the hypothesis of a $B$ meson produced from the $\Upsilon(4S)$ decay. Two variables are used:
\begin{eqnarray}
m_{ES} &=& \sqrt{(E^*_{b})^2 - |\mathbf{p}^*_{B}|^2} \, , \\
\Delta E &=& E^*_B - E^*_b \, ,
\end{eqnarray}
where $E^*_b$ is the beam energy and $(E^*_B,\mathbf{p}^*_B)$ is the $B$ 4-momentum, calculated in the $\Upsilon(4S)$ rest frame. The 
``beam-energy substituted'' mass $m_{ES}$ is also indicated with $m_{bc}$ in Belle's papers. From the 4-momenta of the $X$ decay products
one can reconstruct the $X$ invariant mass and study the line shape.

At hadronic colliders like Tevatron, new resonances are searched for in exclusive final states.
 In particular, final states containing a $J/\psi$ decaying to  $\mu^+\mu^-$ are searched among events collected with
a di-muon trigger. A very large statistics is available in this case, allowing a precise measurement of the
mass (the precision on the width is limited by the detector resolution) and the realization of angular analyses 
for the determination of the $J^{PC}$ quantum numbers.

At the $\tau$-charm factories, the best performances can be obtained for the study of $1^{- -}$ states, produced in large amount 
by setting the center of mass energy at the mass of the resonance, so that the most rare decay channels can be also studied.

\subsubsection{Measurement of $J^{PC}$}
\label{sec:jpc}

In some cases, the determination of the $J^{PC}$ quantum numbers can follow from the production process or the decay channels.
For instance, resonances produced in $e^+e^--$ annihilation (with or without ISR) through a virtual photon can only have
$1^{- -}$ quantum numbers. Conversely, states produced in two-photon events $\epem \to \epem \gamma^* \gamma^* \to \epem X$
can only have C = + and there is a set of selection rules for total angular momentum and parity,
known as ``Yang's theorem''~\cite{yang_theo}, which hold also for resonances decaying to $\gamma \gamma$.

In other cases, an angular analysis and the study of the kinematic spectra is needed in order to extract information 
about the quantum numbers. Considering for example the $J/\psi (\mu^+\mu^-) \pi^+ \pi^-$ decay, the $\pi^+ \pi^-$ pair can be treated as 
a single body, so that the angular dependence is the one of a two body decay. For fixed helicities, it is described by the Wigner function 
$D^{J_i}_{\lambda_i, \lambda_{i,1}  - \lambda_{i,2}}$, where $J_i$ and $\lambda_i$ are the spin and helicity of the decaying particle, and
$\lambda_{i,1}$ and $\lambda_{i,2}$ are the helicities of the decay products. This dependence is reflected by the angular distributions
of the final state particles (two muons and two pions in this example), usually described in terms of a few ``helicity angles'' (given a decay chain 
$X \to A + Y$ with $A \to B + C$, the helicity angle $\theta_A$ is defined as the angle between the $B$ or $C$ direction
in the $A$ rest frame and the $A$ direction in the $X$ rest frame). Additional information can also be extracted from the invariant mass
distribution of the $\pi^+ \pi^-$ pair. These analysis require large data sets and are more efficiently performed at the hadron colliders
than the $B$-Factories. An example can be found in~\cite{Abulencia:2006ma}.

%% file: charmonium.tex
\section{Charmonium}
\begin{figure}[bht]
\begin{center}
\epsfig{file=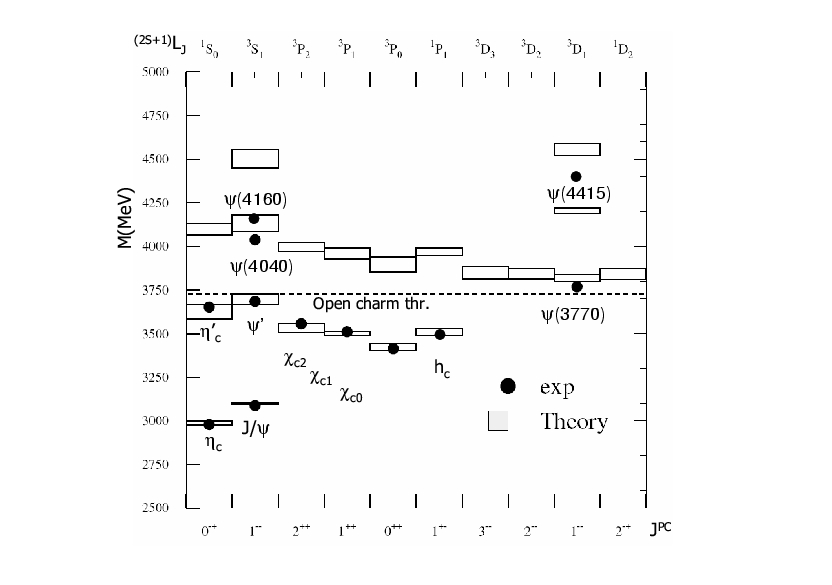,height=6cm}
\caption{\it Charmonium states with 
$L\leq2$. The theory predictions are according to the potential models 
described in Ref.~\cite{Brambilla:2004wf}. 
 }
\label{fig:charmon}
\end{center}
\end{figure}

The presence of heavy quarks allows very good predictions for the 
masses of the regular Charmonium states. 
The heavy quark inside these bound states has in fact  low 
enough energy that the corresponding
spectroscopy is close to the non-relativistic interpretations of 
atoms. The quantum numbers that are more appropriate to characterize a 
state are therefore, in decreasing order of energy splitting among 
different eigenstates, the radial excitation ($n$), the spatial angular 
momentum $L$, the spin $S$ and the total angular momentum $J$. Given 
this set of quantum numbers, the parity and charge conjugation of the 
states are derived by $P=(-1)^{(L+1)}$ and $C=(-1)^{(L+S)}$. 
Fig.~\ref{fig:charmon}  shows the mass and 
quantum number assignments of the well established charmonium states.

All the predicted states below open charm
 threshold have been observed with very good agreement with theory predictions. 
The Charmonium system is therefore an ideal environment to search for exotic states which deviate from regular spectroscopy.

We will first review the experimental observations, reporting both the final states where the states have been observed and those where they have not. Since the easiest quantum number to attribute is $C$, being determined uniquely either by the production or the final state (see Sec.~\ref{sec:jpc}), we will first examine the $C=+$ states (Sec.~\ref{sec:x3872},~\ref{sec:3940}, and ~\ref{sec:otherp}) and then end with the $J^{PC}=1^{--}$ states (Sec.~\ref{sec:1mm}). Next we will discuss the evidences for charged states that play a crucial role in this field.

In the second part of this section we will report the possible interpretations of these states (Sec.~\ref{sec:charmon_interp}) and conclude by summarizing the observations and the open issues(Sec.~\ref{sec:charmon_summary}).

\subsection{The X(3872)}



\label{sec:x3872}
\begin{figure}[bht]
\begin{center}
\epsfig{file=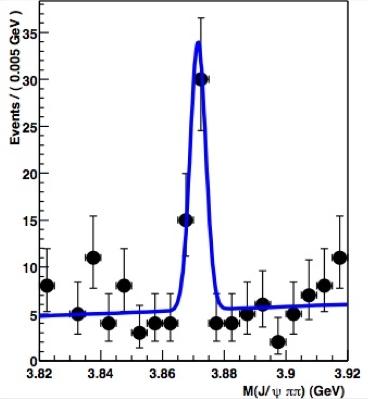,height=6cm}
\epsfig{file=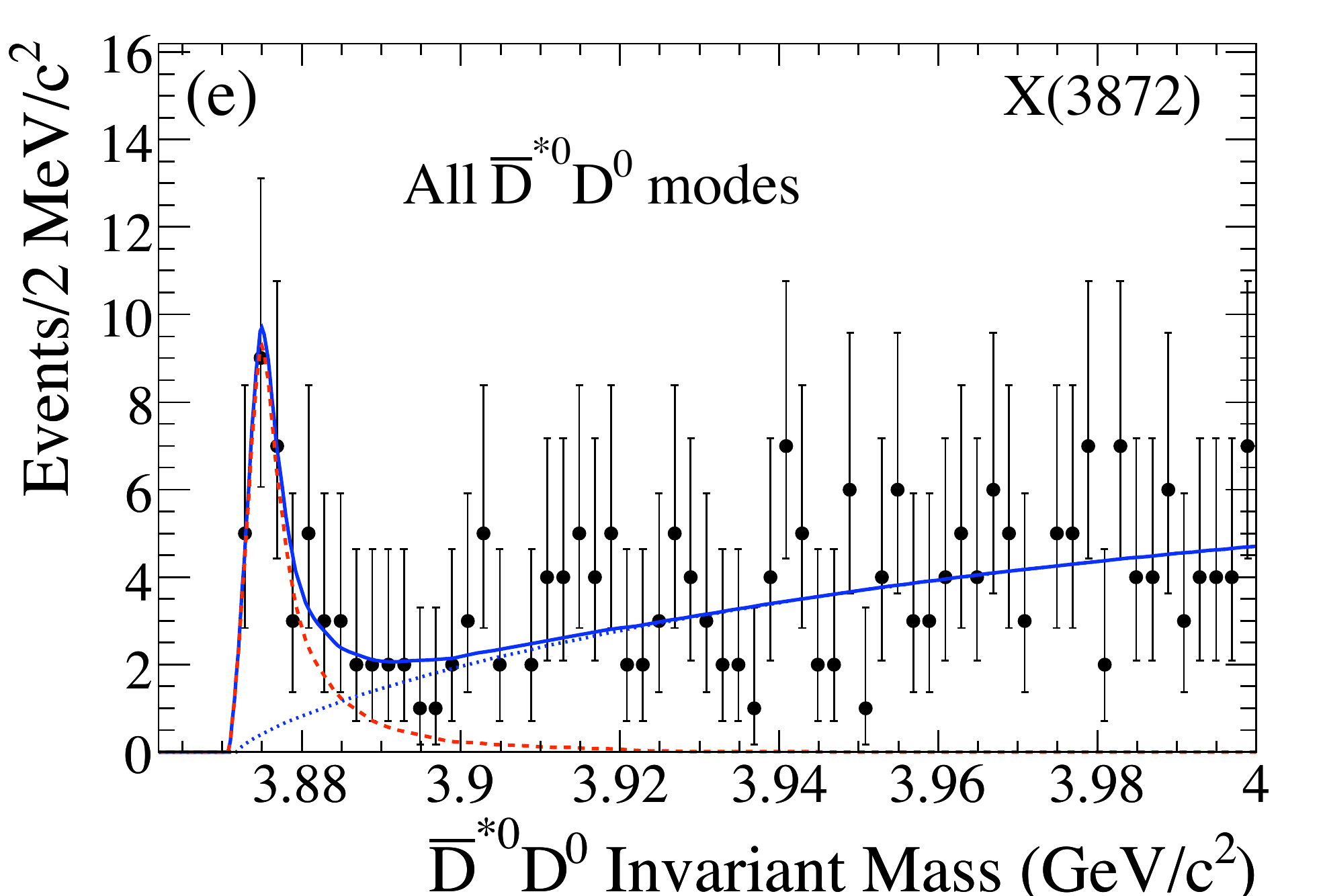,height=6cm} 
\caption{\it Invariant mass spectrum of the $J/\psi\pi^+\pi^-$ system in  the observation paper~\cite{Choi:2003ue} (left) and  of the  $D^*D$ system in Ref.~\cite{Aubert:2007rva}}
    \label{fig:Xjpsipipi}
\end{center}
\end{figure}

The $X(3872)$ was the first state that was found not to fit charmonium spectroscopy.
 It was initially observed by the Belle experiment in \BXK decays and decaying into $J/\psi\pi^+\pi^-$~\cite{Choi:2003ue} and 
 subsequently confirmed both in B decays~\cite{Aubert:2004ns} and in inclusive $p\bar{p}$ production~\cite{Acosta:2003zx,  Abazov:2004kp}  (see the Fig.~\ref{fig:Xjpsipipi}).
This is by far the state for which the most information is available, and it will be here reviewed by topic: quantum numbers, mass and width, and production and decay.
\begin{figure}[bht]
\begin{center}
\epsfig{file=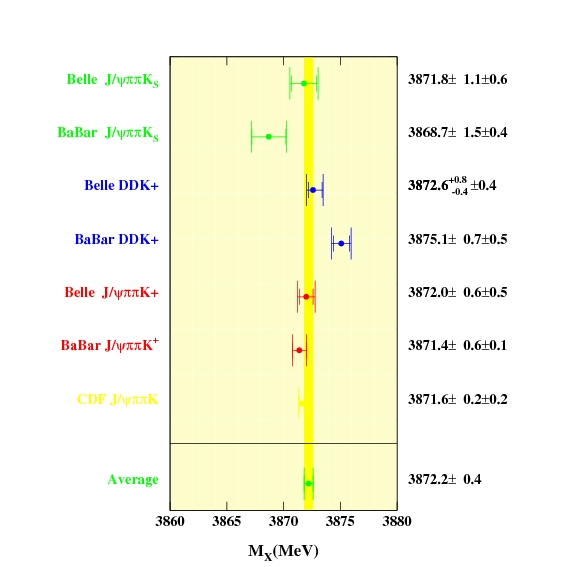,height=7cm} 
 \caption{\it Measured mass of the $X(3872)$ particle. The different 
production modes ($B^0\to XK_S$ and $B^-\to X K^-$)
and the different decay modes ($X\to J/\psi\pi\pi$ and $X\to 
D^{*0}D^0$) are separated.}
    \label{fig:XMass}
    \end{center}
\end{figure}
\subsubsection{Quantum numbers}
The exotic nature of this state was initially signalled by the narrowness of its width ($\Gamma_{X(3872)}=3.0^{+1.9}_{-1.4}\pm0.9$ MeV~\cite{Aubert:2007rva} ) although it could decay strongly into a pair of $D$ mesons. Furthermore, the $\pi^+\pi^-$  invariant mass distribution first~\cite{Choi:2003ue} and a detailed angular analysis next~\cite{Abulencia:2006ma} showed that the dominant decay is $X\to J/\psi\rho$, which would be isospin violating if the $X(3872)$ were a conventional charmonium state.

The above-mentioned angular analysis from the CDF experiment~\cite{Abulencia:2006ma} was able to discriminate among the possible $J^{PC}$ assignments, excluding any other possibility than $J^{PC}=1^{++}$ and $2^{-+}$ .
In the meanwhile there came evidence for the decay $X\to J/\psi\gamma$~\cite{Aubert:2006aj} and not of the decay $X\to \chi_{c1}\gamma$~\cite{Choi:2003ue}, thus confirming
positive intrinsic charge conjugation.

For several years the most favored option has been, since it has $L=0$,   $J^{PC}=1^{++}$. While we were finalizing this review, BaBar published a study of $X\to J/\psi\omega$ where this option is disfavored with respect to the  $J^{PC}=2^{-+}$ one, since it has only 7.1\% probability to match data~\cite{al.:2010jr}. 
Implications of this result are still under study and cannot be reviewed here.

\subsubsection{Mass and width}

\begin{table}[htb]
\begin{center} 
\caption{ 
Measured $X(3872)$ product branching fractions (PBF), separated by production and decay mechanism. 
When more than a publication is present the combination is performed  assuming gaussian uncorrelated errors. The last two columns report the results in terms of absolute $X$ branching fraction ($B_{fit}$) and of 
the branching fraction normalized to $J/\psi\pi\pi$ ($R_{fit}$) as obtained from the global likelihood fit described in the text. Ranges and limits are provided at 68\% and 90\% C.L. respectively.}
\label{tab:xdecays} 
\begin{tabular}{llccc} \hline 
$B$ Decay mode& $X$ decay mode &PBF($\times 10^5$) &$B_{fit}$&$R_{fit}$\\ \hline
$XK^\pm$  & $X\to J/\psi\pi\pi$ & 0.82$\pm$0.09~\cite{Aubert:2008gu,:2008te}& $\left[ 0.035,0.075 \right]$ &N/A\\
$XK^0$  & $X\to J/\psi\pi\pi$ & 0.53$\pm0.13$~\cite{Aubert:2008gu,:2008te}& N/A &N/A\\ 
$XK^\pm$  & $X\to D^{*0}D^0$ & 13$\pm3$~\cite{Aubert:2007rva,Gokhroo:2006bt}& $\left[ 0.54,0.8 \right]$ & $\left[ 7.2,16.2 \right]$\\
$XK^0$  & $X\to D^{*0}D^0$ & 19$\pm6$~\cite{Aubert:2007rva,Gokhroo:2006bt}& N/A & N/A \\ 
$XK$  & $X\to  J/\psi\gamma$ &  0.22$\pm0.05$~\cite{  :2008rn,Abe:2005ix} & $\left[ 0.0075,0.0195 \right]$ & $\left[ 0.19,0.33 \right]$ \\ 
$XK$  & $X\to \psi(2S)\gamma$ & $1.0\pm0.3$~\cite{  :2008rn}& $\left[ 0.03,0.09 \right]$ & $\left[ 0.75,1.55 \right]$ \\ 
$XK$  & $X\to \gamma\gamma$ & $<$0.024 ~\cite{ Abe:2006gn}& $< 0.0004$ & $<0.0078$\\ 
$XK$  & $X\to  J/\psi\eta$ & $<0.77$~\cite{ Aubert:2004fc}& $< 0.098$ & $<1.9$\\ 
$XK$  & $X\to  J/\psi\pi\pi\pi^0$ & -- ~\cite{ Abe:2005ix}& $\left[ 0.015,0.075 \right]$ & $\left[ 0.42,1.38 \right]$\\ 
$XK^*$  & $X\to  J/\psi\pi\pi$ & $<0.34$~\cite{ :2008te}&N/A&N/A\\ \hline
\end{tabular} 
\end{center}
\end{table}

The most recent advances on this topic concern the measurement of the mass of the  $X(3872)$ and the discussion on whether there is more than one state 
with similar mass (as predicted by tetraquark models) or not. The two options being investigated are that either the neutral and charged $B$ mesons decay 
to different linear combinations of the two possible $X$ states or that the two states decay into different final states (in particular $J/\psi\pi\pi$ and $D^{*0}D^0$).

The first possibility (different $B^0$ and $B^\pm$ decays) has been investigated by CDF~\cite{Aaltonen:2009vj}, where the  $J/\psi\pi\pi$  spectrum has been fitted searching for evidence of multiple structures. The negative result of such a search has allowed to establish that the eventual two states would have a mass difference smaller than 3.2 MeV at 90\% C.L.. BaBar and Belle have instead measured the masses of the states observed in $B^0$ and $B^\pm$ decays separately~\cite{Aubert:2008gu,:2008te}, arriving at similar conclusions, $\Delta M=1.2\pm 0.8$ MeV.

Both scenarios of multiple $X$ states have instead been investigated by several measurements of mass of the $X$ state performed by BaBar~\cite{Aubert:2008gu,Aubert:2007rva} and Belle ~\cite{:2008te,:2008su} .
 The summary of all available mass measurements 
is shown in Fig.~\ref{fig:Xjpsipipi} where 
 the measurements are separated by production and decay channel. The current word average is $M = 3872.2 \pm 0.4$ MeV.
There is an indication that the particle decaying into $J/\psi\pi\pi$ is different from the one decaying into 
$D^{*0}\bar{D}^0$, their masses differ by about 3.5 standard deviations.

In addition, BaBar also measured the $X(3872)$ width,
$\Gamma=$($3.0^{+1.9}_{-1.4}\pm 0.9$) MeV~\cite{Aubert:2007rva} , result which will be interpreted in the next section.

\subsubsection{Production and decay}
\begin{figure}[bht]
\begin{center}
\epsfig{file=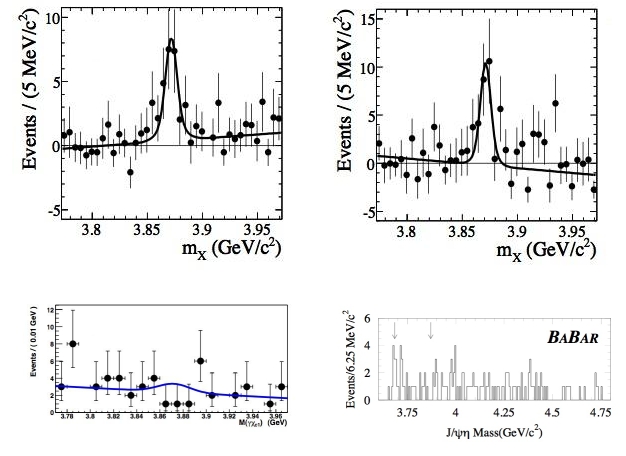,width=12cm} 
 \caption{\it Representative measured invariant mass spectra of the most relevant decay modes where the $X$ mesons has been searched. From left to right, top to bottom:  $J/\psi\gamma$~\cite{:2008rn}, $\psi(2S)\gamma$~\cite{:2008rn}, $\chi_c(1P)\gamma$~\cite{Choi:2003ue}, $J/\psi\eta$~\cite{Aubert:2004fc} }
    \label{fig:Xspectra}
    \end{center}
\end{figure}

The $X$ meson has been searched in several of its possible decay channels ($f$), by looking for $B\to XK$, $X\to f$. 
The searched decay modes and the measured product branching fractions are listed in Tab~\ref{tab:xdecays}.
Representative measured spectra where the signal is present are summarized in Figs.~\ref{fig:Xjpsipipi} and~\ref{fig:Xspectra}.

Several aspects of these measurements can be stressed:
\begin{itemize}
\item albeit with relatively low statistics, the decay branching fraction into $XK$ of the charged and neutral  $B$ decays are consistent:  $BF(B\to XK^0)/BF(B\to XK^\pm)=0.63\pm0.16$ for $X\to J/\psi\pi\pi$   and $BF(B\to XK^0)/BF(B\to XK^\pm)=1.5\pm0.4$ for $X\to  D^{*0}D^0$
\item the $DD^*$ decay mode is the favourite by almost an order of magnitude
\item the  $\psi(2S)\gamma$ branching fraction is comparable with the $J/\psi\gamma$ one.
\item the  $ J/\psi\eta$ decay has been searched for although it would violate charge conjugation. It was actually one of the evidences that lead to the determination of C. 
\end{itemize}

\begin{figure}[bht]
\begin{center}
\epsfig{file=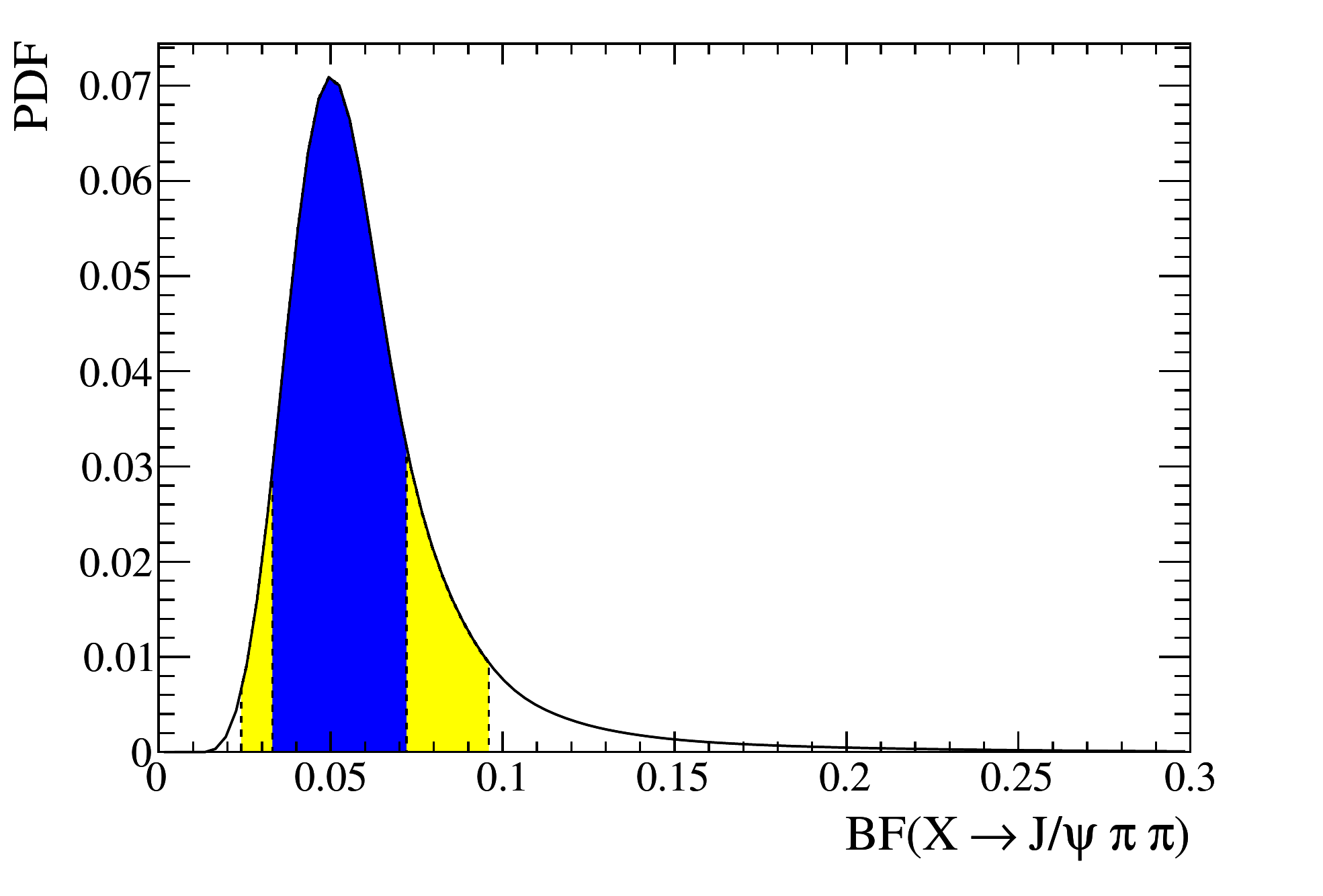,width=6cm} 
\epsfig{file=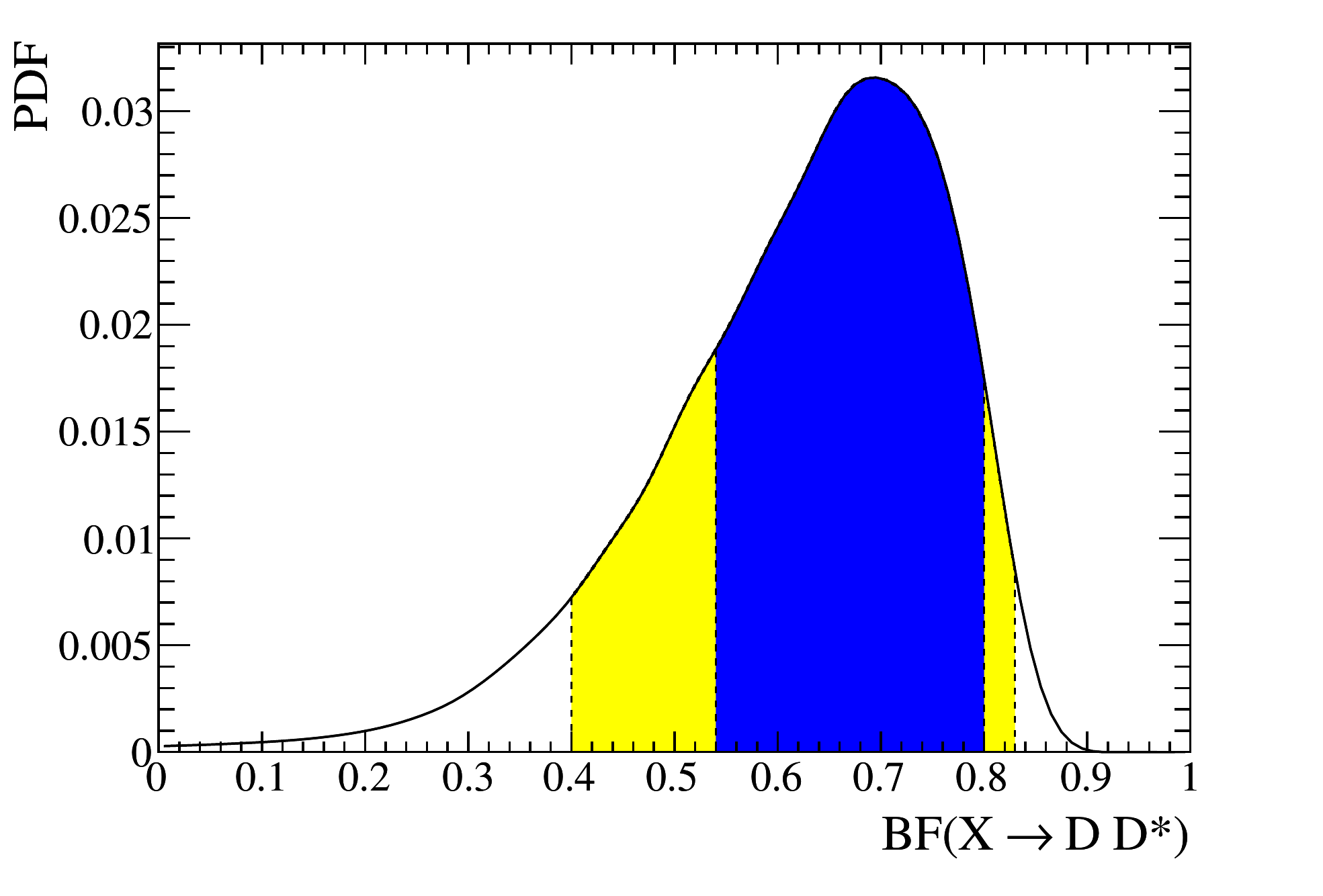,width=6cm} 
\epsfig{file=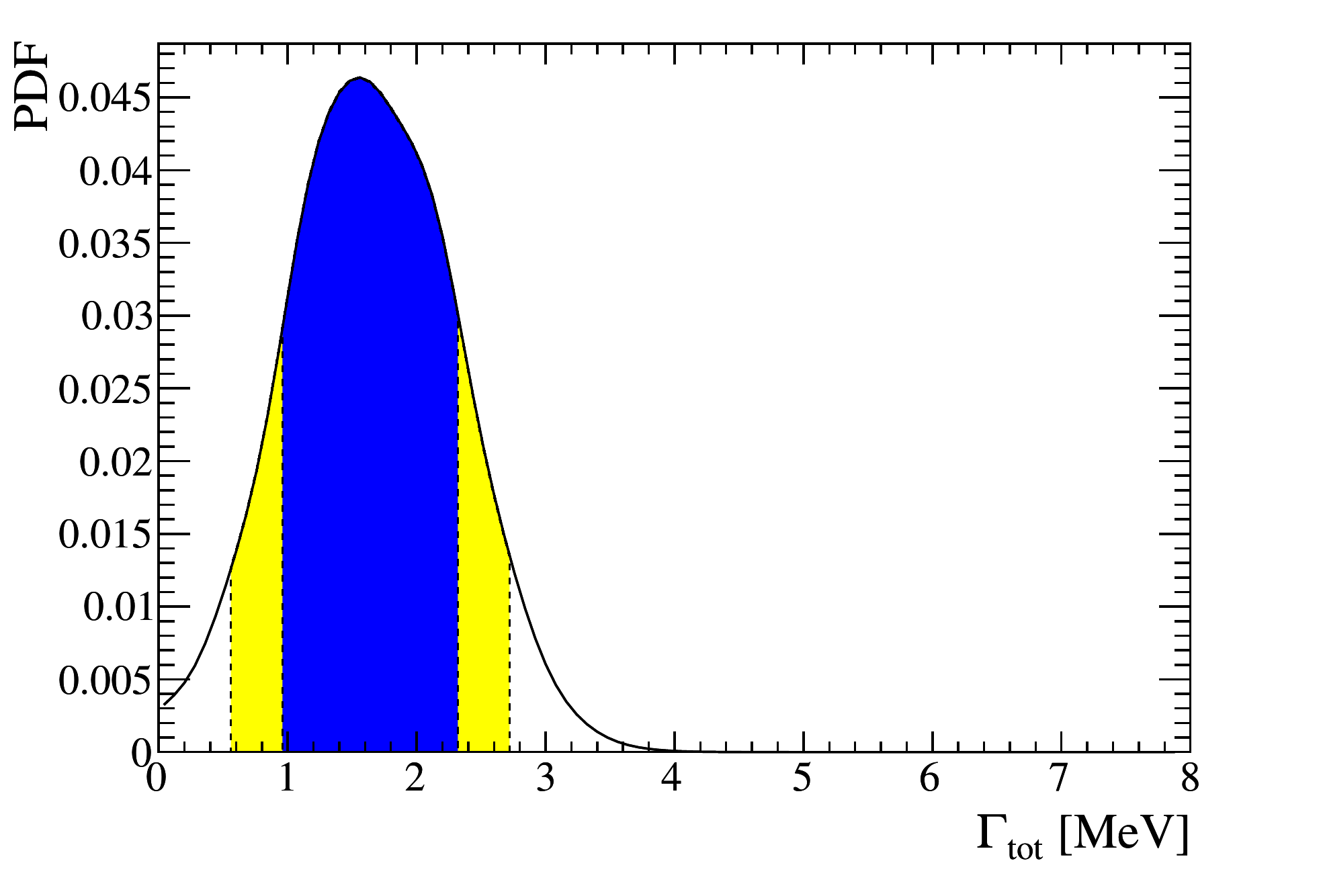,width=6cm} 
\epsfig{file=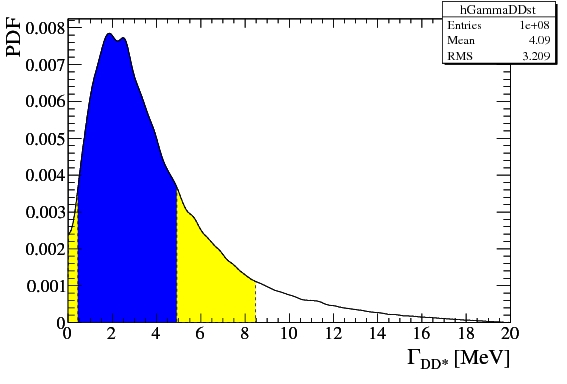,width=6cm} 
 \caption{\it Likelihood function of the  $X$ branching fraction in $J/\psi\pi\pi$ and $D^{*0}D^0$, the total width and the partial width  $\Gamma(X\to D^{*0}D^0)$. See text for a description of the combination method.  The dark (light) filled area corresponds to the 68\% (90\%) C.L. region. }
    \label{fig:xlike}
    \end{center}
\end{figure}

The measured product branching fractions can be translated into absolute branching fractions of the $X$ particle by exploiting the upper limit on $B\to XK$ measured by BaBar from the spectrum of the kaons recoiling against fully reconstructed $B$ mesons~\cite{Aubert:2005vi}, $BF(B^\pm \to K^\pm X(3872))<3.2 \times 10^{-4}$ at 90\% CL. Combining the likelihood from the measurements of the product branching fractions in the observed channels, the  $B\to XK$ upper limit and the measured $X$ width ~\cite{Aubert:2007rva,Choi:2003ue}, with a bayesian procedure we extracted the likelihood for the absolute $X$ BF and the widths in each of the decay modes. Then, we used the probability distributions extracted from this procedure to set limits on the not observed channels. The full shape of the experimental likelihoods was used whenever available, while gaussian errors and poissonian counting distributions have been assumed elsewhere. The 68\% confidence intervals (defined in such a way that the absolute value of the PDF is the same at the upper and lower bound, unless one of them is at the boundary of the physical range) are summarized in Tab.~\ref{tab:xdecays} for each of the decay modes. Fig.~\ref{fig:xlike} shows the likelihoods for the $X$ branching fraction in $J/\psi\pi\pi$ and $D^{*0}D^0$, the total width and the partial width  $\Gamma(X\to D^{*0}D^0)$.

\begin{figure}[bht]
\begin{center}
\epsfig{file=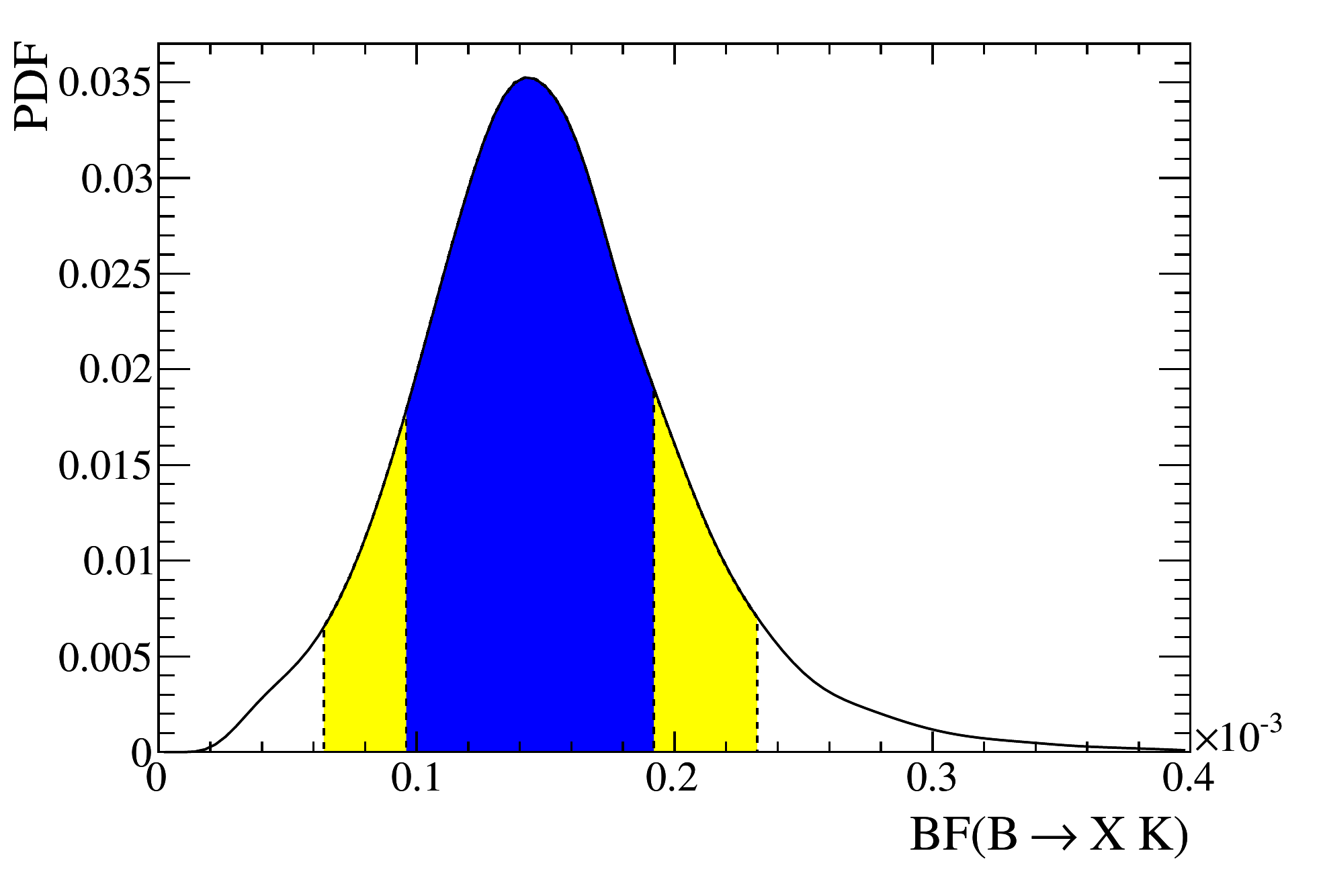,height=7cm} 
 \caption{\it Likelihood function of the  $B\to X$ branching fraction, compared with the measured branching fraction for the corresponding decays with regular charmonia.  See text for a description of the combination method.  The red (yellow) interval corresponds to the 68\% (90\%) C.L. region. }
    \label{fig:bxlike}
    \end{center}
\end{figure}

Finally, information can be obtained also on the production mechanism. The likelihood combination described here allowed also to extract the $B\to XK$ branching fraction. The result (see Fig.~\ref{fig:bxlike})  shows that the X meson is less copiously produced than all other charmonia, for which the corresponding branching fration is at least $5\times 10^{-3}$ .
As far as other production mechanisms are concerned, both the $B\to X K^*$ decays and the $\gamma \gamma$ production have been investigated. The measured $BF(B^0 \to X(3872)K^{*}(892)^0) $ shows that the $X$ meson is produced in association with a $K^*$ less favorably than in association with the $K$, contrarily to the other charmonia. The only search in $\gamma \gamma $ production was performed by CLEO~\cite{Dobbs:2004di}. It did not return a signal and a limit was set: $(2J+1)\Gamma_{\gamma\gamma}B(X -> \pi+ \pi- J/\psi) < 12.9$ eV.

\subsection{The 3940 family}
\begin{table}[htb]
\begin{center} 
\caption{ 
Measured $J^{PC}$, masses, and widths of the "3940 family" of states.}
\label{tab:mass3940} 

\begin{tabular}{lccc} \hline 
State & $J^{PC}$ & Mass (MeV)&Width (MeV)\\ \hline
X(3940)~\cite{Abe:2007sya} &0$^{\pm +}$ & $3942^{+7}_{-6}(stat.)\pm6(sys.)$& $37^{+26}_{-15}(stat.)\pm8(sys.)$\\ 
Y(3940)[Belle]~\cite{Abe:2004zs} & [0,1,2]$^{\pm+}$&$3943\pm13$& $87\pm 22$\\ 
Y(3940)[BaBar]~\cite{Aubert:2007vj} & [0,1,2]$^{\pm+} $&$3914.6^{+3.8}_{-3.4}(stat.)\pm1.9(sys.)$& $33^{+12}_{-8}(stat.)\pm5(sys.)$ \\ 
Y(3915)~\cite{:2009tx} &  [0,1,2]$^{\pm+} $& $3915 \pm3(stat.)\pm2(sys.)$&$17\pm 10(stat.)\pm3(sys.)$ \\ 

Z(3940)~\cite{Uehara:2005qd}&2$^{++}$ &$3926\pm2.7(stat.)\pm1.1 (sys.)$ & $21.3\pm6.8 \pm3.6$\\ \hline
\end{tabular} 
\end{center}
\end{table}

\label{sec:3940}

\begin{figure}[bht]
\begin{center}
\epsfig{file=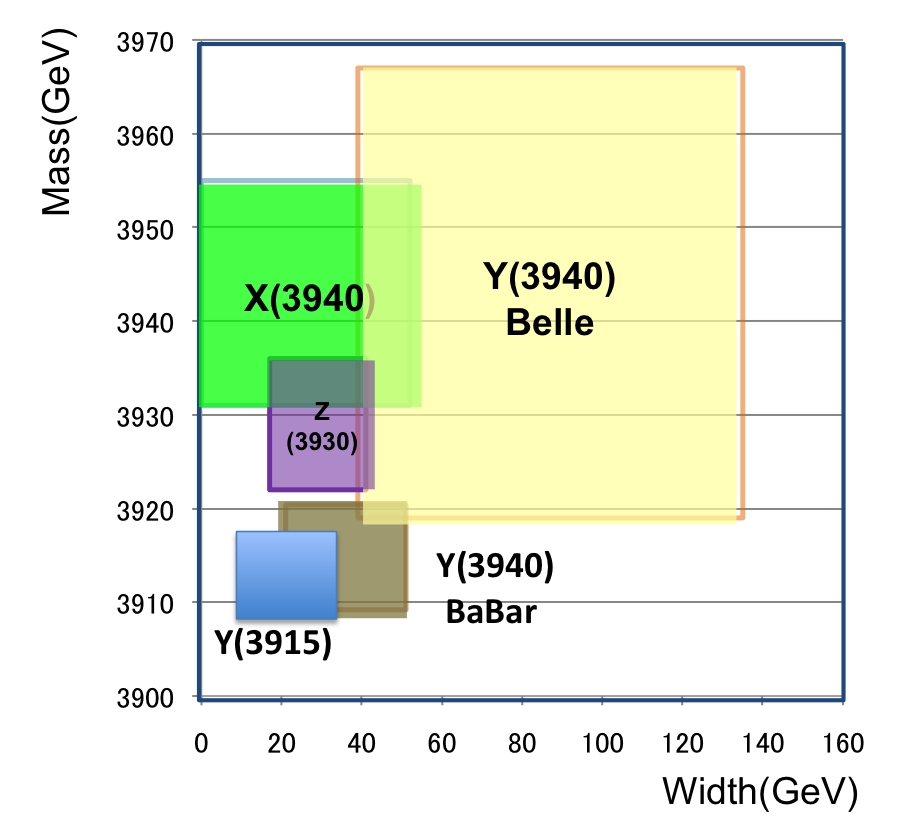,width=8cm} 
 \caption{\it Measured masses and widths of the "3940 family" of states.}
    \label{fig:mass3940}
    \end{center}
\end{figure}

Three different states have been observed in the past years by 
the Belle collaboration with masses close to $3940 \rm {Mev/c}^2$. Their measured masses and widths are summarized in Tab.~\ref{tab:mass3940} and Fig.~\ref{fig:mass3940}.

The first one, 
named  $X$, was observed in continuum events 
(i.e. not in $Y(4S)$ decays) produced in pair with a $J/\psi$ 
meson~\cite{Abe:2007jn}. Subsequently its decay into  $DD^{*}$ has been ascertained by means of a partial reconstruction technique in the same production mechanism~\cite{Abe:2007sya}. The production mechanism constrains it to have positive charge conjugation. Furthermore the fact that in this production mechanism only $J=0$ states are observed makes it likely to be either $J^{PC}=0^{++}$ or $0^{-+}$ . Ref~\cite{Abe:2007jn} was  also able to measure the absolute production rate of this state and searched for $X\to J/\psi\omega$ without evidence of signal. It eventually concluded that $BF(X\to DD^{*})>41\%$ and $BF(X\to J/\psi\omega) < 26\%$ \@ 90\% C.L.. 

\begin{figure}[t]
\epsfig{file=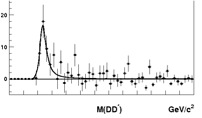,height=5cm}
\epsfig{file=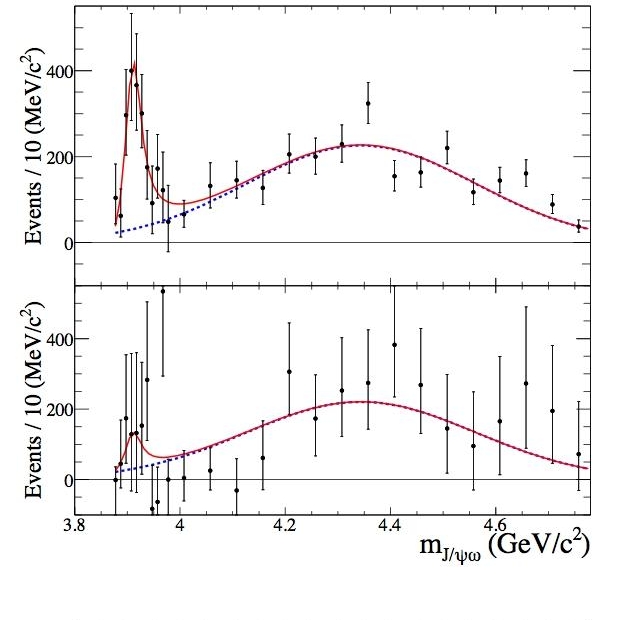,height=5cm}
\epsfig{file=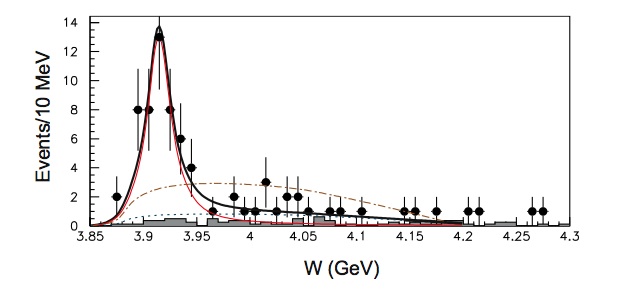,height=4cm}
\epsfig{file=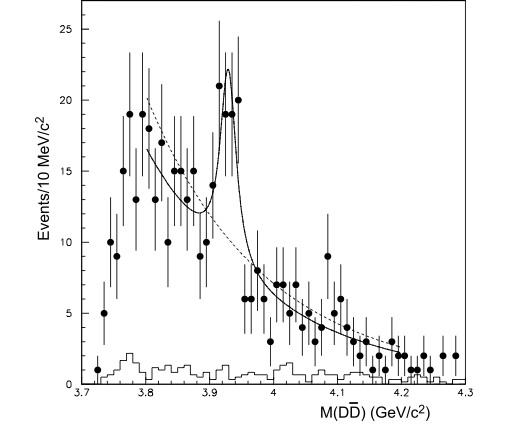,height=4.5cm}
\caption{\it Invariant mass distributions of the most significant observations of the "3940 family" states: $X(3940)\to DD^{*}$, $Y(3940)\to J/\psi\omega$, $Y(3915)\to J/\psi\omega$ and $Z(3940)\to DD$.
}
\label{fig:spectra3940}
\end{figure}

A second one, named 
 $Y$, observed in $B$ decays and decaying into 
$J/\psi\omega$~\cite{Abe:2004zs}, is significantly larger, and only its charge conjugation (C=+) is known, while $J=0,1,2$ and all parities are possible.
The BaBar collaboration confirmed the $Y(3940)\to J/\psi\omega$ 
decay~\cite{Aubert:2007vj}, but measuring a lower mass and 
a width, albeit marginally consistent (see Tab~\ref{tab:mass3940}).  While the experimental differences among the two measurements are still under investigation, a new study has observed a state, named $Y(3915)$
which is consistent with the parameters as measured by BaBar, but produced in $\gamma\gamma$ collisions and decaying into $J/\psi\omega$~\cite{:2009tx} (see  Fig.~\ref{fig:spectra3940}). Under the hypothesis that the $Y(3940)$ and $Y(3915)$ states coincide, this is the first case of two production mechanisms being observed.
The $Y(3940)\to DD^*$ decays have been searched in Ref.~\cite{:2008su}. In the absence of a signal a limit on $BF(Y(3940)\to J/\psi\omega)/BF(Y(3940)\to DD^{*})>0.71$ \@ 90\% C.L. is set. This result supports the hypothesis that this is a different state from the $X(3940)$, where $BF(X(3940)\to J/\psi\omega)/BF(X(3940)\to DD^{*})<0.58$ \@ 90\% C.L..

 The third observed state, named $Z$, is produced in two-photon reactions and decays into 
$D$-pairs~\cite{Uehara:2005qd}. This state has been confirmed in Ref.~\cite{:2010hk}, where the $J^{PC}=2^{++}$ has been ascertained.


Because of their quantum number assignments and their masses these 
states are good candidates for the radial excitation of the $\chi$ 
mesons, in particular the $Z(3940)$ meson could be identified with the 
$\chi_{c2}(2P)$ and the $Y(3940)$ with the $\chi_{c1}(2P)$.
The unclear points are the identification of the $X(3940)$ state and 
the explanation of why the $Y(3940)$ state does not decay preferentially 
in $D$ mesons.

\subsection{Other $C=+$ states}


\label{sec:otherp}

Another set of states has been observed with $C=+$ and masses close to 4150 MeV, creating another interesting cluster of states.

\begin{figure}[hbt]
\epsfig{file=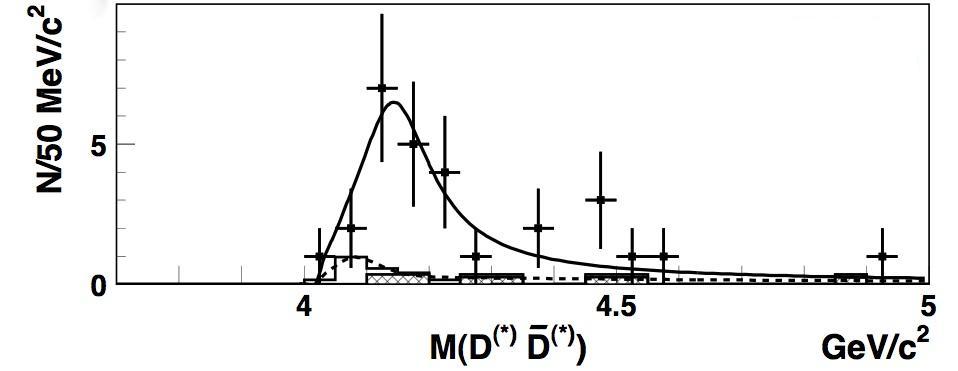,height=5cm}
\epsfig{file=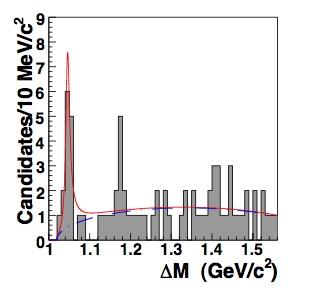,height=5cm}
\epsfig{file=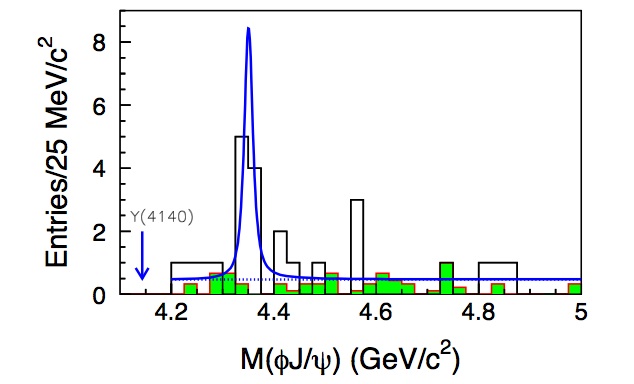,height=5cm}
\caption{\it Invariant mass distributions of the most significant observations of the states with $C=+$ and mass above 4 GeV: $X(4160)\to D^{*+}D^{*-}$, $X(4140)\to J/\psi\omega$, and $X(4350)\to DD$.
\label{fig:spectra4150}}
\end{figure}

The first of such states was observed by Belle produced in continuum events and in pair production with a $J/\psi$ meson and decaying into $D^{*+}D^{*-}$~\cite{Abe:2007sya} (see Fig.~\ref{fig:spectra4150}.
Its measured mass and width are $M=4156^{+25}_{-20}(stat.)\pm15(sys.)$ MeV and $\Gamma=139^{+111}_{-61}(stat.)\pm21(sys.)$ MeV. The production mechanism favors $J=0$ and it enforces $C=+$, making it a good candidate for $\eta_c(3S)$, which would have $J^{PC}=0^{-+}$.

Next comes the report from CDF of an enhancement close to the threshold in the $J/\psi \phi$ invariant mass in
$B\to J/\psi \phi K$ decays~\cite{Aaltonen:2009tz}. The fitted mass of the resonant state,  $Y(4140)$, is $M=(4143.0\pm2.9\pm 1.2)$ MeV with $\Gamma=(11.7^{+8.3}_{5}\pm3.7)$ MeV. While the natural ($L=0$) quantum numbers assignment would be $J^{PC}=0^{++}$, $J^{PC}=1^{-+}$ is also allowed, leaving open the possibility of this state being the lowest lying hybrid. This hypothesis is also suggested by the closeness of the measured mass with the predictions of lattice calculations for the hybrid ground state (see for instance Ref.~\cite{Bernard:1997ib}).
The search for such a state in this production mechanism at $B$-Factories does not have enough statistics to be conclusive. Belle has therefore searched this state in $\gamma\gamma$ production which is expected to be copious in certain models~\cite{Branz:2009yt}. No evidence was found and a limit $\Gamma_{\gamma\gamma}\times BF(\phi J/\psi)<41$ (6) eV for $J^P=0^+$ ($2^+$) were set \@ 90\% C.L. for the $Y(4140)$. 

When performing this search a 3.2$\sigma$ evidence was found with $M=4350.6^{+4.6}_{-5.1}(\rm{stat})\pm 0.7(\rm{syst}))$ MeV and $\Gamma=(13^{+18}_{-9}(\rm{stat})\pm 4(\rm{syst}))$ MeV. Yet another state to place in the spectrum, the $X(4350)$, with C=+, and close to one of the $J^{PC}=1^{--}$ states reported in Sec.~\ref{sec:1mm}.


\subsection{The $1^{--}$ family}


\label{sec:1mm}
The easiest way to assign a value for $J^{PC}$ to a particle is to observe its production via $e^+e^-$ 
annihilation, where the quantum numbers must be the same as the the photon:  $J^{PC}=1^{--}$. $B$ factories
can investigate a large range of masses for such particles by looking for events where the initial state 
radiation brings the $e^+e^-$ center-of-mass energy down to the particle's mass (the so-called 'ISR' events). 
Alternatively, 
dedicated $e^+e^-$ machines, like CESR and $BEP$ scan directly the center-of-mass energies of interest.
\begin{figure}[hbt]
\begin{center}
\epsfig{file=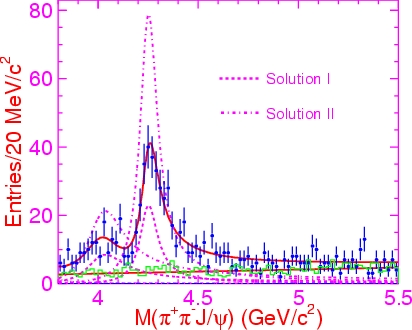,height=4.5cm}
\epsfig{file=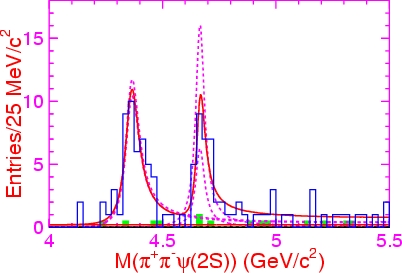,height=4.5cm}
\end{center}
 \caption{\it $J/\psi\pi^+\pi^-$  (left) and $\psi(2S)\pi^+\pi^-$  (right)  invariant mass in ISR production.}
\label{fig:belle1mm}
\end{figure}

The observation of new states in these processes started with the discovery of the $Y(4260)\to 
J/\psi\pi^+\pi^-$ 
by BaBar~\cite{Aubert:2005rm}, promptly confirmed  by CLEO-c both in the same production process~\cite{He:2006kg} and in 
direct production~\cite{Coan:2006rv}. The latter paper also reported evidence for 
$Y(4260)\to J/\psi\pi^0\pi^0$ and some events of $Y(4260)\to J/\psi K^+K^-$. 

While investigating whether the $Y(4260)$ decayed to 
$\psi(2S)\pi^+\pi^-$ BaBar found that such decay did not exist but 
discovered a new $1^{--}$ state, the $Y(4350)$~\cite{Aubert:2006ge}. 
While the absence of $Y(4260)\to \psi(2S)\pi^+\pi^-$ decays could be 
explained if the pion pair in the $J/\psi\pi^+\pi^-$ decay were 
produced 
with an intermediate state that is too massive to be produced with a 
$\psi(2S)$ (e.g. an $f^0$), the absence of $Y(4350)\to 
J/\psi\pi^+\pi^-$ is still to be understood, more statistics might be 
needed in case the $Y(4260)$ decay hides the $Y(4350)$.

Next, Belle has published the confirmation of all these $1^{--}$ 
states~\cite{:2007sj,:2007ea} and at the same time has 
unveiled a new state that was not visible in BaBar data due to the 
limited statistics: the $Y(4660)$. Fig.~\ref{fig:belle1mm} shows the 
published invariant mass spectra for both the $J/\psi\pi^+\pi^-$ and 
the $\psi(2S)\pi^+\pi^-$ decays.

\begin{figure}[bht]
 \begin{center}
\epsfig{file=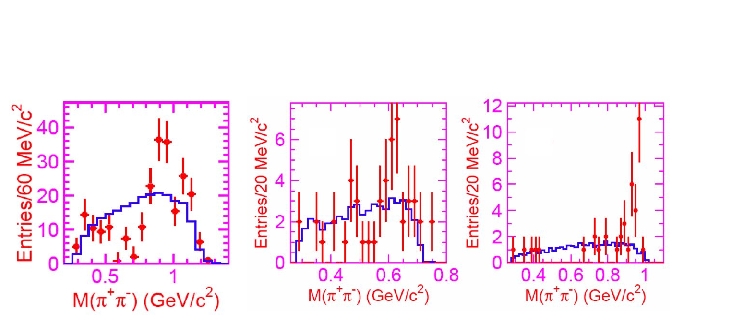,height=6cm}
\caption{\it Di-pion invariant mass distribution in $Y(4260)\to 
J/\psi \pi^+\pi^-$ (left), $Y(4350)\to \psi(2S) \pi^+\pi^-$ (center), 
and $Y(4660)\to \psi(2S) \pi^+\pi^-$ (right) decays. 
}
\label{fig:bellepipiinv}
\end{center}
\end{figure}

Critical information for the unravelling of the puzzle is whether the 
pion pair comes from a resonant state. Fig.~\ref{fig:bellepipiinv} 
shows the di-pion invariant mass spectra published by Belle for all the 
regions where new resonances have been observed. Although the 
subtraction of the continuum is missing, there is some
indication that only the $Y(4660)$ has a well defined intermediate 
state (most likely an $f_0$), while others have a more complex 
structure.

A discriminant measurement between Charmonium states and new 
aggregation forms is the relative decay rate between these decays into 
Charmonium and the decays into two charm mesons. Searches have 
therefore been carried out for $Y\to D^{(*)}D^{(*)}$ 
decays~\cite{Abe:2006fj,collaboration:2007mb,Aubert:2007pa} without any 
evidence for a signal. The most stringent limit is~\cite{Aubert:2007pa} 
$BF(Y(4260)\to D\bar{D})/BF(Y(4260)\to J/\psi\pi^+\pi^-)<1.0 @$ 90\% 
confidence level.
\begin{figure}[hbt]
\begin{center}
\epsfig{file=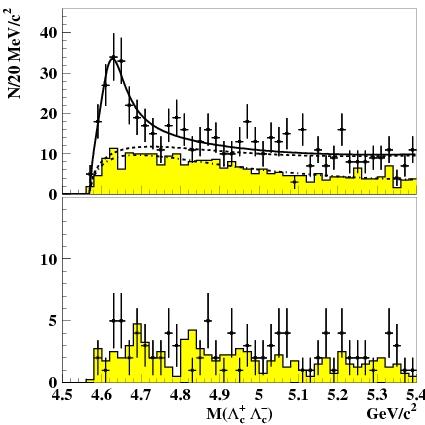,height=5cm}
\epsfig{file=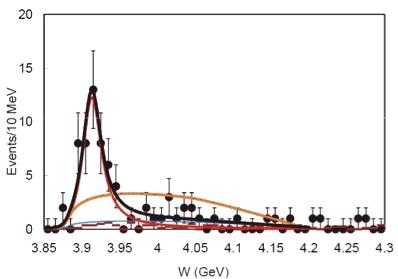,height=5cm}
\caption{\it The $\Lambda_c\bar{\Lambda}_c$
 distribution in ISR (left) and the $ J/\psi\omega$
 distribution in $\gamma\gamma$ events as  measured by Belle. The superimposed lines are the result of 
the fits to the data whose results are reported in the text.
\label{fig:lclc}}
\end{center}
\end{figure}

A distinctive signature of tetraquarks would instead be the observation of decays of these mesons into two baryons. This has lead to the 
search by Belle of resonant structures decaying into $\Lambda_c\bar{\Lambda}_c$ associated to ISR~\cite{Pakhlova:2008vn}. A tantalizing structure appears near threshold (see Fig.~\ref{fig:lclc}, which fitted with a Breit-Wigner returns $M=(4634^{+8+5}_{-7-8})$ MeV and $\Gamma_{tot}=(92^{+40+10}_{-24-12})$ MeV, properties which are close to the one of the $Y(4660)$. A  consistent analysis of the two measured spectra has actually concluded that the two structures come from the same state with a large preference for the baryonic one ($BF(Y(4660)\to\Lambda_c\Lambda_c)/BF(Y(4660)\to\psi(2S)\pi\pi)=(25\pm7)$~\cite{Cotugno:2009ys}).


\subsection{Charged States}


\label{sec:charged}
\begin{figure}[htb]
\begin{center}
\epsfig{file=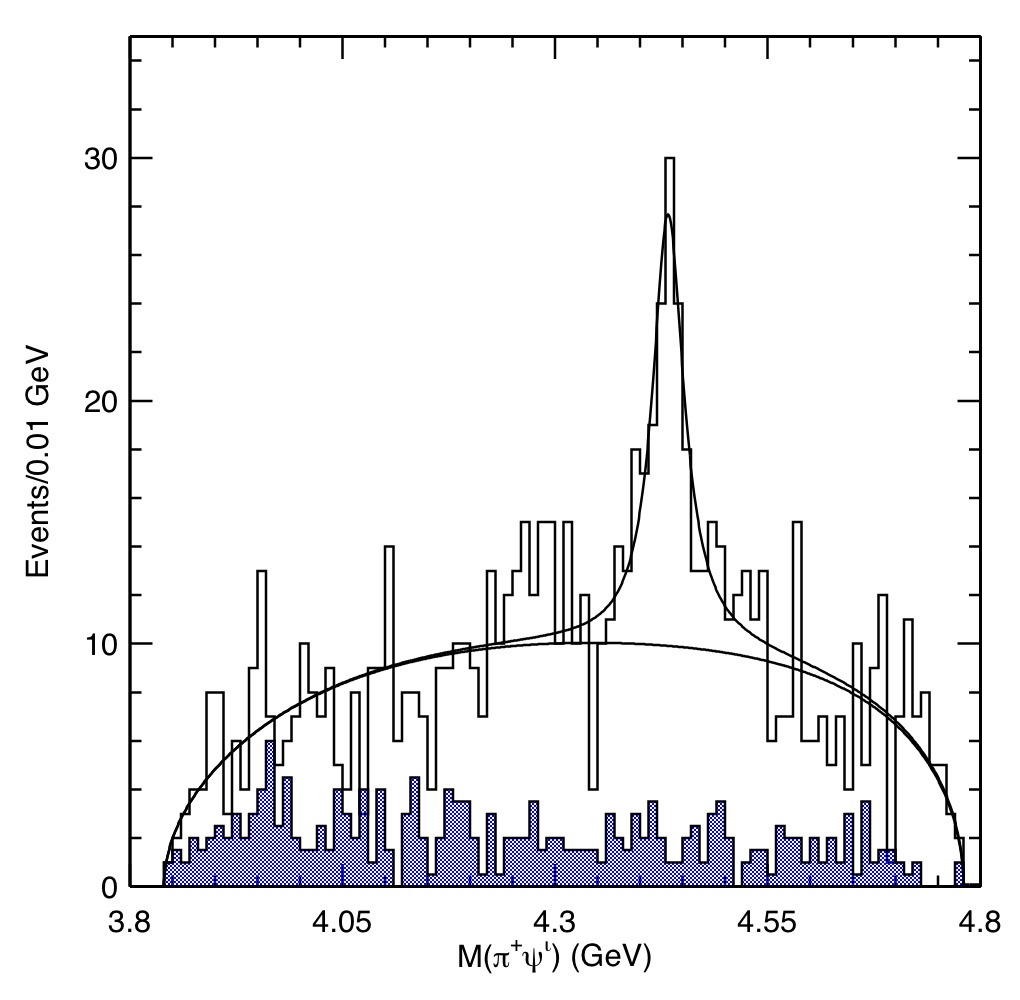, height=6cm} 
\epsfig{file=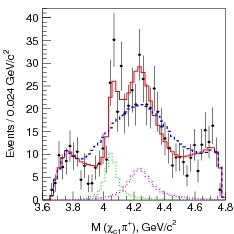, height=6cm} 
\caption{\it The $\psi(2S)\pi^\pm$ (left) and  $\chi_{c1}\pi^\pm$ (right) invariant mass distributions superimposed with the fit result showing the three charged resonances. }
\label{fig:zstates}
\end{center}
\end{figure}

The real turning point in the quest for states beyond the Charmonium was the observation of charged states decaying into charmonium. There is in fact no way to explain such observation without having at least four bound quarks ($c\bar{c}u\bar{d}$). There is currently evidence of three charged states, all seen exclusively by the 
Belle Collaboration: the $Z(4430)$ state
decaying into $\psi(2S)\pi^\pm$~\cite{:2007wg},  and  the $Z_1$ and $Z_2$ states
decaying into $\chi_{c1}(2S)\pi^\pm$~\cite{Mizuk:2008me}.

Unfortunately these states have been observed in B decays in association with a charged kaon, i.e. in three body $B\to X_{cc}\pi K$ decays, where $X_{cc}=\psi(2S)$ or $\chi_{c1}$. Three body decays suffer from interferences between strong amplitudes mediated by different resonances. In these particular cases the $\pi K$ system presents several known resonances that could cause significant effects of reflection. Namely, the decays  $B\to X_{cc}K^*(892)$,   $B\to X_{cc}K^*(1410)$, and in particular their interference constitute irreducible sources of background which are difficult to estimate. The original observation of the $Z(4430)$~\cite{:2007wg} has been therefore object of detailed scrutiny by the BaBar collaboration who performed the search for the same final state by studying in detail the efficiency corrections and the shape of the background taking the latter from the data as much as possible~\cite{:2008nk}.  The search 
resulted into hints of a structure close to Belle's observation, but after accurate estimate of the background results into an exclusion on the product of branching fractions BF($B\to Z(4330)K^+)$BF$(Z(4330)\to\psi(2S)\pi)<3.1\cdot10^{-5}@ 95\%$ C.L. to be compared with BF($B\to Z(4330)K^+)$BF$(Z(4330)\to\psi(2S)\pi)= (4.1^{+1.0}_{-1.4})\cdot10^{-5}$ as reported in Ref.~\cite{:2007wg}. Above all the paper from BaBar has raised the attention to the problems inherent to the analysis of the Dalitz plot, leading to a reanalysis from Belle~\cite{:2009da}. 

The results of the latter analysis, that confirmed the original ones although the errors on the parameters increase significantly, and of the    $B\to \chi_{c1}\pi K$  decays (not yet analyzed with the full three body approach) are shown in Fig.~\ref{fig:zstates}. The fits to the  $\psi(2S)\pi$ and $\chi_{c1}\pi$ invariant mass distributions return  masses  $M=4443^{+24}_{-18}$ MeV, $M=4051\pm14^{+20}_{-41}$ MeV, and  $M=4248^{+44+180}_{-29-35}$ MeV,
and widths $\Gamma=109^{+113}_{-71}$ MeV,  $\Gamma=82^{+21+47}_{-17-22}$ MeV, and  
$\Gamma=177^{+54+316}_{-39-61}$ MeV  for the $Z(4430)$, $Z_1$, and $Z_2$ states respectively.
\begin{figure}[bht]
\begin{center}
\epsfig{file=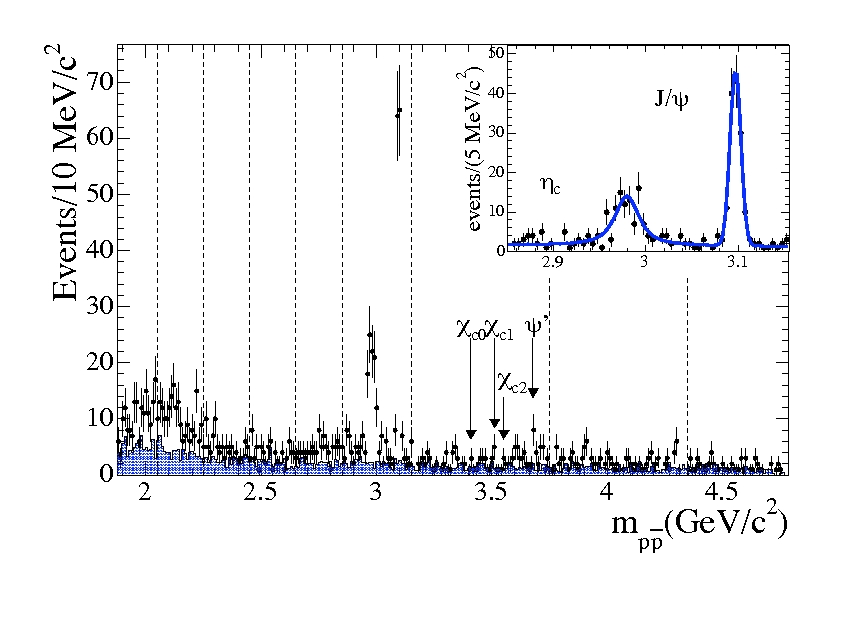,width=5.5cm}
\epsfig{file=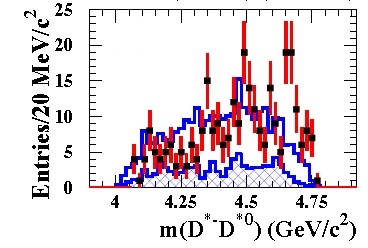,height=4cm} 
 \caption{\it Two more representative spectra that have not been analyzed and that will be discussed in the concluding section of the paper:
 $p\bar{p}$ invariant mass spectrum from $B\to p\bar{p}K$ decays as in Ref.~\cite{Aubert:2005gw} and $D^{*0}D^{*-}$ invariant mass spectrum from $B\to D^{*0}D^{*-}K$ decays as in Ref.~\cite{Aubert:2003jq}}
    \label{fig:charmonbaryons}
    \end{center}
\end{figure}


\subsection{Interpretations}
\label{sec:charmon_interp}

In this section we match the information of the observed states, summarized in the previous sections, with the possible theory interpretations detailed in Sec.~\ref{sec:theory}, we detail the arguments in favour or against the different interpretations and, when possibile, mention the measurements that would help resolve the ambiguities.

\subsubsection{X(3872)}
\label{sec:charmon_interp_X}

$X(3872)$ represents a very peculiar case because it lies very close to the $D\bar{D}^{*}$ threshold, its binding energy with respect to this threshold being compatible with zero.
This characteristic lead many authors to identify the $X$ with a $D\bar{D}^{*}$ molecule at threshold. Since $m_X<m_{D^0}+m_{D^{0*}}$, at least in the $J/\psi\pi^+\pi^-$ channel, the state could be considered as a proper bound state, {\it i.e.} with negative binding energy ($E_{B}=m_x-m_{D^0}-m_{D^{0*}}<0$).

This interpretation, as the others we will review in the following, stand on the hypothesis 
that the $X$ has $J^{PC}=1^{++}$. 
This prejudice came from the fact that
even if CDF stated in \cite{Abulencia:2006ma} it could not distinguish between
$1^{++}$ and $2^{-+}$ , the Belle analysis \cite{Abe:2005iya} seemed to prefer the $1^{++}$
assignment.

The existence of such a bound state has been investigated using the potential models discussed in Sec. \ref{subsec:molecules}. A lot of work has been spent since the first observation of the $X$ to obtain solid and quantitative results from those potential models, but yet there is not a final answer in the literature to the question ``Does a $1^{++}$ $D\bar{D}^{*}$ bound state exist at threshold?''.

Tornqvist was the first to state in \cite{tornqvist2} and later to confirm in \cite{Tornqvist:2004qy}, that a $D\bar{D}^{*}$ bound state for $J^{PC}=0^{-+},1^{++}$ and $I=0$ is likely to exist. From the point of view of one pion exchange the problem is tightly related to the deuteron case. In the deuteron case the tensor term of the potential in Eq. (\ref{ope}) favors the binding allowing for {\it S} to {\it D} wave transitions. Indeed the scalar part of the potential provides only about one third of the total binding energy. It is important to notice that one cannot predict exactly the deuteron binding energy without a detailed knowledge of the very short range interactions, since its value cannot be deduced from the other scales of the problem. Nevertheless once the pion-nucleon coupling constant is known and a regularization scheme for the potential is given, one can use the measured binding energy to deduce the value of the cutoff $\Lambda$ and thus the long distance part of the wave function. The same reasoning applies to the $X$ case. The binding channels for a $D\bar{D}^{*}$ state are $J^{PC}=0^{-+},1^{++}$ with $I=0$. Assuming some definite value for the binding energy, one is able to compute the wave function at long distance whose components are $D^{0}\bar{D}^{0*}$ and $D^{+}D^{-*}$ both in {\it S}-wave and {\it D}-wave.

A slightly different result is found by Swanson in \cite{Swanson:2003tb}, where the combined effects of short-range one-gluon-exchange potential and long-range one-pion-exchange potential are required to obtain the binding. The possibility to include other channels, such as $J/\psi\rho$, $J/\psi\omega$ and $D^{+}D^{-*}$ is investigated.

On the other hand, a completely different result is obtained by Suzuki in \cite{Suzuki:2005ha}. The effective potential between $D$ and $\bar{D}^{*}$ due to one pion exchange is computed from the Born scattering amplitude:
$
\mathcal{A}=(\epsilon'^{*} \cdot q)(\epsilon \cdot q)g^2/(m^2_\pi-q^2-i\epsilon)
$,
\noindent where $q$ is the $\pi$ momentum and thus the difference between the $D$ and $D^{*}$ momenta. 
Since the decay $D^{0*}\to D^0\pi^0$ occurs with a tiny $Q$-value $\Delta=m_{D*}-m_{D}-m_{\pi}\sim8~{\rm MeV}$ the pion is produced practically at rest and thus in the non relativistic limit for the denominator of the pion propagator is:
$m^2_\pi-q^2\simeq -(2m_\pi\Delta-|\mathbf{q}|^2)$. 
When computing the Fourier transform of the scattering amplitude one obtains a three-dimensional delta function plus correction of $\mathcal{O}((2m_\pi\Delta)^2)$, which can be neglected. 
Unlike the one-dimensional $\delta$-function, the three-dimensional $\delta$-function in a Schroedinger equation does not admit bound states. The author concludes that the $X$ cannot be regarded as an analog of the deuteron and furthermore questions that the binding is recovered when the short-distance one-gluon-exchange potential is added. Actually for a spatially extended object, as the $X$ is, it is difficult to believe that the binding is provided by short range forces.
Furthermore the possibility that the binding is due to the coupling with other molecular channels, such as $J/\psi\omega$, seems very unlikely since the coupling to this channel would occur through $D/D^*$ exchange thus resulting in a Yukawa potential with very short range ($\sim$0.1~fm). On the other hand Suzuki considers more likely that the binding proceeds through the coupling to charmonium states, due to $u,d$ quark exchange. However in this case the nature of the state would be essentially that of a charmonium and the dynamics completely different from the molecular one. For a recent review on the one pion exchange binding mechanism see~\cite{Thomas:2008ja}.


Besides the molecular interpretation, alternative explanations have been proposed, such as a regular charmonium assignment \cite{Barnes:2003vb,Eichten:2004uh} and a tetraquark interpretation \cite{Maiani:2004vq,Maiani:2007vr}. 
We will review the characteristics of the different interpretations by topic.

{\it Quantum numbers}.
From the point of view of quantum numbers many considerations were made soon after the first observation of $X$ in 2003 \cite{Pakvasa:2003ea}. The absence of the decay in $D^{0}\bar{D}^{0}$ despite the ample phase space (decay momentum $p_*\sim$500~MeV) suggests unnatural spin parity $J^P=0^{-},1^{+},2^{-},\ldots$ or unnatural spin-charge conjugation parity $J^{PC}=0^{+-},1^{-+},2^{+-},\ldots$. The dipion quantum numbers help discriminating  among the possible $J^{PC}$ of the $X$. A scalar dipion ($J^{PC}=0^{++}$) combined in {\it S}-wave with the $J/\psi$ gives a $1^{--}$ state, in {\it P}-wave a $1^{+-}$ state and in {\it D}-wave a $2^{--}$ state, all with isospin $I=0,2$; a vector dipion ($J^{PC}=1^{--}$) in {\it S}-wave with the $J/\psi$ gives a $1^{++}$ state, while in {\it P}-wave a $2^{-+}$ state, all with isospin $I=1$. Finally a tensor dipion ($J^{PC}=2^{++}$) gives in {\it S}-wave a $2^{--}$ state with isospin $I=0,2$. 
Both the $J^{PC}=1^{++}$ and $2^{-+}$ assignments favor the vector dipion configuration, which would also explain the predominance of the decay into $J/\psi\rho$.



As stated in the beginning, the vicinity of the $X$ mass to the $D\bar{D}^{*}$ threshold has prompted considerations of the possibility that $X$ could be a $D\bar{D}^{*}$ molecule:

\begin{equation}
\label{d0d0}
\frac{|D\bar D ^{*}\rangle\pm|\bar DD^{*}\rangle}{\sqrt{2}}
\end{equation}

\noindent Potential models predict that $I=0$ states are favored with respect to $I=1$ states, in particular they predict $0^{-+}$ and $1^{++}$ bound states. This would imply an equal contribution of the charged and neutral mesons components:

\begin{equation}
|D\bar{D}^{*}\rangle=\frac{|D^{0}\bar{D}^{0*}\rangle+|D^{+}D^{-*}\rangle}{\sqrt{2}}
\end{equation}

\noindent However since the binding energy is much smaller than the mass gap between neutral and charged pairs, one expects substantial isospin breaking in the wave function. Indeed, since the $D^{+*}D^{-}$ threshold ($\sim$3879~MeV) lies $\sim$8~MeV above the $D^{0*}\bar{D}^{0}$ threshold ($\sim$3871~MeV) it should have a smaller weight than $D^{0*}\bar{D}^{0}$, which means that there will be a strong $I=1$ component in the state.
Some new results on the isospin issue have been obtained in~\cite{Gamermann:2009fv,Gamermann:2009uq,Gamermann:2010nz} studying the charged versus neutral $D$ mesons components of the X wave-function.



As for the charmonium option two assignments have been proposed~\cite{Barnes:2003vb,Eichten:2004uh,Quigg:2004vf}: $1D$ and $2P$ respectively for a $J^{P}=2^{-}$ and a $J^{P}=1^{+}$ state, since the parity and charge conjugation quantum numbers of a pure $c\bar{c}$ state are related to spin and orbital angular momentum as $P=(-1)^{L+1}$ and $C=(-1)^{L+S}$. A computation of the radiative decays and of the principal hadronic decay modes allows to exclude some of the states among $1D$ and $2P$, due to the extremely small width of the $X$. The states which remain left are $1^{3}D_{3,2}$, $1^{1}D_{2}$, $2^{3}P_{1}$ and $2^{1}P_{1}$. Since the $X$ has even charge conjugation, there remain two possibilities: $1^{1}D_{2}$ and $2^{3}P_{1}$, respectively for $2^{-+}$ and $1^{++}$  $J^{PC}$ quantum numbers.

The main limit of the charmonium assignment, which seemed to be totally ruled out, is the isospin quantum number. The observation of the decays $X\to J/\psi\pi^+\pi^-\pi^0$ and $X\to J/\psi\pi^+\pi^-$ with almost equal branching ratios indicates that $X$ contains  an equal amount of $I=1$ and $I=0$ components. A standard $c\bar c$ state cannot account for this, since it is a pure isoscalar state. Colangelo and collaborators  argued that the isospin violation is not so severe \cite{Colangelo:2007ph} due to the different phase space volumes available respectively for $J/\psi\pi^+\pi^-$ and $J/\psi\pi^+\pi^-\pi^0$ final states. This is due to the different total decay widths of $\omega$ and $\rho$. Taking this into account the $I=1$ amplitude is approximately five times smaller than the $I=0$ one. This smaller isospin violating effect could be explained with the difference in mass between the charged and neutral $D$-mesons as intermediate states in the $X$ decays.

Finally the tetraquark. In this approach the $1^{++}$ diquark-antidiquark bound state, with symmetric spin contribution:

\begin{equation}
X_q=[cq]_{S=1}[\bar c\bar q]_{S=0}+[cq]_{S=0}[\bar c\bar q]_{S=1}
\end{equation}

\noindent is a candidate to explain $X$, since it is expected to be narrow like all diquark-antidiquark systems below the baryon-antibaryon threshold. Furthermore unnatural spin parity forbids the decay to $D^{0}\bar{D}^{0}$, while the decay to $J/\psi$ plus vector meson is allowed with conservation of the spin of the heavy quark pair. The isospin quantum number is related to the finer structure of the $X_q$ state. The two flavor eigenstates $X_u$ and $X_d$ mix through self energy diagrams, which annihilate a $u\bar u$ pair and convert it into a $d\bar d$ pair through intermediate gluons. In the basis $\{X_u,X_d\}$ the annihilation diagrams contribute equally to all the entries of the mass matrix, while the contribution of the quark masses is diagonal. The resulting $2\times 2$ mixing matrix is:

\begin{equation}
\begin{pmatrix}m_u+\delta & \delta \\ \delta & m_{d}+\delta\end{pmatrix}
\end{equation}

\noindent At the scale determined by the $c\bar c$ pair the annihilation term $\delta$ is expected to be small and thus the mass eigenstates should coincide with flavor eigenstates to a rather good extent. One can put:

\begin{equation}
\begin{split}
X_{low}&=\cos\theta X_u+\sin\theta X_d\\
X_{high}&=-\sin\theta X_u+\cos\theta X_d
\end{split}
\end{equation}

\noindent where, in the limit $\delta<<m_{u,d}$, $\cos\theta\sim$1 and $\sin\theta\sim\theta$, giving an almost maximal isospin breaking. The mass difference between the two states is: $M(X_{high})-M(X_{low})=(m_d-m_u)/\cos2\theta$. 
Indeed, as already mentioned, the possibility that there were two different states $X(3872)$ and $X(3875)$ is not excluded from experimental data.
An alternative mechanism to explain the isospin violation in the tetraquark picture is the possibility of a $\omega-\rho^0$ mixing, as proposed in~\cite{Terasaki:2009in}.

{\it Mass and width}.
Beside the vicinity of the $X$ mass to the $D^{0}\bar{D}^{0*}$ threshold, one of the characteristics of the $X$ is its extremely narrow width, that needs to be accounted for.

In the molecular picture the mass of the $X$ is directly related to the $D^0\bar{D}^{0*}$ threshold.
For what concern the width of the state the main arguments follow from the {\it low-energy universality} \cite{Braaten:2003he} of such a loosely bound molecule. Indeed an {\it S}-wave $D^{0}\bar{D}^{0*}$ molecule has an extremely small binding energy $E_B$, which results in an unnaturally large scattering length $a=\sqrt{2\mu E_B}$. Therefore $E_B=$0.25~MeV gives $a\sim\,8~{\rm fm}$, which is larger than the typical range of strong interactions. Thanks to this feature the molecule has properties that depends on $a$, {\it i.e.} universal properties,  insensitive to the details of the interaction between mesons. Low-energy universality implies also that the asymptotic form of the $D^{0}\bar{D}^{0*}$  bound state wave function is known and can be expressed in terms of $a$: $\psi(r)=(1/\sqrt{2\pi a})e^{-r/a}/r$. In addition to the $D^0\bar{D}^{0*}$ component there will be contributions from the other $1^{++}$ hadronic states $H$ with nearby threshold. In general one can write:

\begin{equation}
|X\rangle=Z^{1/2}_{DD^*}\left(|D\bar{D}^{*}\rangle+|\bar{D}D^{*}\rangle\right)+\sum_{H}Z^{1/2}_{H}|H\rangle
\end{equation}

Low-energy universality implies that as the scattering length $a$ increases, the probabilities for states other than $D^0\bar D^{0*}$ decrease as $Z_H\sim1/a$ and in the limit $a\to \infty$ the state becomes a pure $D^{0}\bar{D}^{0*}$ molecule. 
It is important here to notice that the $X$ represents a really particular case of bound state, since its binding energy is much smaller, {\it i.e.} compatible with zero, than the natural energy scale $m^2_\pi/\mu\sim$20~MeV.

In this framework one can compute the partial widths in the different decay channels. As for the constituent decay channels one has:

\begin{equation}
\begin{split}
\Gamma(X\to D^0\bar D^0 \pi^0)&=Z_{DD*}C_\pi\Gamma(D^{0*}\to D^{0}\pi^0)\\
\Gamma(X\to D^0\bar D^0 \gamma)&=Z_{DD*}C_\gamma\Gamma(D^{0*}\to D^{0}\gamma)
\end{split}
\end{equation}

\noindent where $C_\pi$ and $C_\gamma$ take into account the interference effect between the $D^{0*}$ and $\bar{D}^{0*}$ decays. This interference effect was computed for the first time by Voloshin in \cite{Voloshin:2003nt} and is able to account for an enhancement factor of the width of $2$ at maximum. As for the short-distance decay modes ($J/\psi\rho$, $J/\psi\omega$, $J/\psi\gamma$, $\psi(2S)\gamma$), since $Z_H\sim 1/a$, the main contribution will come from the $D^{0}\bar{D}^{0*}$ component of the wave function. One can estimate that these decay rates scale like $1/a$ as $a\to \infty$. In this way the molecular picture accommodates the narrowness of the $X$. 
We will discuss some problems of this picture in the last paragraphs of this section.

As for the charmonium assignment potential models predict masses which are smaller than 3872~MeV for the $1D$ case, while larger than 3872~MeV for the $2P$ one  \cite{Barnes:2003vb}. The $1^1D_2$ state should have a quite large radiative branching fraction to $\chi_c\gamma$: $\mathcal{B}(1^1D_2\to\chi_c\gamma)\sim 0.5$, which instead has not been observed until now. On the other hand the $2^3P_1$ charmonium is predicted to have branching fractions of a few percents to $J/\psi \gamma$ and $\psi(2S)\gamma$, which seems in agreement with Tab.~\ref{tab:xdecays}. However, in \cite{Eichten:2004uh} a study of the effect of the open charm thresholds on the charmonium properties gives very large partial width into $D^{0}\bar{D}^{0*}$ for $2^{3}P_{1}$ which are not compatible with the narrowness of the $X$. The $1^{1}D_{2}$ state on the contrary is predicted to have a large branching fraction to $h_c\gamma$, which has not yet been looked for. The experimental search for this decay mode would be thus of crucial interest.
The interplay between the open-charm mesons and the  charmonium states has been indicated as a possible solution for the identification of the $X$. 
Many authors computed the effect of virtual or real $D$-mesons loops in shifting the masses of ordinary $c\bar c$ states, for a recent review see~\cite{Barnes:2010gs}. 
Of particular interest for the $X$ case is a study by Kalashnikova~\cite{Kalashnikova:2005ui}, in which a bound state with $J^{PC}=1^{++}$ nearly at threshold is obtained from the interaction of the  $2^3P_1$ charmonium level with $D\bar{D}^{*}$ mesons. For an update on this subject see~\cite{Coito:2010cq}.

In the tetraquark picture, as explained in \cite{Maiani:2004vq,Maiani:2007vr}, the mass of the $1^{++}$ candidate state can be written in terms of the chromo-magnetic coupling and the $[cq]$ diquark mass.


\noindent Taking for the couplings the values obtained in \cite{Maiani:2004vq} and assuming as an input the mass of the $X$ one can deduce the mass of the $[cq]$ diquark: $m_{[cq]}=$1933~MeV. From this value the entire spectrum of $[cq][\bar c\bar q]$ can be computed \cite{Maiani:2004vq}. 
The finer structure of the $X_q$ states has been already discussed and leads to the prediction of two different states, which can be identified, as a first approximation, with the two flavor eigenstates $X_u$ and $X_d$. 
The observation in 2006 of a state decaying to $D^{0}\bar D^{0}\pi$ with mass 3875~MeV favored the assignment: $X_u=X(3875)$, decaying mainly into $J/\psi\pi^+\pi^-$ and $X_d=X(3872)$ decaying into $D^{0}\bar D^{0}\pi$. 
The mass ordering of these two neutral states seems to be reversed, since the $u$ quark is lighter than the $d$ quark and thus one would expect $X_u$ to be lighter than $X_d$. However the quarks which form the diquarks in the $X_u$ have the same electric charge and thus a consistent consideration of the electrostatic energy can perhaps change the order of the masses.
Besides these two neutral states, two charged states arise as a natural prediction of the tetraquark picture $X^+=[cu][\bar c\bar d]$ and $X^-=[cd][\bar c \bar u]$. The lack of any observation of these two states constitutes the main drawback to the tetraquark assignment.

{\it Production and decay}.
The production of a $D^{0}\bar D^{0*}$ molecule necessarily proceeds through the production of a $D$ meson pair. This production mechanism has been extensively studied in $B$ decays by Braaten and collaborators in \cite{Braaten:2004fk}. 
For a bound state at threshold standard quantum mechanics predicts that the wave function in momentum space has a Lorentzian shape. From this information one can extract the elastic scattering amplitude $\mathcal{A}(D^{0}\bar{D}^{0*}\to D^{0}\bar{D}^{0*})$ and the transition amplitude $\mathcal{A}(D^{0}\bar{D}^{0*}\to X)$ as the residue at the pole of the elastic scattering amplitude.

This allows to compute the partial width of $B^{\pm}\to XK^{\pm}$ and the differential distribution $d\Gamma(B^{\pm}\to D^{0}\bar{D}^{0*}K^{+})/dM_{DD^{*}}$ as functions of model dependent parameters, describing the formation process of the $X$ as a coalescence process between $D^{0}$ and $\bar{D}^{0*}$ mesons. The distribution $d\Gamma(B^{\pm}\to D^{0}\bar{D}^{0*}K^{+})/dM_{DD^{*}}$ shows a peak more and more evident near the threshold as the binding energy is decreased.  In this computation the authors make use of an effective field theory description of the interaction between the two $D$-mesons. The effective field theory method has been first introduced in~\cite{Voloshin:2003nt} and then used in~\cite{Braaten:2003he,Fleming:2009kp} to compute some properties of the $X$ meson, relying on the fact that as far as the exchanged momentum between the $D$'s is small, they can be described with elementary fields. In this approach a coupling to the regular charmonium state has also been implemented~\cite{Braaten:2003he}.

In \cite{Braaten:2004ai} the exclusive production of $X$ in B decays is studied. The most relevant result of this paper is the prediction of a deviation from the standard charmonium case of the ratio $\mathcal{R}=\mathcal{B}(B^{0}\to X K^{0})/\mathcal{B}(B^{+}\to X K^{+})\sim 0.1$ (in the charmonium case $\mathcal{R}\sim \tau(B^{0})/\tau(B^{+})\sim~0.9$). The BaBar collaboration has measured \cite{Aubert:2008gu} the two products of branching ratios: $\mathcal{B}(B^{+}\to K^{+}X)\times \mathcal{B}(X\to J/\psi\pi^+\pi^-)$ and $\mathcal{B}(B^{0}\to K^{0}X)\times \mathcal{B}(X\to J/\psi\pi^+\pi^-)$. These measures give $\mathcal{R}=0.6\pm0.2$. Instead for a tetraquark state $\mathcal{R}=0.5\pm 0.3$~\cite{Maiani:2007vr}.

In the molecular picture the decay into $J/\psi+{\rm light}\,\,{\rm hadrons}$ has been accounted for by including the possibility for inelastic channel in the $D^{0}\bar{D}^{0*}$ scattering introducing a complex scattering length \cite{Braaten:2005ai}. The authors find factorization formula for the decay width separating the long distance term (the transition amplitude $X\to D^{0}D^{0*}$) characterized by a typical distance of $a\simeq 8~{\rm fm}$, from the short distance one, acting on distances of about $1/m_{\pi}$. The first term can be computed as before as the residue at the pole of the elastic scattering amplitude and does not contains any unknown parameter, while the second one is related to the coupling of the $X$ to the hadronic channel considered. This is estimated from the wave-function component associated to each hadronic channel and can be borrowed from the potential model results. 

Finally studies of the long-distance $D^0\bar{D}^0\pi^0$ decay mode has been carried out to obtain the relative partial width and the modification of the line shapes of the resonance due to the vicinity to the threshold. 
The study of the line shapes in this channel has been useful to distinguish between a bound state below threshold and a virtual state above threshold.
Indeed the temporary agreement between Belle and BaBar data on the evidence for an higher mass value of the peak in the $D^0\bar{D}^{0*}$ invariant mass spectrum, lead many authors to solve this problem identifying the $X$ with a virtual state.
In~\cite{Hanhart:2007yq} it was performed an analysis on the Belle data~\cite{Gokhroo:2006bt}. This analysis concluded indeed that the resonance observed at $3875~{\rm MeV}$ in the $D^0\bar{D}^{0*}$ channel could be related to the one observed at $3872~{\rm MeV}$ in the $J/\psi\pi\pi$ channel only if the $X$ is taken to be a virtual state. The consequences of this conclusion for the isospin breaking have been considered in~\cite{Voloshin:2007hh}.
In a subsequent analysis~\cite{:2008su}, with higher statistics, the Belle collaboration claimed that the resonance peak in the $D^0\bar{D}^{0*}$ channel appears at $\sim3872~{\rm MeV}$.
In \cite{Stapleton:2009ey} the authors perform a re-analysis of the new data from Belle~\cite{:2008su} together with the ones from BaBar~\cite{Aubert:2007rva} on the $X$ decays to $D^{0}\bar{D}^{0}\pi$. 
There are two main novelties in this work: 
(1) they include a finite width for the $D^{0*}$ constituent;
(2) they consider the fact that the invariant mass of a $D^{0}\pi^{0}$ pair coming from the decay of a $D^{0*}$ in the $X$ is not equal to the mass of a $D^{0*}$ meson 
since the energy of a $D^{0*}$ meson bound inside the molecule differs from its mass by about the $X$ binding energy.
They derive the line shapes of the resonance in the two final states and fit the two data sets separately. They obtain the position of the peak, its width
and some ratio of branching fractions. The fit to the $D^{0}\bar{D}^{0}\pi^{0}$ decay mode returns a real scattering length, which means that the transition
amplitude to the inelastic channels (in this case $J/\psi\pi^{+}\pi^{-}$) is null. From the same fit they obtain that $\mathcal{B}(X\to D^{0}\bar{D}^{0}\pi^{0})/\mathcal{B}(X\to J/\psi\pi^{+}\pi^{-})=0.004$, which instead
implies a strong predominance of the inelastic decay channel ($J/\psi\pi^{+}\pi^{-}$) over the elastic one ($D^{0}\bar{D}^{0}\pi^{0}$). The authors adduce this inconsistency to an improper treatment of the experimental resolution.

The same analysis has been repeated and upgraded in \cite{Kalashnikova:2009gt,Hanhart:2010wh}.
In particular~\cite{Kalashnikova:2009gt} included the $\psi^{'}\gamma$ decay mode. The results are consistent with $X$ being a true bound state.
The line shapes have been studied also in~\cite{Baru:2010ww},  where a general discussion about near to thresholds pole in the {\it S}-matrix is carried on, and in~\cite{Artoisenet:2010va}, where the authors state that the $X$ resonance is the result of a fine-tuning between the regular $2^3P_1$ charmonium state and the $D^0\bar{D}^{0*}$ threshold.

Other decay modes have been considered in the hypothesis of $X$ being a loosely bound $D^0\bar{D}^{0*}$ molecule.
In particular Voloshin and Dubynskiy~\cite{Voloshin:2004mh,Dubynskiy:2007tj} studied the transition $X\to \chi_{cJ}\pi^0$, considering the $c\bar c$ pair inside the $X$ to be in a spin-triplet configuration, since it is favored by even charge conjugation.
In~\cite{Voloshin:2005rt} the $X\to D\bar{D}\gamma$ decay mode is taken into account, since the photon spectrum would provide information on both the long and short distance component of the molecular state.

Concerning the production mechanism in $p\bar p$ collisions, 
as we have already stated in Sec.\ref{subsec:molecules}, the production of a loosely-bound {\it S}-wave molecule must proceeds through the production of $D$-mesons pairs with suitably small relative three-momentum to account for the tiny binding energy. 
While in the $B$-mesons decays the $D$-meson pair is produced practically at rest ($M_B-M_K-M_D-M_{D*}\sim$900~MeV), it seems odd that a such a loosely bound molecule could be produced promptly in a high energy hadron collision environment. 
Indeed one can learn some general features of the $X$ as a molecule without any knowledge on the interaction details. 
The scattering length can be interpreted as the spatial extension of the hadron itself and is estimated to be $\sim8{\rm fm}$, much more than that of an ordinary hadron. Using a minimal uncertainty principle one is able to extract the relative momentum spread of the pair of mesons bound inside the $X$: $\Delta k\simeq {\rm 15~MeV}$. The central value for the relative momentum is of the order of the decay momentum of the $D$-mesons: $k_{0}\simeq {\rm 30~MeV}$.

\begin{figure}[bht]
\begin{center}
\epsfig{file=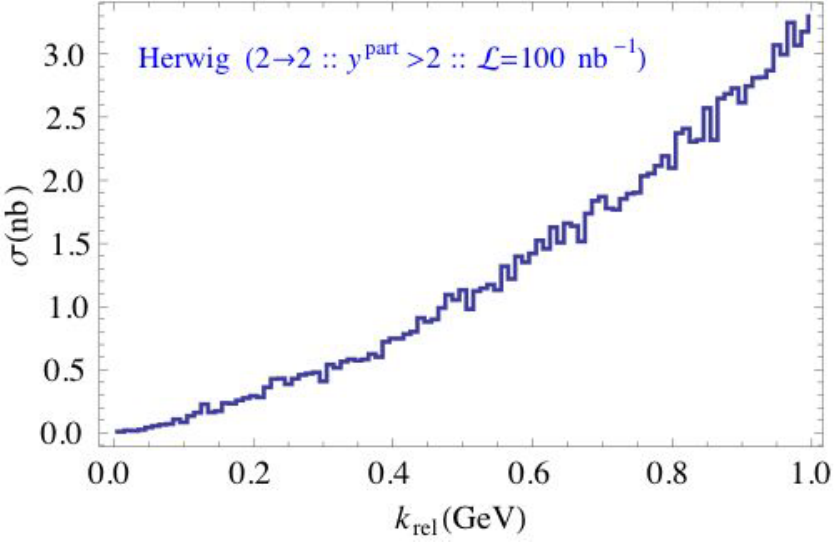,height=4cm}
\epsfig{file=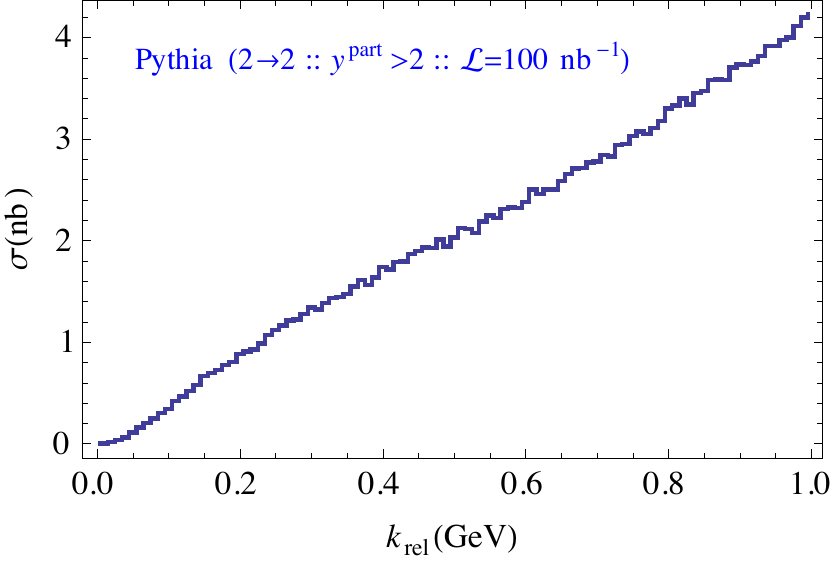,height=4cm}
\caption{Left: the integrated cross section obtained with  Herwig as a function of the relative momentum of the mesons in the $D^{0}\bar{D}^{0*}$ molecule. This plot is obtained after the generation of $55\times 10^{9}$ events with parton cuts $p_{\perp}^{{\rm part}}>2$~GeV and $y^{{\rm part}}>6$. The cuts on the final $D$ mesons are such that the molecule produced has a $p_\perp>5$~GeV and $y<0.6$. Right: same plot but using Pythia.}
\label{fig:xsectX}
\end{center}
\end{figure}

Starting from these considerations \cite{Bignamini:2009sk} computed the maximum theoretical value for the prompt production cross section of the $X$ in $p\bar p$ collisions at the Tevatron. The authors make use of a Schwartz inequality, which allows to write the theoretical prompt production cross section as an integral over the differential cross section for the production of a pair of $D^{0}\bar{D}^{0*}$ mesons with fixed relative three-momentum. The integral extends up to $k_0+\Delta k\sim$50~MeV. The differential distribution for the cross section can be computed using standard MC event generators such as Herwig and Pythia. The results are shown in Fig.~\ref{fig:xsectX}. The maximum integrated cross section amounts to $\sigma^{{\rm th}}_{{\rm max}}\simeq{\rm 0.085~nb}$ for both the MC event generators.


This estimate has been compared with a lower limit for the experimental cross section obtained by the CDF collaboration properly taking into account the rapidity and transverse momentum cuts:
the CDF collaboration recently performed an analysis \cite{Abulencia:2006ma,cdfnote,Aaltonen:2009dm,Abe:1997jz} to distinguish the fraction of $X$ and $\psi(2S)$ produced promptly from that originated from $B$-decays and found that only the $(16\pm 4.9({\rm stat.})\pm 2.0 ({\rm syst.}))\%$ of the $X$ produced in $p\bar p$ collisions comes from $B$ decays. Using the corresponding $\psi(2S)$ result one can obtain an estimate of the product $\sigma^{{\rm exp}}_{{\rm prompt}}(p\bar p \to X+{\rm all})\times \mathcal{B}(X\to J/\psi \pi^+\pi^-)=(3.1\pm 0.7){\rm nb}$. From the results on $\mathcal{B}(X\to J/\psi\pi^+\pi^-)$, shown in Sec. \ref{sec:x3872} one can conclude that $\sigma^{{\rm exp}}(p\bar p \to all)\leq 33{\rm nb}$. Thus the upper bound on the theoretical cross section is $\sim$300 times smaller than the lower limit on the experimental cross section, putting in trouble the molecular interpretation of the $X$.

Nevertheless, in a subsequent paper by Braaten and Artoisenet \cite{Artoisenet:2009wk} the effect of final state interactions (FSI) between the $D^0$ and $\bar{D}^{0*}$ mesons has been taken into account to reconcile the experimental and theoretical estimates. According to the authors of \cite{Artoisenet:2009wk} FSI should be able to rescatter higher relative momentum pairs to lower relative momentum pairs, thus allowing to consider $k_{max}\sim 2m_\pi$ and thus to integrate a larger cross-section. Furthermore they introduce an enhancement factor in the cross-section $\sim 1/(k^2+a^2)$, as stated in the Migdal-Watson theorem. In this way they succeed in reconciling the theoretical and experimental values.

This result is strictly related to the use of the Migdal-Watson theorem, which formalizes the FSI mechanism. There are two main conditions that have to be met in order to safely apply the theorem. First, the relative momentum between the two rescattering particles has to be smaller than the inverse of the range of strong interactions, namely $200~{\rm MeV}$. Second, it is necessary that no more than two hadrons match the first condition. Reference \cite{Bignamini:2009fn} showed that the latter is not satisfied. Furthermore going up to relative momentum of $\simeq 2m_{\pi}$ one is no allowed to ignore higher partial wave scattering beside the {\it S}-wave one. However if FSI works in the direction of an enhancement \cite{Bignamini:2009fn} predicts that if any $X_{s}=D_{s}\bar{D}^{*}_{s}$ exists it should have an observable cross section at the Tevatron, namely $\sigma^{{\rm th}}_{\rm max}(p\bar p\to X_s +{\rm all})\sim (2\pm 1){\rm nb}$. It could decay to $J/\psi K K$. This can be used as a test table for our understanding of FSI and of meson-meson interaction at the same time.

Finally another possible production mechanism for the molecular $X$ has been proposed in~\cite{Dubynskiy:2006cj,Voloshin:2006pz} to be $e^+e^-\to \gamma X$.

As for the charmonium assignment one could refer to \cite{Braaten:1996dn} for a complete review
of charmonium production in high energy collisions. Concerning the $J^{PC}=2^{-+}$ possibility a study of gluon fragmentation to $D$-wave charmonia is contained in \cite{Cho:1994qp}.

In these final paragraphs we discuss some new considerations about the $D^{0}\bar{D}^{0}\pi^{0}$ partial decay width and the radiative decay modes.
We think that these two arguments may offer an additional way to test the supposed molecular nature of the $X$.


From the analysis described in Sec. \ref{sec:x3872}, which summarizes all the experimental informations available up to now, we are able to extract a probability density function (PDF) of the partial width of the $X\to D^{0}D^{0*}\to D^{0}\bar{D}^{0}\pi$ (see Fig. \ref{fig:xlike}) and consequently estimate the probability $\Gamma(D^{0}\bar{D}^{0*})<$100~KeV to be $\sim$ 0.9\%.



This result can put in some trouble the molecular interpretation of the $X$. Indeed, due its small binding energy one would expect the $D^{0*}$ and $\bar{D}^{0*}$ mesons bound inside the $X$ to decay as if they were free, that is with the decay widths measured for free mesons. In the following we will refer to the decay $D^{0*}\to D^{0}\pi$, but the same considerations are valid for the charge conjugate mode. We will discuss in the end the effect of interference between the charge conjugate modes.

\begin{figure}[bht]
\begin{center}
\epsfig{file=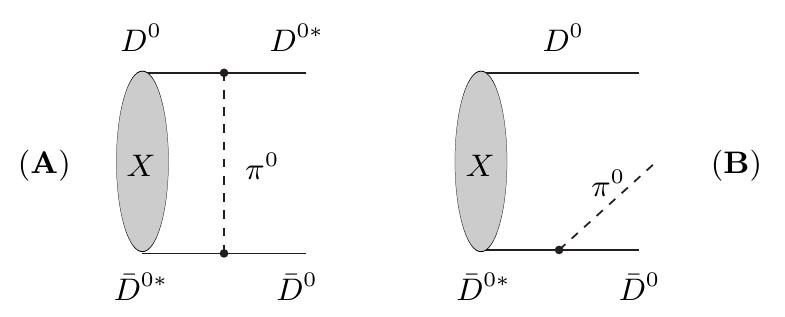,height=3.5cm}
\end{center}
\caption{Feynman diagrams for one pion exchange between the two mesons bound in the molecule (A) and for the decay mode $X\to D^{0}\bar{D}^{0}\pi$ (B).}
\label{fig:diagrams}
\end{figure}

The measured partial width for $D^{0*}\to D^{0}\pi$ is $\Gamma(D^{0*}\to D^{0}\pi)=70~{\rm KeV}$. Nevertheless when considering this decay mode for a meson bound with a $D^0$ meson to form the $X$ we find ourselves in a rather peculiar situation: the process $D^{0*}\to D^{0}\pi$ is the same process which is responsible for the binding of the molecule. It has been claimed by many authors that the one pion exchange potential is sufficiently attractive in this channel to admit a bound state exactly at threshold.
Thus it can occur that a would-be-real-pion is reabsorbed by the other meson keeping the molecule bound and making it living longer than its constituents. The relevant diagrams are reported in Fig.\ref{fig:diagrams}.

The effect of this reabsorption process can thus influence the partial width of the state in this decay mode. 
To obtain a qualitative estimate one can compare the average formation time $T_f$ of the would-be-real-pion with the average time which elapses between the exchange of two subsequent pions $T_{coll}$ between the mesons. If $T_f > T_{coll}$ there is some probability that the pion is reabsorbed in the pion clouds which surrounds the $D^0$ meson, otherwise on average the reabsorption will not occur. This semiclassical analysis, albeit naive, is the only model independent possibility.

The avreage formation time is proportional to the perpendicular component of the pion four momentum $k_{\perp}$ in the rest frame
of the $X$. From the two body decay kinematics we know that $k_{\perp}<k_{*}$, the decay momentum in the $D^{0*}$ rest frame, thus:

\begin{equation}
T_f \sim \frac{1}{k_{*}}
\end{equation}

\noindent Since the $D^{0*}$ is not on shell $k_{*}=\sqrt{\lambda(q^2,m^{2}_{D},m^{2}_{\pi})}/2\sqrt{q^2}$. The {\it off-shellness} $q^2$ of the $\bar{D}^{0*}$ meson is due to the exchange of pions with the $D^{0}$ inside the molecule. However, as the binding energy is extremely small the pions exchanged are soft, in the sense that a hard pion would break the molecule. 




In the soft-pion regime one finds that:

\begin{equation}
q^2=(p-k)^2=\frac{M^2_{{\small X}}}{2}-m^2_D+2\left((k^0)^2-(k^3)^{2}-{\bf k}^{2}_{\perp}\right)
\end{equation}

\noindent where $p$ is the four momentum of the two $D$-mesons. Furthermore we can neglect the temporal and longitudinal component 
of the pion four momentum:

\begin{equation}
q^2\sim\frac{M^2_{{\small X}}}{2}-m^2_D-2{\bf k}^{2}_{\perp} \Rightarrow q^2_{min} \lesssim q^2 \lesssim  q^2_{max}
\end{equation}

\noindent where $q^2_{min}\equiv M^2_{{\small X}}/2-m^2_D-2(k^{max}_{\perp})^2$ and $q^2_{max}\equiv M^2_{{\small X}}/2-m^2_D$.
When one takes into account the fact that  $T_f$ is a decreasing function of $q^2$ in the region of interest ($q^2>(m_D+m_{\pi})^2$) 
a range for the average formation time can be computed:

\begin{equation}
1.5 \times 10^{-23} \, \sec\, \lesssim T_f\lesssim\, 1.8 \times 10^{-23} \, \sec 
\end{equation}

\noindent which is of the same order of magnitude of the typical time of strong interactions.\\
\noindent We now give an estimate of $T_{coll}$ in a semiclassical approach. We describe the $D^0\bar{D}^{0*}$ {\it S}-wave molecule
as two point-like particles which move in a one dimensional box of length $\sim~10{\rm fm}$, the size of the molecule.
The two particles exchange pions, {\it i.e.} interact, when their distance is smaller than $1~{\rm 1fm}$, the range of strong interactions.
To account for the quantum nature of the system we implement the uncertainty principle at each step, choosing the relative momentum
from a Gaussian with mean value equal to the decay momentum $k_*\simeq$ 40~MeV and spread $\Delta k\sim 1/R$, where $R$ is the relative distance.
The initial position of the particle is chosen from a Gaussian with zero mean value and $\Delta R=10~{\rm fm}$.
We put a cutoff on the relative momentum $\Lambda\sim 200/300~{\rm MeV}$.
The time evolution last for the average life-time of the $X$ which is $\tau = \hbar/\Gamma(X) \sim 2.2\times 10^{-22}\sec$. 
The evolution is repeated for several initial configurations. The outcome is stable against changes of the time step length and of the momentum cutoff 
and gives $T_{coll}\sim 10^{-22}\sec$. This means that $T_{f}<T_{coll}$ and thus one expects that $\Gamma(X\to D^0\bar{D}^{0}\pi^0)\simeq \Gamma(D^{0*}\to D^0\pi^0)$.
Any deviation from this prediction would seem unjustified in the molecular picture.

We finally consider radiative decays. To describe the decays of the $X$ in the tetraquark picture we need to introduce three amplitudes. 
The decay of a diquark-antidiquark bound state into a pair of mesons can occur through the exchange of a quark and an antiquark belonging respectively to the diquark and the antidiquark. 
There are indeed three different flavor configurations:  the exchange of two light quarks $\mathcal{A}\left([cq][\bar c\bar q]\to[c\bar q][\bar{c} q]\right)\equiv A_1$, the exchange of two heavy quarks $\mathcal{A}\left([cq][\bar c\bar q]\to[\bar{c} q][c \bar{q} ]\right)\equiv A_2$ and the exchange of a light quark and a heavy quark $\mathcal{A}\left([cq][\bar c\bar q]\to[q\bar{q}][c \bar{c} ]\right)\equiv A_3$. $A_1$ and $A_2$ account for $X\to D^{0}\bar{D}^{0*}$, while $A_3$ accounts for $X\to J/\psi \pi\pi$. 

\begin{figure}[bht]
\begin{center}
\epsfig{file=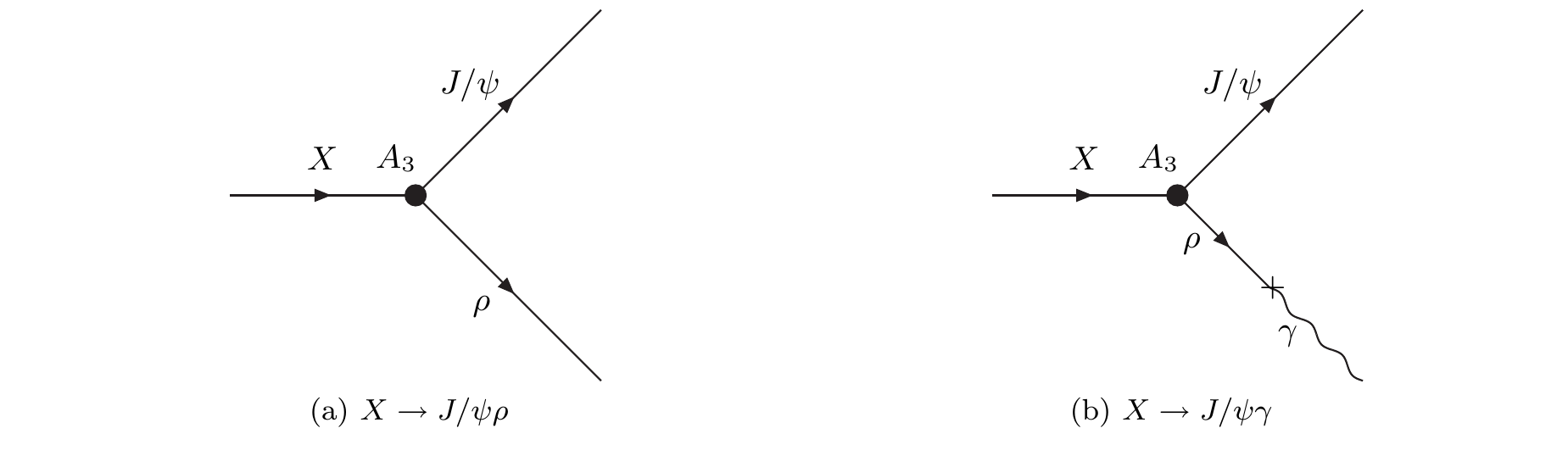,height=4.0cm}
\caption{Radiative and hadronic decay of the $X(3872)$ described with the same contact vertex $A_3$. The radiative decay proceeds indeed through the hadronic transition $X\to J/\psi \rho$.}
\label{fig:radiative}
\end{center}
\end{figure}

It has been experimentally confirmed that the pion pair in the decay into $J/\psi\pi^+\pi^-$ comes mainly from a $\rho$. Thus we can write the partial width for this decay as:

\begin{equation}
\begin{split}
\Gamma(X\to J/\psi\pi\pi)&=\frac{1}{8\pi}\frac{m^3_{\rho}}{M_{X}}|\mathcal{A}_{3}|^2\int\;ds\;\frac{1}{s}\frac{\lambda^{1/2}\left(s,m^{2}_{\pi},m^{2}_{\pi}\right)}{\lambda^{1/2}\left(m^2_{\rho},m^{2}_{\pi},m^{2}_{\pi}\right)}\left[\frac{1}{\pi}\frac{m_{\rho}\Gamma_{\rho}}{(s-m_{\rho}^2)^2+(m_\rho \Gamma_\rho)^2}\right]\\
&\left[\frac{|\vec{p}_{\rho}|}{\sqrt{m^{2}_{\rho}+|\vec{p}_{\rho}|^2}\left(\sqrt{s+|\vec{p}_{\rho}|^2}+\sqrt{m^{2}_{J/\psi}+|\vec{p}_{\rho}|^2}\right)}\right]_{|\vec{p}_{\rho}|=\lambda^{1/2}\left(M^{2}_{X},s,m^{2}_{J/\psi}\right)/2M_{X}}
\end{split}
\end{equation}

\noindent Since the amplitude for the radiative decay $X\to J/\psi \gamma$ proceeds through the annihilation of a pair of light quarks into a photon, the hadronic part of the amplitude is the same as in the decay $X\to J/\psi \pi\pi$, Fig.\ref{fig:radiative}. Exploiting the vector meson dominance one can write:

\begin{equation}
\langle J/\psi \gamma|X \rangle=\langle\gamma|\rho\rangle\frac{1}{m^{2}_{\rho}}\langle J/\psi \rho|X \rangle=\frac{f_{\rho}}{m^2_{\rho}}\mathcal{A}_{3}
\end{equation}

\noindent The partial decay width can thus be written as:

\begin{equation}
\Gamma(X\to J/\psi\gamma)=2|\mathcal{A}_{3}|^2 \left(\frac{f_{\rho}}{m_{\rho}^2}\right)^2\frac{1}{8\pi\; M^2_{X}} \frac{\sqrt{\lambda(M^{2}_{X}, m^{2}_{\psi}, 0)}}{2M_{X}}
\end{equation}

\noindent Using $f_{\rho}=0.152~\rm{GeV}^2$ \cite{Polosa:2000ym} one obtains $\Gamma(X\to J/\psi\gamma)/\Gamma(X\to J/\psi\pi\pi)\sim0.84$, which is in rather a good agreement with the experimental value reported in Tab.~\ref{tab:xdecays}: $\mathcal{B}(X\to J/\psi \gamma)/\mathcal{B}(X\to J/\pi\pi)\sim 0.3 \pm 0.1$.

We can further exploit the result obtained for the width of $X\to J/\psi\gamma$ in order to give an estimate of the decay width into $J/\psi\gamma\gamma$.
We can indeed compute the transition matrix element in terms of $A_3$ exploiting the coupling of the $J/\psi$ to the photon:

\begin{equation}
\langle \gamma \gamma|X \rangle=\langle\gamma|J/\psi\rangle\frac{1}{m^{2}_{J/\psi}}\langle J/\psi \gamma|X \rangle=\frac{f_{J/\psi}}{m^2_{J/\psi}}\langle J/\psi \gamma|X \rangle=\frac{f_{J/\psi}}{m^2_{J/\psi}}\frac{f_{\rho}}{m^2_{\rho}}\mathcal{A}_{3}
\end{equation}

\noindent The partial decay width can thus be written as:

\begin{equation}
\label{2gamma}
\Gamma(X\to \gamma\gamma)=\frac{4}{3}|\mathcal{A}_{3}|^2 \left(\frac{f_{J/\psi}}{m^2_{J/\psi}}\right)^2\left(\frac{f_{\rho}}{m_{\rho}^2}\right)^2\frac{1}{8\pi\; M^2_{X}} \frac{\sqrt{\lambda(M^{2}_{X}, 0, 0)}}{2M_{X}}
\end{equation}

\noindent Using $f_{J/\psi}=1.254~{\rm GeV}^2$ \cite{Deandrea:2003pv} one obtains $\Gamma(X\to \gamma\gamma)/\Gamma(X\to J/\psi\pi\pi)\sim0.02$, which has to be compared with the experimental value in Tab.~\ref{tab:xdecays}:  $\mathcal{B}(X\to\gamma \gamma)/\mathcal{B}(X\to J/\pi\pi)<0.01$. The inconsistence of the theoretical prediction with respect to data is not dramatic if one takes into account the very strong assumptions made to derive Eq.~(\ref{2gamma}).


\subsubsection{The 3940 family and other $C=+$ states}

\label{sec:charmon_interp_3940}

\underbar{$X(3940)$ and $X(4160)$}\\
The possible assignment of $X(3940)$ with the charmonium $\eta_c(3S)$ 
with quantum numbers $J^{PC}= 0^{-+}$ would be supported by the following facts: 
\begin{itemize} 
\item the non observation of the $D \bar D$ decay,
\item the lower states $\eta_c(1S)$ and $\eta_c(2S)$ are also produced 
through double charmonium production,
\item the predicted width for a $3^1S_0$ state with a mass of $3943$~MeV 
is $\sim 50$~MeV~\cite{Eichten:2005ga}, 
in reasonable agreement sith the observed width.
\end{itemize}
However the predicted mass is below the potential model estimate of 
$\sim 4050$~\cite{bgs2005}. Moreover the observation of the $X(4160)$, 
compatible with $J^{PC} = 0^{-+}$ and at the same distance from the predicted 
mass of $3^1S_0$ as the $X(3940)$, makes the picture more complicated. In fact 
the next radial charmonium excitation $\eta_c(4S)$ is predicted to be close 
to $4400$~MeV, so that it seems unlikely to accomodate both $X(3940$ and 
$X(4160)$ as charmonium states. Additional important tests would be the study 
of the angular distribution of the $D \bar D^*$ final state and the 
possible observation in $\gamma \gamma \to D \bar D^*$~\cite{Godfrey:2008nc}. 

Another interpretation which can not be excluded apriori is that of the 
$X(3940)$ as a $0^{-+}$ hybrid charmonium, which can not decay 
to $D \bar D$~\cite{Close:2007ny}. Spin dependent splittings 
place two states, $0^{-+}$ and $1^{-+}$, below the vector $1^{--}$ with 
equal mass gaps of the order of 
100~MeV~\cite{Barnes:1982zs,Barnes:1982tx,Chanowitz:83,Merlin:87}. 
If the $1^{--}$
is identified with the $Y(4260)$, then it is not excluded that 
$X(3940)$ is its lighter $0^{-+}$ partner. 
\\
\underbar{$Y(3940)$ and $Y(4140)$}\\
The molecular nature of the $Y(3940)$ has been considered as a possibility
by a number of authors~\cite{Liu:2009ei,Branz:2009yt,Zhang:2009vs,Branz:2010qw}. 
According to this picture, the hadronic wave function would be 
$\frac{1}{\sqrt{2}}(|D^{*+} D^{*-}> + |D^{*0} \bar{D^{*0}}>)$, with 
$J^{\rm PC} = 0^{++}$ or $2^{++}$. In this scenario the $Y(3940)$ would 
have a molecular partner, the $Y(4140)$, with composition 
$D_s^{*}\bar{D}_s^{*}$, quantum numbers     
$J^{PC}=0^{++}$ or $2^{++}$ and a similar binding energy of about 
80 MeV \cite{Albuquerque:2009ak,Ding:2009vd,Zhang:2009st}. 
This idea is supported by the fact that the mass difference  
between these two mesons is approximately the same as the mass difference       
between the $\phi$ and $\omega$ mesons: $m_{Y(4140)}-m_{Y(3940)}\sim m_\phi-    
m_\omega\sim 210$~MeV. However, with a meson exchange mechanism to 
bind the two charmed     
mesons, it seems natural to expect a more deeply bound system in the case that  
pions can be exchanged between the two charmed mesons, as in the $D^*D^*$,      
than when only $\eta$ and $\phi$ mesons can be exchanged, as in the $D_s^*      
D_s^*$ system~\cite{Nielsen:2009uh}. The molecule picture predicts 
that decays proceed via rescattering with decays to hidden and open charm 
states equally probable, so that decays to $D \bar D$ and $D \bar D^*$ 
are foreseen. Another prediction of the molecular hypothesis is that 
the constituent mesons can decay independently, leading to 
$D_s^{*+} D_s^- \gamma$ and $D_s^{*-} D_s^+ \gamma$~\cite{Liu:2009ei,Branz:2008ha,Liu:2009pu}. 
Predictions for the radiative decays 
$Y(3940)/Y(4140) \to \gamma \gamma$ in 
Ref.~\cite{Branz:2010qw,Branz:2009} yield similar results for the 
$J^{\rm PC}$ $0^{++}$ and $2^{++}$ assignments. 
In addition, a $D^{*+} D_s^{*-}$ molecule is predicted with mass of about 
$4040$~MeV, decaying to $J/\psi \rho$~\cite{Mahajan:2009pj}. 

The conventional $c \bar c$ assignment of the $Y(4140)$ is very unlikely 
since, being its mass above open charm threshold, it should have a large 
width. It was also shown in \cite{Liu:2009iw} that the $Y(4140)$ 
probably can not be the second radial excitation 
of any of the $P$-wave charmonium states: $\chi_{cJ}^{''}~(J=0,~1)$. 
If it were the case, the branching ratio of the hidden charm decay,       
$Y(4140)\to J/\psi\phi$, would be much smaller than                             
the experimental observation \cite{Eichten:2008}.

The $[c s][\bar c \bar s]$ tetraquark hypothesis for 
the $Y(4140)$ has been investigated by several 
authors~\cite{Mahajan:2009pj,Stancu:2009ka,Liu:2009ei}. However, 
in this scenario, it would decay with similar widths, of the order 
of 100~MeV, to hidden and open charm final states. 

With the present experimental information, also other explanations 
for the $Y(4140)$ can not be excluded, such as hybrid 
charmonium~\cite{Rosner:2007mu,Mahajan:2009pj,Wang:2009uh} 
(its mass lyes in the range predicted by QCD for hybrids) or 
rescattering of $D_s D_s^*$~\cite{vanBeveren-Rupp-1} or the opening up of a new 
final state channel~\cite{vanBeveren-Rupp-2}. 
Ref. \cite{Molina-Oset} performed a study of the vector 
vector interaction in the 
framework of the hidden gauge formalism, finding 
three resonances with poles close to the 
masses of $Y(3940)$, $Z(3930)$ and $X(4160)$ and 
quantum numbers $J^{PC} = 0^{++}$, $2^{++}$ and $2^{++}$ 
respectively. In Ref. \cite{Liang-Molina-Oset} their 
radiative open charm decays have been investigated. 
\\
\underbar{$X(4350)$}\\
The possible quantum numbers for a state decaying into $J/\psi \phi$ 
are $J^{PC}= 0^{++}$, $1^{-+}$ and $2^{++}$, with the second possibility 
being exotic. Even with the modest information available for this state, 
different interpretations already appeared in the literature: 
a $[c s][\bar c \bar s]$ tetraquark with $J^{PC}= 2^{++}$~\cite{Stancu:2009ka}, 
a molecular $D_s^{*+} D_{s0}^{*-}$ state~\cite{Zhang:2009,Albuquerque:2009}, 
an excited 
$P$ wave charmonium state $\Xi_{c2}''$~\cite{Liu:2009} and a 
mixed charmonium-$D_s^* D_s^*$ state~\cite{Wang:2009}. 
In Ref.~\cite{Albuquerque:2010} 
a QCD sum rules study has been performed to test whether the 
$X(4350)$ can be an exotic $J^{PC}= 1^{-+}$ $D_s^* D_{s0}^*$ 
(or $D^* D_{0}^*$) molecular state. The mass value obtained is 
$5.5 \pm 0.19$~GeV ($4.92 \pm 0.08$), thus inconsistent with the 
experimental mass value. 
\\
\underbar{$Z(3930)$}\\
The measured properties of $Z(3930)$ 
are consistent with expectations for the previously unseen 
$2^3P_2$ charmonium state, $\chi_{c2}'$~\cite{Eichten:2005ga,Barnes:2003vb}, for 
which the predicted mass and width are $3972$~MeV and $28.6$~MeV (assuming 
the observed mass), respectively~\cite{Swanson:2005tq,bgs2005,Eichten:2005ga}. 
Also the measured two-photon production rate is consistent with the one of 
the $\chi_{c2}'$~\cite{Barnes1992}. Additional confirmations of this 
interpretation would come from the observation of the 
$Z(3930) \to D {\bar D^{*}}$ decay with a branching ratio $\sim 25$\% 
and the from the radiative decay $Z(3930) \to \psi(2S) \gamma$ with 
a partial width of the order of $100$~KeV~\cite{bgs2005,Eichten:2005ga,Godfrey:2008nc}. 

\subsubsection{The $1^{--}$ family}
\label{sec:charmon_interp_1mm}
The three $1^{--}$ states, $Y(4260)$, $Y(4360)$ and $Y(4660)$ are characterized 
by large total widths and by charmonium decay modes. 
Since their masses are higher than the $D^{(*)}\bar{D}^{(*)}$
threshold, if they were $1^{--}$ charmonium states
they should decay mainly to $D^{(*)}\bar{D}^{(*)}$. However, the
observed $Y$ states do not match the peaks in $e^+e^-\to D^{(*)\pm}D^{(*)
\mp}$ cross sections. Furthermore, the first available charmonium 
state is $\Psi(3D)$, for which the quark models predict a mass 
of about $4500$~MeV, which does not fit any of the new states. 
Therefore, the masses and widths of these three new $Y$ states seem 
to be inconsistent with any of the $1^{--}$ $c\bar{c}$ 
states~\cite{kz,zhu,seth}. 
An interesting interpretation is that the $Y(4260)$ is a charmonium 
hybrid~\cite{closey,zhuy1,kou-pene}. 
Its mass lies infact in the ball park predicted by lattice 
QCD\cite{lattice} and flux tube model. Actually, 
recent lattice simulations \cite{latticenew}  and QCD string models 
calculations \cite{kala} 
predict that the lightest charmonium hybrid has a mass of about 4400 MeV, 
which is 
closer to the mass of the $Y(4360)$, thus not excluding the interpretation 
of the latter as an hybrid. A prediction of the hybrid hypothesis
is that the dominant open charm decay mode would be a meson pair with one
$S$-wave $D$ meson $(D,~D^*,~D_s,~D_s^*)$ and one $P$-wave $D$ meson $(D_1,~
D_{s1}$) \cite{kokoski,closey,Close:1994hc}. 
In the case of the $Y(4260)$ this suggests dominance
of the decay mode $D\bar{D}_1$. Therefore, a large $D\bar{D}_1$ signal 
could be understood as a strong evidence in favor of the hybrid 
interpretation for the $Y(4260)$. Up to now the Belle experiment 
did not find evidence for such a 
signal~\cite{Pakhlova-hybrid-1,Pakhlova-hybrid-2}.
In the case of the $Y(4360)$ and $Y(4660)$,
since their masses are well above the $D\bar{D}_1$ threshold, their decay rates
into $D\bar{D}_1$ should be very large if they were charmonium hybrids. 
Another prediction of the hybrid scenario is the existence of partner 
states: the flux tube model predicts a multiplet of states nearby in 
mass with conventional quantum numbers ($0^{-+}$, $1^{+-}$, $2^{-+}$, 
$1^{++}$, $1^{--}$) and states with exotic quantum numbers ($0^{+-}$, 
$1^{-+}$, $2^{+-}$)~\cite{godfrey}. 

In Ref.~\cite{Close:2009ag} a possible interpretation of the 
$Y(4260)$ and $Y(4360)$ as $D_1 D^*$ molecular states has been proposed: 
two $S-$wave mesons could be bound via pion exchange. A distinctive 
decay channel would be $D \bar D 3 \pi$. 
Further molecular interpretations of the  $Y(4260)$ bound by meson exchange 
has been considered in Refs.~\cite{Ding:2008gr,ywm}.

As already stressed, a 
crucial information for understanding the structure of these states is 
whether the pion pair comes from a resonance state. From the di-pion 
invariant mass spectra, there is some 
indication that only the $Y(4660)$ has a well defined intermediate state 
consistent with $f_0(980)$. Due to this fact and the 
proximity of the mass of the $\psi'-f_0(980)$ system with the mass of the 
$Y(4660)$ state, in Ref.~\cite{Guo:2008zg},
the $Y(4660)$ was considered as a $f_0(980)~\psi'$ bound state. 
If this interpretation of the $Y(4660)$ is correct, heavy quark spin symmetry
implies that there should be a $\eta^\prime_c-f_0(980)$ bound state 
\cite{ghm2}. This state would decay mainly into $\eta^\prime_c\pi\pi$, and  
the authors of Ref.~\cite{ghm2} predicted the mass of such a state to be
$4616^{+5}_{-6}$ MeV. The enhancement at $M = 4630$~MeV 
in the $\Lambda_c^+ \Lambda_c^-$ distribution has been suggested 
to be a FSI effect of the 
$Y(4660) \to \Lambda_c^+ \Lambda_c^-$ \cite{Guo:2010tk}.  
The $Y(4660)$ was also suggested to be a baryonium state \cite{qiao}, 
a canonical  5 $^3$S$_1$ $c\bar{c}$ state \cite{dzy}, 
and a tetraquark with a $[cs]$-scalar-diquark and a 
$[\bar{c}\bar{s}]$-scalar-antidiquark in a $2P$-wave state 
\cite{efg}.

In the case of $Y(4260)$, in Ref.~\cite{Maiani:2005pe} it was considered as a 
$[cs]$-scalar-diquark $[\bar{s}\bar{c}]$-scalar-antidiquark in a $P$-wave state. 
Studying the uncertainty in the determination of the orbital 
term, a mass of $M=(4330\pm70\ {\rm MeV})$ was estimated, in nice 
agreement with the mass of $Y(4260)$ 
but also consistent with the mass of $Y(4360)$. 
However, from the $\pi\pi$ mass 
distribution, none of these two states,  
$Y(4260)$ and $Y(4360)$ has a decay
with an intermediate state consistent with $f_0(980)$ and, therefore, it is 
not clear that they should have an $s\bar{s}$ pair in their structure.
Besides, in Ref.~\cite{efg} the authors estimate that the mass of a
$[cs]_{S=0}[\bar{c}\bar{s}]_{S=0}$ tetraquark in  a $P$-wave state
would be 200 MeV higher than the $Y(4260)$ mass. The authors of 
Ref.~\cite{efg} found that a more natural interpretation for the $Y(4260)$
would be a $[cq]_{S=0}[\bar{c}\bar{q}]_{S=0}$ tetraquark in  a $P$-wave state.
Other interpretations for the $Y(4260)$ appeared in the literature: 
a baryonium $\Lambda_c-\bar{\Lambda}_c$ 
state \cite{qiao2}; an $S$-wave threshold effect \cite{Rosner:2006vc}; a resonance
due to the interaction between the three, $J/\psi\pi\pi$ and $J/\psi K\bar{
K}$, mesons \cite{MartinezTorres:2009xb}; 
a 4$S$ charmonium state \cite{estra};  
an $S$-wave molecule 
$\rho^0 \chi_{c1}$ \cite{Liu:2005} 
or $\omega \chi_{c1}$ \cite{Yuan:2005dr}; an enhancement (not a true resonance) 
connected with the opening of the $D_s^* D_s^*$ threshold and the coupling to 
the $J/\psi f_0(980)$ and $J/\psi \sigma(600)$ 
channels \cite{vanBeveren:2009fb,vanBeveren:2009jk,vanBeveren:2010mg}
The three $Y$ states were also interpreted as 
non-resonant manifestations of the Regge zeros \cite{eefy}.

\subsubsection{Charged States}
\label{sec:charmon_interp_charged}
As already stressed, the real turning point
in the discussion about the structure of the new observed charmonium states
was the observation by Belle Collaboration of a charged state decaying into
$\psi'\pi^+$, produced in $B^+\to K\psi'\pi^+$ \cite{:2007wg}.
Since the minimal quark content of this state is $c\bar{c}u\bar{d}$, this
state is a prime candidate for a multiquark meson. \\

\underbar{$Z^+(4430)$}\\

There are many theoretical interpretations for the $Z^+(4430)$ structure. 
Since its mass is close to the $D^*D_1$ threshold, Rosner \cite{Rosner:2007mu}
suggested that it is an $S$-wave threshold effect, while others considered it
to be a strong candidate for a $D^*D_1$ molecular state
\cite{Meng:2007fu,lmnn,Liu:2007bf,Liu:2008xz,Ding:2008mp,blnrs}.

Considering the  $Z^+(4430)$ as a loosely bound $S$-wave $D^*D_1$ molecular
state, the allowed angular momentum and parity are $J^P=0^-,~1^-,~2^-$,
although the $2^-$  assignment is probably suppressed in the $B^+\to Z^+K$
decay by the small phase space. Among the remaining possible $0^-$ and $1^-$
states, the former will be more stable as the latter can also decay to
$DD_1$ in $S$-wave. Moreover, one expects a bigger mass for the $J^P=1^-$
state as compared to a $J^P=0^-$ state. The molecule explanation predicts 
that the $Z^+(4430)$ decays into 
$D^* \bar D^* \pi$~\cite{Meng:2007fu} through the decay of 
its consituents and $\psi(2S) \pi$ via 
rescattering~\cite{godfrey}. 

There is also a quenched lattice QCD calculation that finds attractive
interaction for the $D^*D_1$ system in the $J^P=0^-$ channel \cite{laz}.
The authors of Ref.~\cite{laz} also find positive scattering lenght. Based
on these findings, they conclude that althoug the interaction between the two 
charmed mesons is attractive in this channel, it is unlikely that they can 
form a genuine bound state right below the threshold. 

Other possible interpretations are tetraquark 
state \cite{Maiani:2005pe,glp,Bracco:2008jj,Branz:2010sh}, a cusp
in the $D^*D_1$ channel \cite{bugg},  a baryonium state \cite{qiao}, a 
radially excited $c\bar{s}$ state \cite{mms}, 
or a hadro-charmonium state \cite{volre}.
The tetraquark hypothesis implies that
the $Z^+(4430)$ will have neutral partners decaying into $\psi'\pi^0/\eta$ 
or $\eta_c(2S) \rho^0/\omega$. According to the tetraquark picture 
the decay channels should be $D \bar D^*$, $D^* \bar D^*$, $J/\psi \pi$, 
$J/\psi \rho$, $\eta_c \rho$ and $\psi(2S) \pi$, but not $D \bar D$ 
because of its spin-parity~\cite{Ding:2007ar}
Further discussions about its production and decay can be found 
in Refs.~\cite{Rosner:2007mu,cz1,cz2,cz3} and \cite{biz}, respectively.
The tetraquark model also predicts a second nearby state with mass 
$\sim 4340$~MeV, decaying into $\psi' \pi^+$~\cite{Maiani:2007}.\\

\underbar{$Z_1^+(4050)$ {\sl and} $Z_2^+(4250)$}\\

Due to the                                                                      
closeness of the $Z_1^+(4050)$ and $Z_2^+(4250)$ masses to the                  
$D^{*}\bar{D}^*(4020)$ and $D_1\bar{D}(4285)$ thresholds, these states could    
be interpreted as molecular states or threshold effects. However, 
since the mass of $Z_1^+(4050)$ is above $D^{*}\bar{D}^*$ threshold, 
the molecular interpretation is disfavoured, even if studies 
present in the literature give contradictory results: in 
Ref.~\cite{Liu:2005ay}, using a meson exchange model, strong attraction 
for the $D^{*}\bar{D}^*$ system with $J^{P}=0^+$ is found, while 
using a boson exchange model,   
the author of Ref.~\cite{Ding:2009zq} concluded that the interpretation of            
$Z_1^+(4050)$ as a $D^{*}\bar{D}^*$ molecule is not favored. 
In the case of $Z_2^+(4250)$, using a meson     
exchange model, it was shown in Ref.~\cite{Ding:2008gr} that its interpretation as a   
$D_1\bar{D}$ or $D_0\bar{D}^*$ molecule is disfavored.

\subsection{Summary and outlook}


\label{sec:charmon_summary}

A large number of new Charmonium states have been measured and the knowledge of their characteristics has significantly improved. Fig.~\ref{fig:charmoniumSummary} shows the observed states overlaid on the regular charmonium states as predicted by the potential models~\cite{Brambilla:2004wf}.
The most likely $J^{CP}$ assignment is shown. Charged states are labelled with the most likely $J^P$ assignment. Colours identify the grouping of the states within the previous paragraphs: red for the individual states (Secs.~\ref{sec:x3872} and \ref{sec:otherp}), green for the 3940 family (Secs.~\ref{sec:3940}), blue for the states around $M = 4140$~MeV (Sec.~\ref{sec:otherp}), purple for the $1^{--}$ states{Sec.~\ref{sec:1mm}), and orange for the charged states (Sec.~\ref{sec:charged}),
Black states are traditionally considered as regular charmonium, although $\psi(4040)$ 
seems to behave as an exotic particle since it does not match any potential model and is close in mass to a charged state with the same $J^P$. In summary of the observed states above open charm threshold only the $X(4160)$, $Z(3940)$, $Y(3940)$, and $\psi(3770)$ are good candidates for regular charmonium: $\eta_c(3S)$, $\chi_{c,2}(2P)$, $\chi_{c,0}(2P)$, and $\psi(1D)$ respectively. 

\begin{figure}[bht]
\begin{center}
\epsfig{file=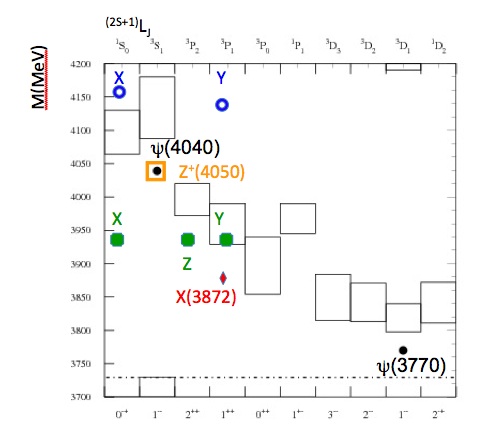,height=5.6cm}
\epsfig{file=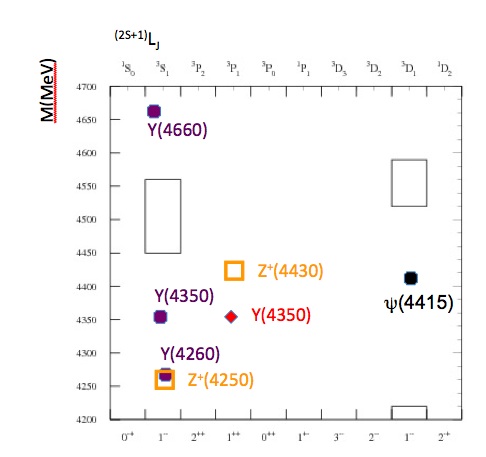,height=6cm}
\caption{\it Observed states with hidden charm above the open charm threshold. The theory predictions are according to the potential models 
described in Ref.~\cite{Brambilla:2004wf}. 
 }
\label{fig:charmoniumSummary}
\end{center}
\end{figure}

Besides observing an increasing number of exotic states we are still far from having clear assignments for each of the states between different possible interpretations. Moreover, a new spectroscopy implies the existence of a large number of states, whose absence would have to be justified. 
A perception of the status of the global picture and an indication on where to search can be obtained from a comparison between possible spectra and observed states.

Expected tetraquark spectra were derived, under assumptions which lead to uncertainties $O$(100~MeV), in Sec.~\ref{sec:th:spectra}. Fig.~\ref{fig:tetraspectraexp0} and Fig.~\ref{fig:tetraspectraexp1} show a comparison between the expected and observed spectra for the first two radial excitations of $J=0^+,1^-$, and $1^+$ spectra and the observed states. Besides the $X(3872)$ which is assumed to be a tetraquark when building the model and that would naturally be constituted by two states close in mass, the following states have a match within the 100 MeV : $Y(4350)$, $Z(4430)$, $X(4160)$, $Y(4260)$, $Y(4350)$ and $Y(4660)$. 
Such a small number of matched states is opposed to the large number of needed states: 18 (27) for each of the $J=0 (1)$ multiplets. While it can be argued that production and decay mechanism can differentiate between the states and that experimental sensitivity differs significantly between final states (for instance $D_s$ mesons are much more difficult to detect than $D$ mesons), one striking observation is that no attempt is made to search for the strange states, that would decay into charmonium plus a kaon.

\begin{figure}[bht]
\begin{center}
\epsfig{file=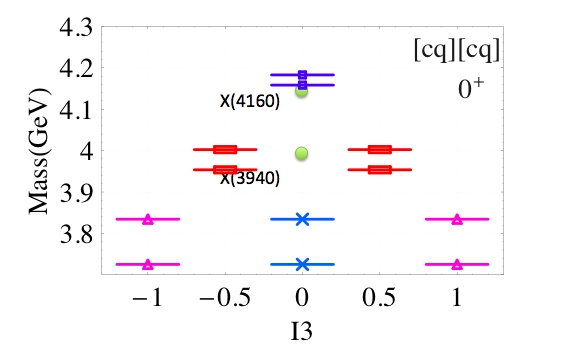,height=4.5cm}
\epsfig{file=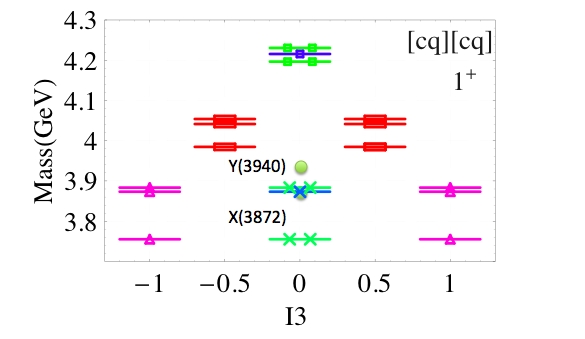,height=4.5cm}
\epsfig{file=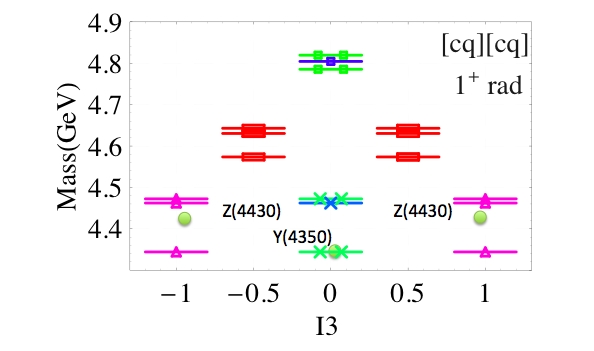,height=4.5cm}
\caption{\it Expected tetraquark spectra as derived in Sec.~\ref{sec:th:spectra} for the first and second radial excitations of states with $J^P=0^+,1^+$ with the observed states superimposed.  Color coding is the following: $[cl][\bar{c}\bar{l}]$, where $l=u,d$ are purple triangles (charged), light green double crosses (neutral with $C=+$), and light blue crosses  (neutral with $C=-$); $[cl][\bar{c}\bar{s}]$ and charge conjugates are red boxes; $[cs][\bar{c}\bar{s}]$ are dark green double squares($C=+$), and dark blue  squares ($C=-$).
 }
 \label{fig:tetraspectraexp0}

\end{center}
\end{figure}

\begin{figure}[bht]
\begin{center}
\epsfig{file=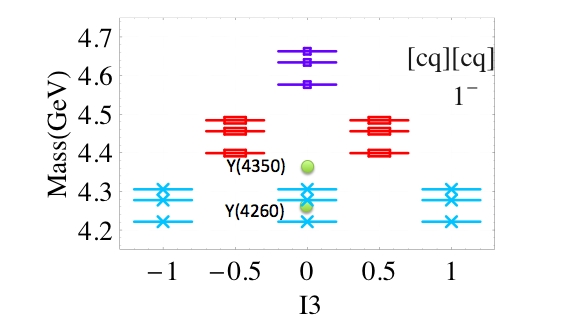,height=4.5cm}
\epsfig{file=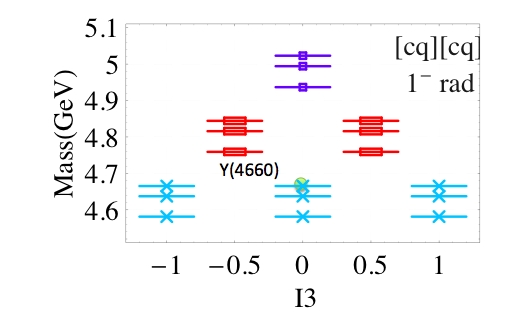,height=4.5cm}
\caption{\it Expected tetraquark spectra as derived in Sec.~\ref{sec:th:spectra} for the first and second radial excitations of states with $J^P=1^-$ with the observed states superimposed.  Color coding is the following: $[cl][\bar{c}\bar{l}]$, where $l=u,d$ are blue crosses,  $[cl][\bar{c}\bar{s}]$ and charge conjugates  red boxes, and $[cs][\bar{c}\bar{s}]$ are purple squares. Different values of $C$ are degenerate in the model and therefore they are not separated in the plot.
 }
\label{fig:tetraspectraexp1}
\end{center}
\end{figure}

Predictions in the case of molecules are more difficult. It is easy to classify the masses around which a molecule could possibly lay by computing all sums of the masses of two mesons with correct quantum numbers (see Fig.~\ref{fig:molespectraexp}). Here all pairs of either a charmonium  and a light meson, neglecting the scalar nonet which is here treated as a tetraquark itself and states with a width larger than 50 MeV, or two open charm mesons. Each case should be considered separately to estimate the production cross sections and the binding energies.Since the molecules have masses lower than the sum of the constituent mesons it is interesting to search for their decays in the final states at lower masses. 

Among the observed state the  $X(3872)$ is the only state that matches a threshold within 10 MeV.
\begin{figure}[bht]
\begin{center}
\epsfig{file=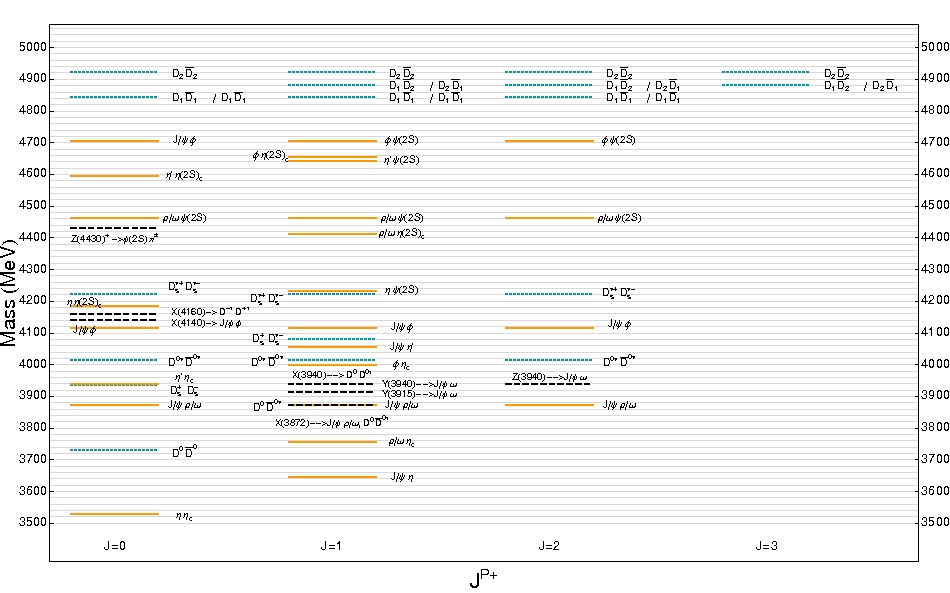,height=7cm}
\epsfig{file=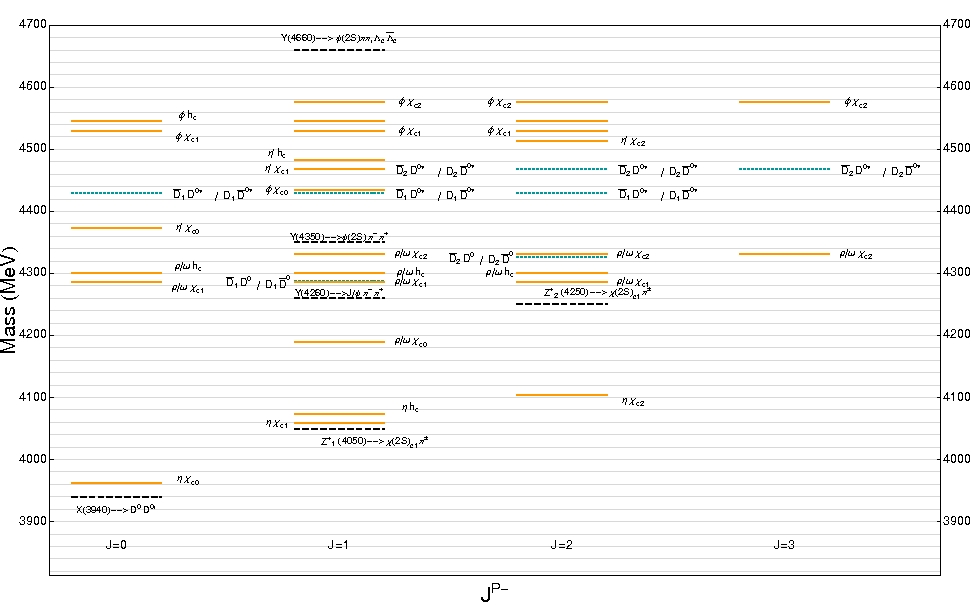,height=7cm}
\caption{\it Sum of the masses of any pair made by either a charmonium and a light meson (blue dotted lines) or two open charm mesons (orange full line) for $C=+$(top) and $C=-$(bottom).
The assigned $J^{PC}$ are computed assuming $L=0$ and null charge. Black dashed lines represent the observed exotic charmonium states.
 }
\label{fig:molespectraexp}
\end{center}
\end{figure}

%% file: bottomonium.tex
\section{Bottomonium }

The discovery of exotic charmonium resonances that do not fit in the
standard charmonium scheme suggests the presence of their bottom
companions, with similar properties.

The interest on this field has been boosted by the Belle discovery of an anomalously large
$\Upsilon(nS) \pipi$ ($n=1,\,2$)  production around the $\Upsilon(5S)$
resonance~\cite{belle_Y5sYpipi}. 
Since the $\Upsilon(nS) \pipi$ channel is the equivalent, in the
bottom sector, of the
$\psi(nS) \pipi$ channels preferred by some of the exotic charmonia, the
interpretation of this result as the effect of an exotic bottomonium
with a mass around the $\Upsilon(5S)$ peak is quite natural~\footnote{One could
also hypothesize the presence of an exotic component mixed with the
$\Upsilon(5S)$ to form a single resonance.}.

The current predictions for tetraquarks with bottom content have been already presented in
Sec.~\ref{sec:theory:tetraquarks}. Here we report the
searches performed by Belle and BaBar (Sec.~\ref{sec:bottom:searches}) and review some interpretations of the
experimental results (Sec.~\ref{sec:bottom:interpretations}).



\subsection{The search for exotic bottomonium}
\label{sec:bottom:searches}

Searches for exotic bottomonia have been performed at the $B$-Factories
exploiting \emph{inclusive} and \emph{exclusive} techniques. In the
inclusive searches, structures are searched in the $\epem \to
hadrons$ cross section, looking in particular at the hadronic ratio $R
= \sigma(\epem \to hadrons) / \sigma(\epem \to \mumu)$. In the
exclusive analysis, specific final state channels are searched for,
with a particular attention devoted to the bottom equivalents of the
decay channels preferred by the exotic charmonia, in particular $\Upsilon(nS) \pipi$.

\subsubsection{Spectroscopy from the Belle $\Upsilon(5S)$ campaign}
\label{sec:bottom:belle_Y5s}

\begin{figure}
\begin{center}
\epsfig{file=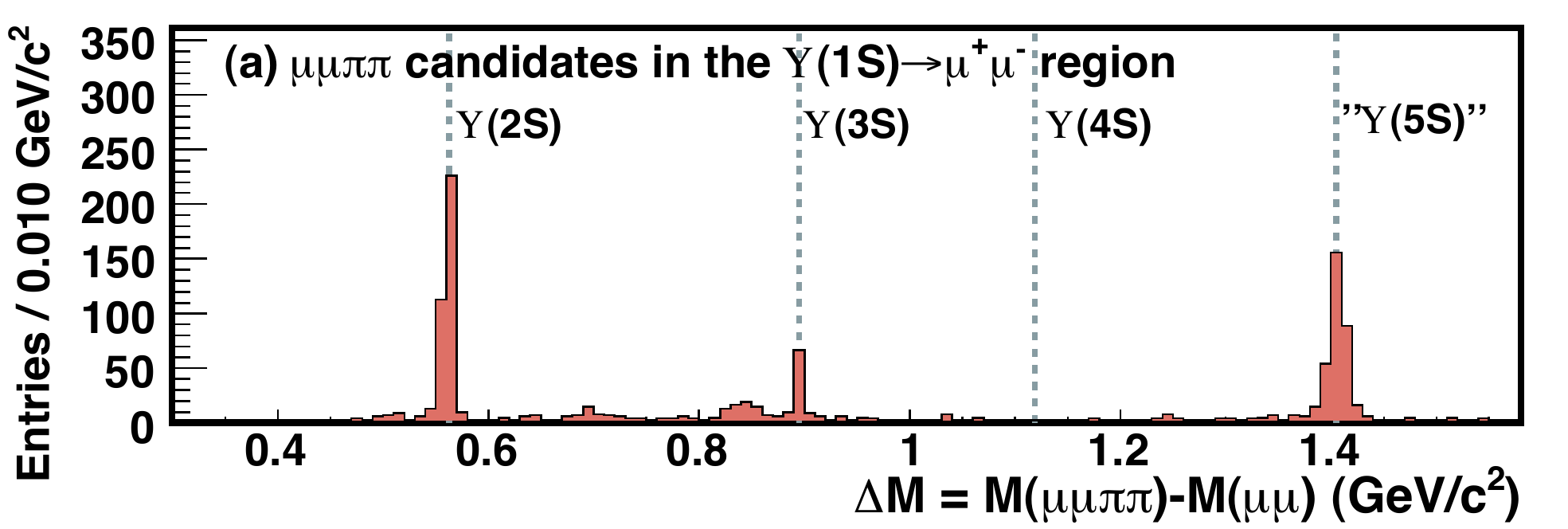,width=12cm}
\epsfig{file=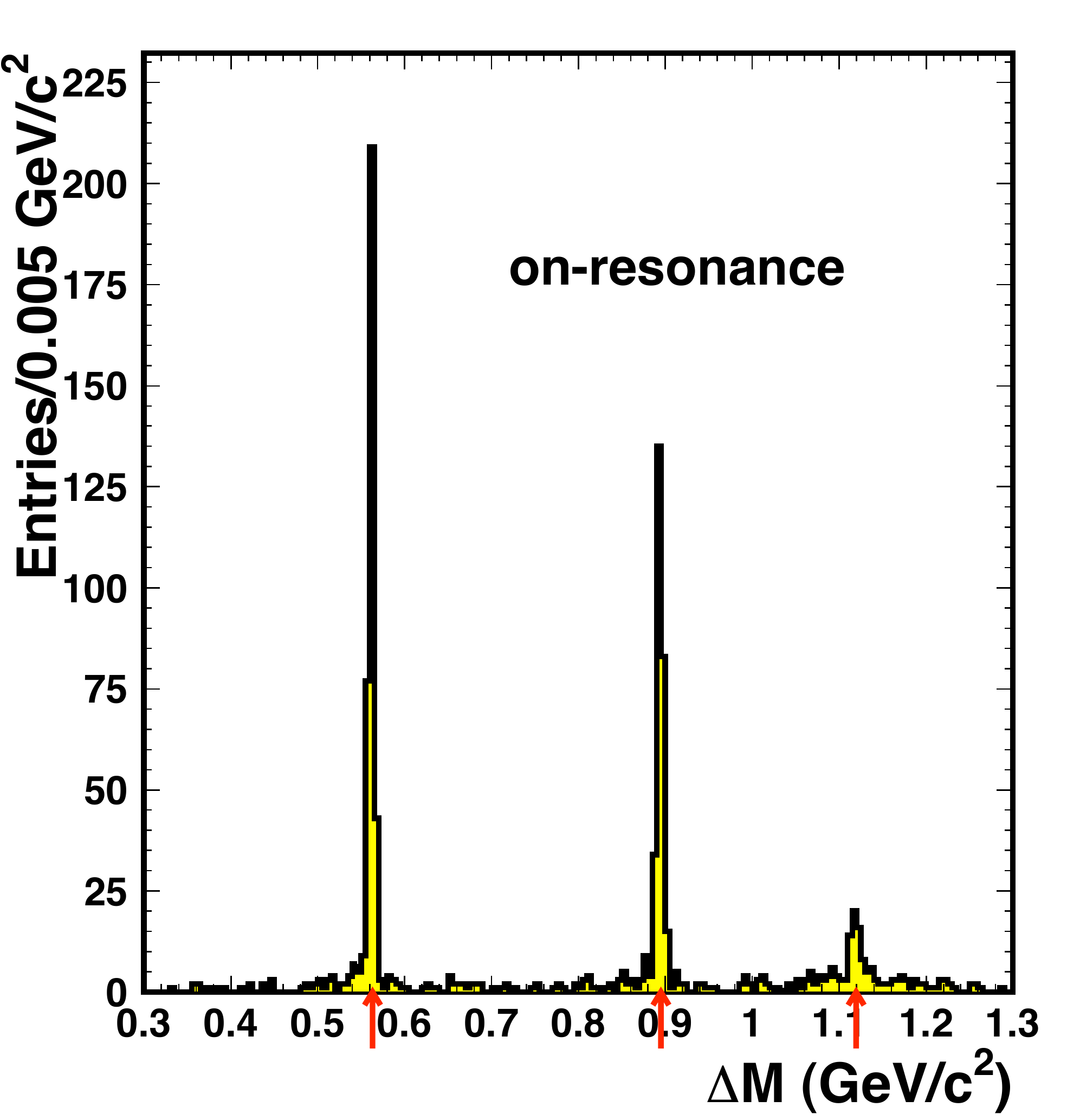,height=5cm}
\caption{Distribution of $\Delta M = M(\mumu\pipi) - M(\mumu)$ obteined at the center of mass energy of the $\Upsilon(5S)$ (top), compared
with the distribution of the same variable at the $\Upsilon(4S)$ (bottom), where the three arrows indicate, from the left, the $\Upsilon(2S)$,
$\Upsilon(3S)$ and $\Upsilon(4S)$ peak.}
\label{fig:bottom:upspipi}
\end{center}
\end{figure}

We already mentioned the observation, made by the Belle collaboration,
of an anomalously large $\Upsilon(nS) \pipi$ around the
$\Upsilon(5S)$~\cite{belle_Y5sYpipi}.
This analysis is based on the first data ($21.7~\invfb$) collected at the center of mass energy of the 
$\Upsilon(5S)$ resonance. The authors look for the $\Upsilon(nS) \pipi$ final states, possibly produced in
association with an ISR photon, by selecting events with four tracks,
two of which are identified as muons and have an invariant mass consistent
with the $\Upsilon(nS) \to \mumu$ hypothesis. If a resonance of
mass $M$ decays into $\Upsilon(nS) \pipi$, the
distribution of the invariant mass difference $\Delta M = M(\mumu\pipi)
- M(\mumu)$ is expected to peak at $\Delta M = M -
M_{\Upsilon(nS)}$. A sharp peak corresponding to $M \sim
M_{\Upsilon(5S)} \sim 10.865$ GeV is found in both $\Upsilon(1S) \pipi$ and
$\Upsilon(2S) \pipi$ channels, which is inconsistent with the
naive expectations from the scaling of the $\Upsilon(4S)$ rates~\cite{belle_Y4sYpipi}, as shown in Fig.~\ref{fig:bottom:upspipi}.

Subsequently, the Belle collaboration performed an energy scan with
$\sqrt{s}$ between 10.83 and 11.02 GeV, in six steps, for a total integrated luminosity
of $\sim 7~\invfb$~\cite{belle_scan}. In this case, the authors look
for events around $\Delta M = \sqrt{s} - M_{\Upsilon(nS)}$ (i.e. without
ISR production), in order to measure the $\Upsilon(2S) \pipi$ rate
in each of the six energy points and determine the resonance shape. 
The analysis confirmed the large rate observed in the previous test, 
adding also the observation of the $\Upsilon(3S) \pipi$
channel. The combined fit of the resonance shapes with a single
Breit-Wigner PDF, interfering with a flat continuum, gives a mass of $10.8884^{+0.0027}_{-0.0026} \pm
0.0012$ GeV and a width of $30.7^{+8.3}_{-7.0} \pm 3.1$ MeV, that differ by about 3$\sigma$
and 5$\sigma$ respectively from the values quoted for the $\Upsilon(5S)$ 
in the PDG~\cite{PDG}. This observation triggerd the hypothesis of two near but separated
structures around 10.88 GeV.

\subsubsection{The BaBar scan above the $\Upsilon (4S)$}

Before the works by Ali \emph{et al.}, Ebert \emph{et al.} and Wang \emph{et al.} were published, a naive scaling 
of the exotic charmonium masses to the bottom case suggested the 
presence of exotic bottomonia in the region between the $\Upsilon(4S)$ (10.58 GeV)
and the $\Upsilon(6S)$ candidate ($\sim 11$ GeV). In particular, it is 
well known that the masses for the regular bottomonia can be obtained, 
at a good approximation, by scaling the masses of the regular charmonia
by the mass difference between the $\Upsilon(1S)$ and the $J/\psi$.
If it is assumed to work also for the exotic mesons, the masses of the 
$Y(4260)$, $Y(4350)$ and $Y(4660)$ can be scaled to ``predict'' three 
exotic $1^{--}$ bottomonia at 10.62, 10.71 and 11.02 GeV, that could
be directly produced in $e^+e^-$ collisions. The BaBar collaboration decided 
to investigate this possibility by performing an energy scan between 
10.54 and 11.20 GeV. Steps of 5 MeV were performed, by collecting about
25~\invpb per step, for a total of 3.3~\invfb. An additional data set of 600~\invfb
was also collected in 8 steps around the $\Upsilon(6S)$ candidate (10.96 to 11.10 GeV).
This scan overtakes by a factor 30 in statistics and a factor 4 in step fineness the previous 
one, performed at Cornell~\cite{cusb_scan,cleo_scan}.
The search for exotic bottomonia can be performed in these data set exploiting 
both inclusive and exclusive reconstruction. 

At present, only the results of
the inclusive analysis have been published~\cite{babar_scan}. Hadronic final
states are selected by requiring at least three tracks, with a reconstructed energy 
greater than 4.5 GeV, while event shape variables are used to reject the QED
background. Di-muon events are selected by looking or two tracks, collinear
within $10^\circ$, with an invariant mass of at least 7.5 GeV. The \bbbar$(n \gamma)$
contribution $R_{b}$ to the total hadronic ratio is measured, by subtracting 
from the total hadronic yield the contribution from continuum ($\epem \to \qqbar$, $q = u,d,s,c$) and 
two-photon ($\epem \to \epem \gamma^* \gamma^* \to \epem X_h$) events,
estimated below the \BB threshold and scaled according to the expected
$\sqrt{s}$ dependence, while the ISR ($\epem \to \Upsilon(nS) \gamma_{ISR}$) 
contribution is included in $R_{b}$. The result of the analysis is showed in 
Fig.~\ref{fig:xsec_exotics}. The authors also performed a simple fit in the 
region between 10.80 and 11.20 GeV, including two Breit-Wigner resonances
and a flat continuum background, allowing a partial interference between the
three components: $\sigma = |A_{nr}|^{2} + |A_r + A_{10860}BW(M_{10860},\Gamma_{10860}) + A_{11020}BW(M_{11020},\Gamma_{11020})|^{2}$. 
The measured $\Upsilon(5S)$ mass and width are $10.876 \pm 0.002$ GeV and $43 \pm 4$ MeV 
respectively, but are affected by large systematic uncertainties related to choice of the fit model. 
Recently, the Belle collaboration updated the analysis of~\cite{belle_scan}, by including a measurement of 
$R_{b}$ and a fit that replicates the BaBar fit and confirms these results. At present, the Belle's analyses provide 
a discrepancy of only 2.2$\sigma$ (corresponding to 9 MeV) and 1.4$\sigma$ in mass and width
between the inclusive and the exclusive measurement.


\subsection{Interpretations of bottomonium results}
\label{sec:bottom:interpretations}

The interpretation of the results obtained so far in the search for exotic bottomonia is 
still quite controversial.  On one side, the observation of a large $\Upsilon(nS) \pipi$
rate around the $\Upsilon(5S)$ is in contrast with a naive rescaling of the corresponding
$\Upsilon(4S)$ rate, and can suggest the presence of an exotic resonance lying near the
$\Upsilon(5S)$; this hypothesis is also strengthened by the study of the invariant dipion mass spectra 
and the angular distributions of the observed decay~\cite{Ali_Ypipi}. On the other hand,
although a well established model for this decay of the standard bottomonium is still
missing, some works have shown that it is possible in principle to bring back
the Belle observation into the standard scheme~\cite{meng_Ypipi,simonov_Ypipi}.
In the Meng and Chao's work a rescattering model is used, where a $\Upsilon(5S)$ 
first decays into a \BB pair, then a $B$ meson exchange produces a lower $\Upsilon$
and a scalar resonance ($\sigma$ or $f_0$)  decaying into \pipi. Instead, in
the Simonov and Veselov's paper, an effective Lagrangian for \qqbar, $\qqbar \pi$ and $\qqbar \pi \pi$
production is introduced. In both cases, parameters are fixed independently
of the $\Upsilon(5S)$ measurements (from charmonium or lower $\Upsilon$'s)
and a large $\Upsilon(5S) \to \Upsilon(nS) \pipi$ rate is predicted, in rough agreement
with the experimental results.

The observation of a significant difference between the measured mass and width
of the $\Upsilon(5S)$ in inclusive and exclusive decays is also not surprising even
in the standard framework. A proper description of the hadronic ratio $R$ near the
open charm and bottom thresholds requires to take into account the mixing of the
pure $\Upsilon(nS)$ states, introduced by their coupling with the OZI-allowed states 
$D^{(*)}D^{(*)}$ and $B^{(*)}B^{(*)}$~\cite{eichten_charmonium, tornqvist_bottomonium}.
Such a kind of treatment, for instance, allowed  to predict the two structures
observed by BaBar between 10.60 and 10.75 GeV without introducing any exotic
resonance. An important conclusion of these studies is that the total hadronic cross 
section is not expected to be a simple superposition of Breit-Wigner resonances, and
significant differences can arise between the inclusive and exclusive shapes of the 
cross section. Hence simple fits based on a superposition of Breit-Wigner
resonances are not well grounded and the results are difficult to interpret,
and in absence of a consistent treatment of the inclusive and exclusive 
processes within the same theoretical framework, no conclusion can be drawn
from the small differences observed in the shapes.

Recent efforts to better understand the hadronic ratio, in particular in the region 
between the $\Upsilon(5S)$ and the $\Upsilon(6S)$, have been also performed 
by Ali \emph{et al.}~\cite{Ali_spectrum} and by van Beveren and Rupp~\cite{vanBeveren}.
In particular, in the Ali \emph{et al.} paper the exotic resonance responsible for 
the large $\Upsilon(nS) \pipi$ production is identified with a possible peak
noticed in the BaBar $R_b$ measurement around 10.91 GeV, although from
an experimental point of view it doesn't seem to have enough statistical 
significance to claim an observation. Instead, in van Beveren and Rupp's
paper, a quite uncommon interpretation of the BaBar result is presented:
the peak at 10.58 GeV, usually identified with the $\Upsilon(4S)$ resonance,
is interpreted as a threshold effect, while the $\Upsilon(4S)$ is found at 10.735 GeV; 
moreover, the peaks around 10.88 and 11.02 GeV, usually assigned to the 
$\Upsilon(5S)$ and $\Upsilon(6S)$, are identified with the $\Upsilon(3D)$ 
and the $\Upsilon(5S)$, respectively.

\begin{figure}
\begin{center}
\epsfig{file=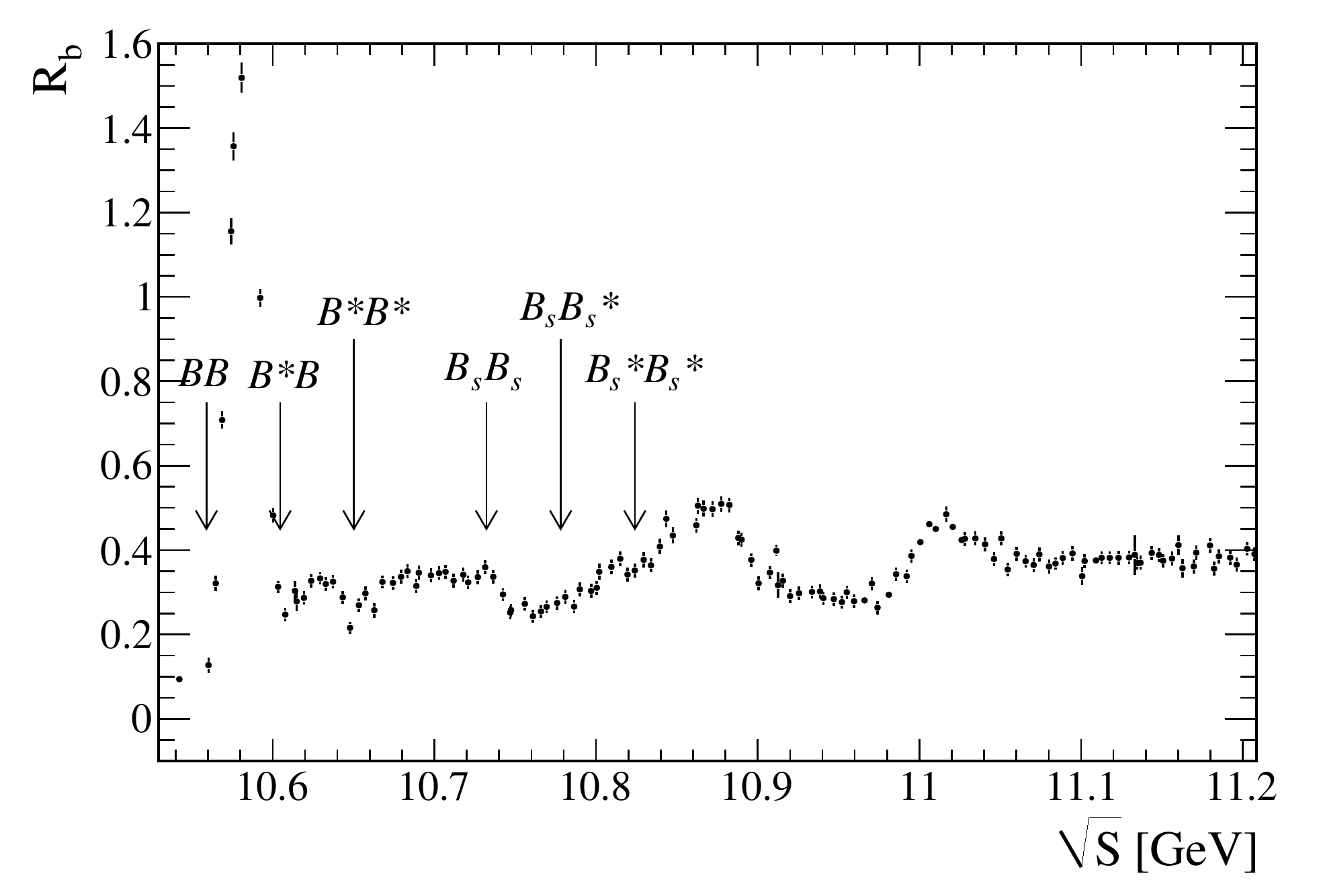,width=6cm}
\epsfig{file=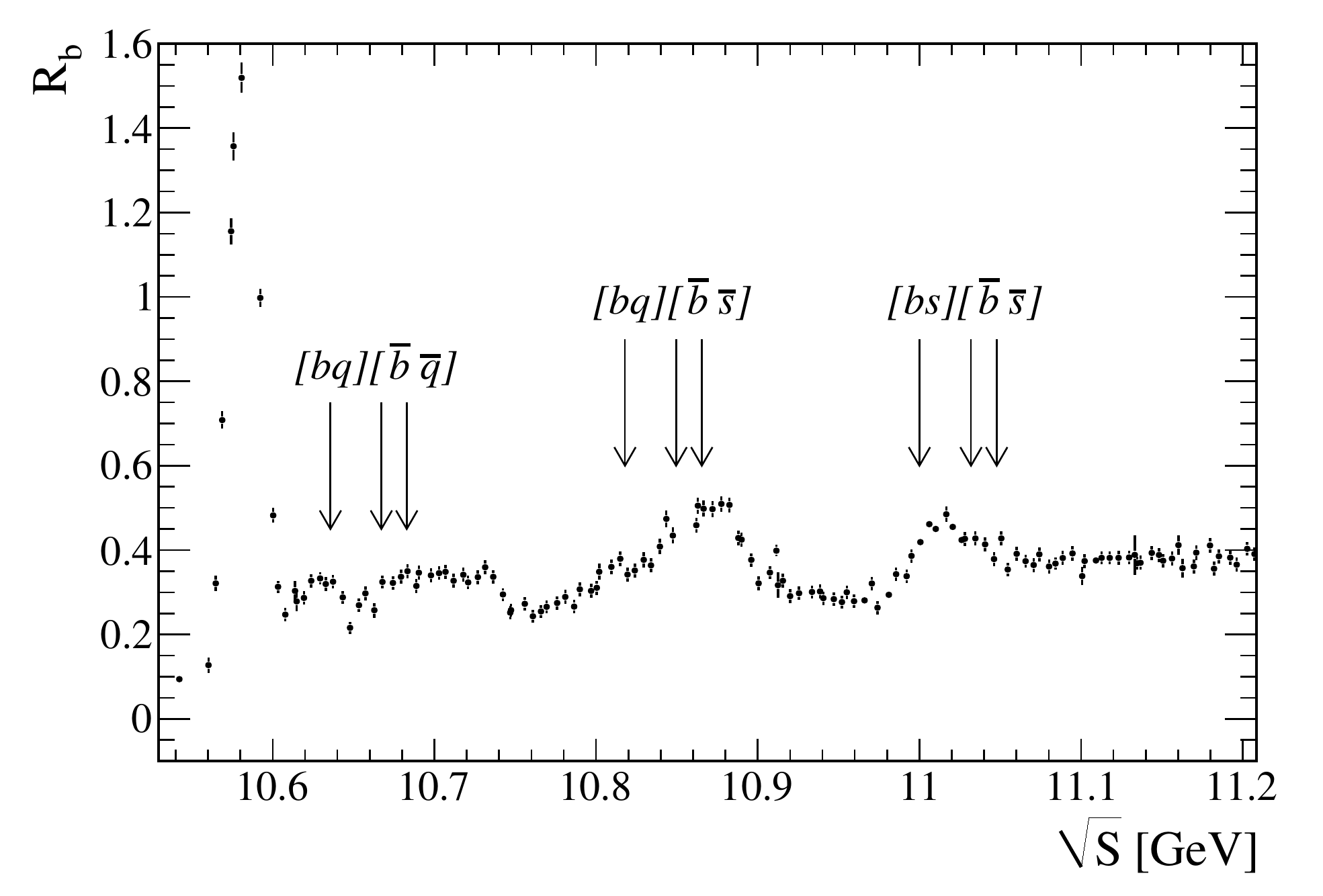,width=6cm}
\caption{$B \overline B$ thresholds (left) and expected tetraquark $1^{--}$ states (right), superimposed to the $R_b$ spectrum measured by the BaBar
collaboration.}
\label{fig:xsec_exotics}
\end{center}
\end{figure}

Finally, in Fig.~\ref{fig:xsec_exotics} we illustrate the $B \overline B$ thresholds in whose neighborhood molecular states
could be expected, and the tetraquark states predicted in this work (see Fig.~\ref{fig:bqbql1}), superimposed to the $R_b$ spectrum measured
by the BaBar collaboration.

%% file: conclusion.tex
\section{Outlook and conclusions }
It is now almost seven years since the first heavy quarkonium exotic states has been observed and the number of exotic states has significantly increased. In order to explain all these states one needs to go beyond the assumption that mesons can only be made of a quark and an anti-quark and consider the possibility to be observing bound states of four quarks or two quarks and valence gluons. These novel states of aggregation are not a new idea and it was indeed already in the foundations of the quark model~\cite{GellMann:1964nj}. Also, the hypothesis that the mesons of the scalar nonet are actually tetraquarks  can be dated back to the '70s~\cite{Jaffe:1976ig}. But only now, with the presence of several exotic candidates with heavy quarks the possibily to have a global picture is becoming real.

Unfortunately, such a spectroscopy made of a larger number of constituents (four quarks for tetraquarks and molecules and two quarks and gluons for hybrids) is much reacher than standard spectroscopy and therefore a significantly larger number of states could potentially exist. To the aim of building this global picture, a complete set of theoretical predictions and a systematic experimental search is needed. The work in the theoretical path is progressing although  strong interactions are extremely hard to compute, even in presence of heavy quarks, and therefore predictions are extremely hard and uncertain. From the experimental point of view the amount of work ahead is if possible even larger. 

\begin{sidewaysfigure}[bth]
\begin{center}
\epsfig{file=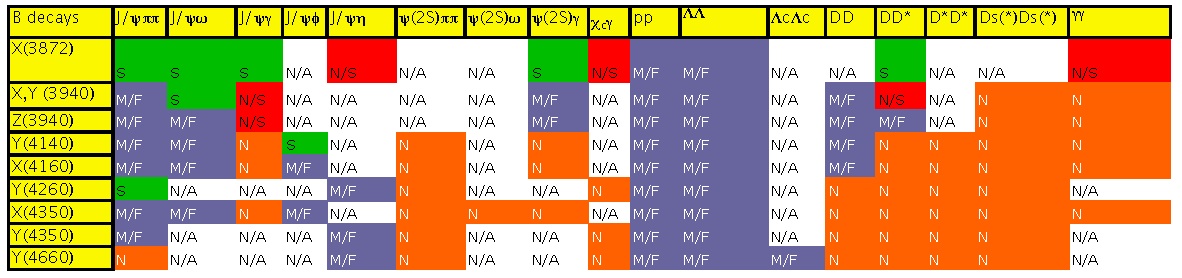,height=5cm}
\epsfig{file=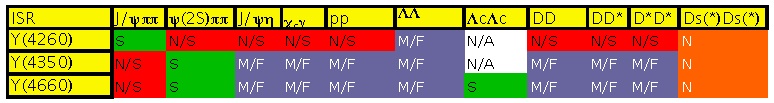,height=2.5cm}
\caption{\it Status of the searches of the new states  in the processes $B\to X K$ (top) and $e^+e^-\to X\gamma_{ISR}$, $X\to f$, for several final states $f$. 
Final states where each exotic states were observed ("S") or excluded ("N/S") are indicated. A final states is marked as "N" if the analysis has not been performed in a given mass range and with "M/F" if the spectra are published but a fit to a given state has not been performed. Finally "N/A" indicates that  quantum numbers forbid the decay and "N/F" if an analysis is experimentally too challenging.}
\label{fig:obsmatr1}
\end{center}
\end{sidewaysfigure}

\begin{sidewaysfigure}[bht]
\begin{center}
\epsfig{file=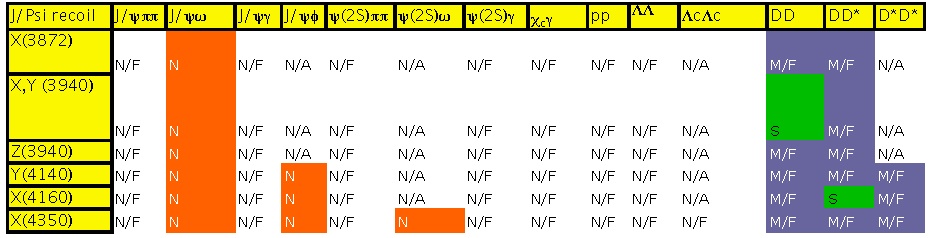,height=5cm}
\epsfig{file=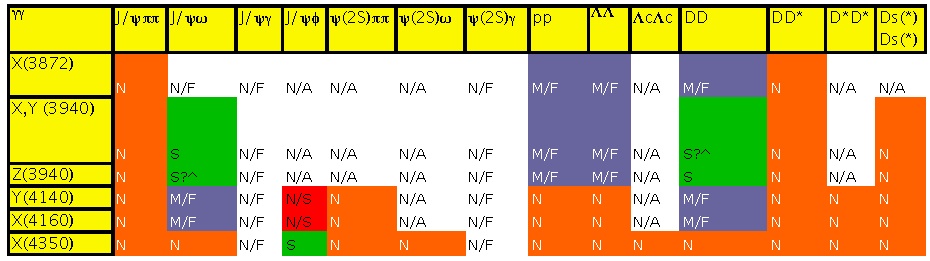,height=5cm}
\caption{\it Status of the searches of the new states in  collisions $e^+e^-\to X J/\psi$(top) and $\gamma \gamma\to X$ (bottom), $X\to f$, for several final states $f$. 
Symbols are explained in the caption of Fig.~\ref{fig:obsmatr1}.}
\label{fig:obsmatr2}
\end{center}
\end{sidewaysfigure}

\begin{sidewaysfigure}[bht]
\begin{center}
\epsfig{file=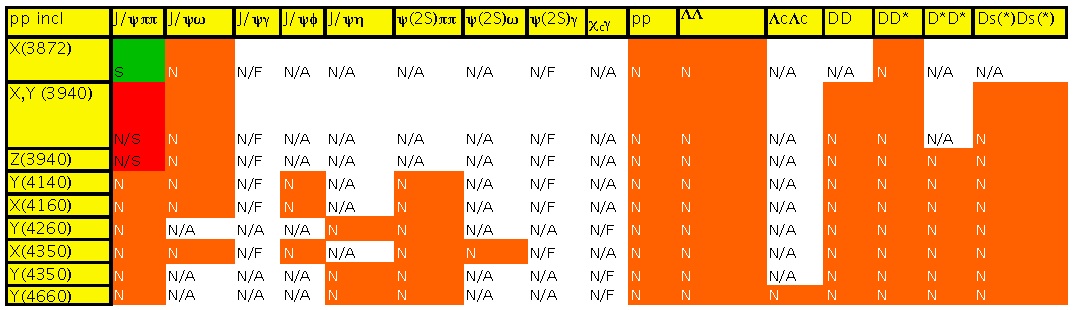,height=6cm}
\epsfig{file=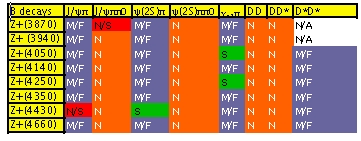,height=5cm}
\caption{\it Status of the inclusive searches of the new neutral states in $p\bar{p}$ collisions (top) and of the  new charged states (bottom) for several final states. 
Symbols are explained in the caption of Fig.~\ref{fig:obsmatrch}.}
\label{fig:obsmatrch}
\end{center}
\end{sidewaysfigure}

The status of the experimental observations is extremely fragmented. Pictorial summaries, separated by production mechanisms, are in Figs.~\ref{fig:obsmatr1}-\ref{fig:obsmatrch}.
The exotic states have been observed always in only one production mechanism and often in only one final state. None of the states has been searched systematically in all final states. Sometimes the analysis of a given final state for a given production mechanism is missing, but often it has either been performed only in a limited mass range or the invariant mass spectrum has been published without a fit to the possible new state. This is mostly due to the fact that some of the analyses did not show a significant signal themselves and they were published before a new state was observed. 
As an example, the invariant mass spectra of a charmonium and a photon in Fig.~\ref{fig:Xspectra} are focussed on the $X(3872)$ region and no information is available outside it. On the other side in the same Fig.~\ref{fig:Xspectra}, the  $J/\psi\eta$ invariant mass spectrum is published, but not fitted for all the possible new states. Whatever the cause is, the critical point is that building a global picture requires having either an observation or a limit in each final state for each candidate new state: limits allow to quantify the level at which a decay mode has not been observed and can show whether a signal could have been observed in a final state or the efficiency, the branching fraction or the background level would make it unobservable. It is also worth remarking that although finding new decay modes of these states might not necessarily cast light on the nature of these particles, they would in any case concur to the evidence of the existence of the states that suffer from lack of statistics.

Looking into this "observational" tables in detail, $B$ decays (top of Fig.~\ref{fig:obsmatr1}) are the most studied, but there is a significant amount of missing fits ("M/F"). Particularly severe is the lack of analysis of the baryonic spectra ($p\bar{p}$, shown in Fig.~\ref{fig:charmonbaryons}, and $\Lambda{\bar{\Lambda}}$) since baryonic decays are a signature of tetraquark states. Some other modes have never been studied, mostly because the number of expected events is very low. Nonetheless the study of $B\to \psi(2S)\pi\pi K$ decays should be relatively clean, while $D^*\bar{D}^*$ and above all $D_s^{(*)}\bar{D}_s^{(*)}$ suffer from the low branching fractions of the observed states.

States produced in conjunction with an initial state radiation (ISR) photon  (bottom of Fig.~\ref{fig:obsmatr1})  have an unambigous $J^{CP}$ assignment and are therefore less analyses are needed to establish their properties. Nonetheless it is striking to see that a large fraction of analyses have been carried out exclusively for the first observed exotic state, the $Y(4260)$. It can also be noticed that no search is published involving $D_s^{(*)}$ mesons: while the efficiency is expected to be very low, background should be low as well and surprises can always arise. Finally the $Y(4660)$ has been object of one of the combined analyses we are advocating here~\cite{Cotugno:2009ys}: two states apparently different, observed in $\psi(2S)$ and $\Lambda_c{\bar{\Lambda_c}}$ final states, if fitted under the same ansatz were found to be consistent with being the same and interesting ratios of branching fractions were measured.

On the recoil of a $J/\psi$ and in $\gamma\gamma$ interactions  ( Fig.~\ref{fig:obsmatr2} )  only $C=+$ neutral states can be observed. This restricts the number of final states of interest. Also, the low multiplicity of these decays and the large missing momentum in the case of $\gamma\gamma$ decays makes these analyses experimentally challenging. On the other side $C=+$ states are the least known and reinforcing the evidence of the signals would help.
It is also interesting to notice that, mostly due to statistics, the recoil to any other particle but the $J/\psi$ has not been investigated. Given the selection rules the recoil to $\chi_{c0}$ or $\chi_{c2}$ would be very interesting. 

Searches at hadron colliders  (top of Fig.~\ref{fig:obsmatrch}) have the advantage to have very large samples. On the other side backgrounds are very high and therefore final states with too high multiplicity and above all with neutral particles are not at reach.  Finally, it is hard to extract information on production cross sections and therefore branching fractions. Nonetheless, it is clear that a systematic search in $p\bar{p}$ collisions is likely to clarify the picture significantly and it is a pity that it is still missing. This holds also and in particular for the charged states -- for instance the longly debated $Z^+(4430)$ should give a large signal in the $J/\psi\pi^+$ spectrum -- and in the bottomonium -- where no other experiment is in the position of studying inclusive $\Upsilon(nS)\pi^+\pi^-$ and  $\Upsilon(nS)\pi^\pm$ spectra.

 Concerning the searches of charged exotic states, the most striking signature of states made of more than two quarks, very few searches have been conducted in $B$ decays. We believe that for each exotic neutral spectrum the corresponding charged state should be searched for completeness and, as shown in the bottom of Fig.~\ref{fig:obsmatrch} only five combinations of final states and exotic states has been searched for. As an example, no information has been extracted from Fig.~\ref{fig:zstates} on the charged partner of the $X(3872)$, $Z(3870)$ in our table, which has long been pointed out as a critical state to search for.  Moreover among the four quark bound states there must be states that contain a single $s$ quark (see for instance Fig.~\ref{fig:tetraspectraexp0} and Fig.~\ref{fig:tetraspectraexp1}), their mostly distinctive signature being a charmonium plus a charged kaon. These states could be searched in $B$ decays in association with a $s\bar{s}$ state or inclusively in $p\bar{p}$ collisions.
 
 Concerning the bottomonium states, the searches are extremely limited and there no confirmed evidence of exotic states yet. The search potential of $B$-Factories is limited by the fact that bottomonium at masses higher than the open bottom threshold can only be produced in conjunction with initial state radiation and that in addition the accessible mass range is full of threshold openings (see Sec.~\ref{sec:bottom:interpretations}). This restricts the search to $J^{PC}=1^{--}$ states with significant signatures in exclusive final states. Besides the lack of information about most of the possible final states, the real contribution could come from hadron colliders, where the search could be inclusive over $\Upsilon(nS)X$ final states (where X is made of charged tracks) and extended to exotic states with open bottom, eventually even with open charm.
 
 In conclusion, seven years of discoveries of exotic particles need to be followed by  a systematic study of the possible new spectroscopy to be able to give a global and definite picture. To this aim still a lot can be extracted from existing data of $B$-Factories and Tevatron. Nonetheless in the present generation of experiments statistics is extremely low and we will not be able to have a convincing picture until the results of ultra-high intensity machines, LHC and the Super Flavour Factories (SuperB and/or SuperBelle) , will be available.